\documentclass[12pt]{article}

\newcommand{\VersionInformation}{}  
\InputIfFileExists{Debug}{}{}

\usepackage{amsmath}
\usepackage{amsthm}
\usepackage{amsfonts}          
\usepackage{amssymb}           
\usepackage[mathscr]{eucal}
\usepackage{graphicx}
\usepackage{color}
\usepackage{hypbmsec}

\usepackage{rotating}
\usepackage{slashbox}
\usepackage[isu,bf]{caption}
\newlength{\xtrawidth}
\newlength{\xtraheight}
\usepackage[all]{xy}           

\usepackage{xr-hyper}

\newcommand{\partA}{Part~A}
\newcommand{\partB}{Part~B}

\newcommand*{\xautoref}[2][\xrlab]{\autoref{#1-#2}}

\ifx\NoBackreferences\EMPTYMACRO
  \usepackage[backref,linktocpage,bookmarks]{hyperref}
\else
  \usepackage[linktocpage,bookmarks]{hyperref}
\fi

\ifx\DEBUG\EMPTYMACRO
\else
  \usepackage{showlabels}
\fi

\newcommand{\lrstack}[3][t]{
  \ensuremath{
    \begin{array}[#1]{c}
      \multicolumn{1}{l}{\displaystyle{#2}\quad} \\[0.5em] 
      \multicolumn{1}{r}{\displaystyle\quad{#3}}
    \end{array}
  }}
\def\clap#1{\hbox to 0pt{\hss#1\hss}}
\def\mathllap{\mathpalette\mathllapinternal}
\def\mathrlap{\mathpalette\mathrlapinternal}
\def\mathclap{\mathpalette\mathclapinternal}
\def\mathllapinternal#1#2{%
\llap{$\mathsurround=0pt#1{#2}$}}
\def\mathrlapinternal#1#2{%
\rlap{$\mathsurround=0pt#1{#2}$}}
\def\mathclapinternal#1#2{%
\clap{$\mathsurround=0pt#1{#2}$}}	

\makeatletter
  \def\adots{\mathinner{\mkern2mu\raise\p@\hbox{.}
      \mkern2mu\raise4\p@\hbox{.}\mkern1mu
      \raise7\p@\vbox{\kern7\p@\hbox{.}}\mkern1mu}}
\makeatother

\newcommand{\eqdef}{%
  \mathrel{\lower.1mm
    \hbox{$\stackrel{\lower.424ex\hbox{\scriptsize def}}{=}$}}
}
\newcommand{\Q}{\ensuremath{{\mathbb{Q}}}}
\newcommand{\R}{\ensuremath{{\mathbb{R}}}}
\newcommand{\C}{\ensuremath{{\mathbb{C}}}}

\newcommand{\Z}{\mathbb{Z}}
\newcommand{\CP}{\ensuremath{\mathop{\null {\mathbb{P}}}}\nolimits}

\newcommand{\CY}{Calabi-Yau}
\newcommand{\CYm}{\CY{} manifold}

\newcommand{\MW}{Mordell-Weil}
\newcommand{\MWgrp}{\MW{} group}

\newcommand{\even}{\ensuremath{\mathrm{ev}}}
\newcommand{\odd}{\ensuremath{\mathrm{odd}}}

\newcommand{\ptset}{\ensuremath{\{\text{pt.}\}}}
\newcommand{\iunit}{\ensuremath{\mathrm{i}}}
\newcommand{\Cunits}{\ensuremath{\C^\times}}
\newcommand{\free}{\ensuremath{\text{free}}}
\newcommand{\tors}{\ensuremath{\text{tors}}}

\DeclareMathOperator{\diff}{d\!}
\DeclareMathOperator{\Span}{span}
\DeclareMathOperator{\Pic}{Pic}

\DeclareMathOperator{\tr}{tr}

\DeclareMathOperator{\Id}{id}

\DeclareMathOperator{\img}{img}

\DeclareMathOperator{\Hom}{Hom}

\DeclareMathOperator{\Sing}{Sing}
\DeclareMathOperator{\Li}{Li}

\newcommand{\Rep}[1]{\ensuremath{\mathbf{\underline{#1}}}}
\newcommand{\barRep}[1]{\ensuremath{\overline{\Rep{#1}}}}

\newcommand{\textdef}[1]{{\it #1}}

\newcommand{\Xt}{{\ensuremath{\widetilde{X}}}}
\newcommand{\Xb}{{\ensuremath{\overline{X}}}}
\newcommand{\Ct}{{\ensuremath{\widetilde{C}}}}
\newcommand{\ZZZ}{\ensuremath{{\Z_3\times\Z_3}}}
\newcommand{\Lsheaf}{\ensuremath{\mathscr{L}}}
\newcommand{\Osheaf}{\ensuremath{\mathscr{O}}}
\newcommand{\OsheafXt}{\ensuremath{\mathscr{O}_{\Xt}}}

\newcommand{\dual}{\ensuremath{\vee}}

\newcommand{\At}{{\ensuremath{\widetilde{A}}}}
\newcommand{\Ab}{{\ensuremath{\overline{A}}}}

\newcommand{\LSss}{Leray-Serre spectral sequence}
\newcommand{\CLss}{Cartan-Leray spectral sequence}

\newcommand{\Htate}{\ensuremath{\widehat{H}}}
\newcommand{\CPambient}{\ensuremath{\CP^2\times \CP^1 \times \CP^2}}
\newcommand{\dP}[1]{\ensuremath{dP_{#1}}}
\newcommand{\FPtimes}{\underline{\times}}
\newcommand{\Xhat}{{\ensuremath{\Hat{X}}}}
\newcommand{\Bhat}{{\ensuremath{\Hat{B}}}}

\newcommand{\Fprepotential}{\mathscr{F}}

\newcommand{\Fprepot}[1]{\ensuremath{\Fprepotential_{{#1},0}}}
\newcommand{\FprepotNP}[1]{\ensuremath{\Fprepot{#1}^\text{np}}}

\newcommand{\FprepotXNP}{\FprepotNP{X}}

\newcommand{\FprepotXtNP}{\FprepotNP{\Xt}}

\newcommand{\Kahler}{K\"ahler}
\newcommand{\Kcone}{\ensuremath{\mathcal{K}}}

\newcommand{\ThetaEeight}{\ensuremath{\Theta_{E_8}}}
\newcommand{\mathemph}[1]{\textcolor{red}{\mbox{\boldmath $#1$}}}

\newcommand{\isolongrightarrow}{\ensuremath{\stackrel{\sim}{\longrightarrow}}}
\newcommand{\hooklongrightarrow}{\lhook\joinrel\longrightarrow}

{\end{list}}

{\everymath{\displaystyle\everymath{}}\array}%
{\endarray}

\newtheorem{theorem}{Theorem}

\newtheorem{example}{Example}

\def\IZ{\mathbb{Z}}

\include{pst-plot}
\psset{unit=3 pt}
\def\putlab#1)#2#3{\put#1){\makebox(0,0)[#2]{\small #3}}} 
\def\putlin#1,#2,#3,#4,#5){\put#1,#2){\line(#3,#4){#5}}}
\def\putvec#1,#2,#3,#4,#5){\put#1,#2){\vector(#3,#4){#5}}}
\def\putcx#1,#2){\put#1,#2){\circle*{1.4}}}

\newcommand{\GrantsAcknowledgements}{This research was supported in
    part by the Department of Physics and the Math/Physics Research
    Group at the University of Pennsylvania under cooperative research
    agreement DE-FG02-95ER40893 with the U.~S.~Department of Energy
    and an NSF Focused Research Grant DMS0139799 for ``The Geometry of
    Superstrings'', in part by the Austrian Research Funds FWF grant
    number P18679-N16, in part by the European Union RTN contract
    MRTN-CT-2004-005104, in part by the Italian Ministry of University
    (MIUR) under the contract PRIN 2005-023102 ``Superstringhe, brane
    e interazioni fondamentali'', and in part by the Marie Curie Grant
    MERG-2004-006374.}

\newcommand*{\xrlab}{B}
{
  \makeatletter
  \newcommand{\hbs@tocstring}{}
  \newcommand{\hbs@bmstring}{}  
  \externaldocument[\xrlab-]{Bmodel}  
}

\begin{document}

\begin{titlepage}
  \vspace*{-2cm}
  \VersionInformation
  \hfill
  \parbox[c]{5cm}{
    \begin{flushright}
      hep-th/yymmnnn
      \\
      UPR 1177-T
      \\
      DISTA-2007
      \\
      TUW-07-07
    \end{flushright}
  }
  \vspace*{\stretch{1}}
  \begin{center}
     \Huge 
     Worldsheet Instantons and Torsion Curves\\ 
     Part A: Direct Computation
  \end{center}
  \vspace*{\stretch{2}}
  \begin{center}
    \begin{minipage}{\textwidth}
      \begin{center}
        \large         
        Volker Braun${}^1$,
        Maximilian Kreuzer${}^2$,
        \\
        Burt A. Ovrut${}^1$, and
        Emanuel Scheidegger${}^3$
      \end{center}
    \end{minipage}
  \end{center}
  \vspace*{1mm}
  \begin{center}
    \begin{minipage}{\textwidth}
      \begin{center}
        ${}^1$ 
        Department of Physics,
        University of Pennsylvania,        
        \\
        209 S. 33rd Street, 
        Philadelphia, PA 19104--6395, USA
      \end{center}
      \begin{center}
        ${}^2$ 
        Institute for Theoretical Physics,
        Vienna University of Technology, 
        \\
        Wiedner Hauptstr. 8-10, 1040 Vienna, 
	Austria 		
      \end{center}
      \begin{center}
        ${}^3$
        Dipartimento di Scienze e Tecnologie Avanzate, 
        Universit\`a del Piemonte Orientale
        \\
        via Bellini 25/g, 15100 Alessandria, Italy, 
        and INFN - Sezione di Torino, Italy
      \end{center}
    \end{minipage}
  \end{center}
  \vspace*{\stretch{1}}
  \begin{abstract}
    \normalsize 
    As a first step towards studying vector bundle moduli in realistic
    heterotic compactifications, we identify all holomorphic rational
    curves in a Calabi-Yau threefold $X$ with $\Z_3\oplus\Z_3$ Wilson
    lines. Computing the homology, we find that
    $H_2(X,\Z)=\Z^3\oplus\Z_3\oplus\Z_3$. The \emph{torsion} curves
    complicate our analysis, and we develop techniques to distinguish
    the torsion part of curve classes and to deal with the non-toric
    threefold $X$. In this paper, we use direct A-model computations
    to find the instanton numbers in each integral homology class,
    including torsion. One interesting result is that there are
    homology classes that contain only a single instanton, ensuring
    that there cannot be any unwanted cancellation in the
    non-perturbative superpotential.
  \end{abstract}
  \vspace*{\stretch{5}}
  \begin{minipage}{\textwidth}
    \underline{\hspace{5cm}}
    \centering
    \\
    Email: 
    \texttt{vbraun}, \texttt{ovrut@physics.upenn.edu}, 
    \texttt{Maximilian.Kreuzer@tuwien.ac.at},
    \texttt{esche@mfn.unipmn.it}
  \end{minipage}
\end{titlepage}

\tableofcontents

\section{Introduction}
\label{sec:intro}

The goal of this paper is to count world sheet instantons on a certain
Calabi-Yau threefold $X$. Now that in itself was essentially solved by
mirror symmetry a long time ago~\cite{Candelas:1990rm}, but here there
is an important subtlety that does not appear in the most simple
Calabi-Yau constructions. This subtlety is the appearance of
\emph{torsion} curve classes in the degree-$2$ homology of $X$. In
particular\footnote{In the following, $\Z_3\eqdef \Z/3\Z$ always
  denotes the integers mod $3$. Similarly, we write $(\Z_3)^n=\oplus_n
  \Z_3 = \Z_3\oplus\cdots\oplus\Z_3$ for the Abelian group generated
  by $n$ generators of order $3$.},
\begin{equation}
  H_2\big( X, \Z \big) = \Z^3 \oplus \Z_3 \oplus \Z_3
  ,
\end{equation}
which contains the \emph{torsion\footnote{Not to be confused with the
    completely unrelated torsion tensor of a connection.} subgroup}
$\Z_3\oplus\Z_3$.  There are already a few known examples of such
Calabi-Yau manifolds with torsion curves~\cite{Aspinwall:1995rb,
  Batyrev:2005jc, gross-2005, Ferrara:1995yx, Aspinwall:1995mh}, but
the proper instanton counting has never been done before.

Still, the question remains: Why should we be interested in this?  We
are really interested in instanton corrections to the heterotic MSSM
constructed in~\cite{Braun:2006me, Braun:2006ae, Braun:2006th}, in
particular to the superpotential for bundle moduli. Classically, there
is no superpotential generated for the vector bundle moduli if the
bundle is at a smooth point in its moduli space (see
also~\cite{MR1797016} for a non-smooth example). If there were no
potential generated for the vector bundle moduli then there would be
no hope of stabilizing all moduli, a phenomenological disaster. As is
well known, only genus $0$ instantons (rational curves) contribute to
the superpotential, and we will exclusively consider these in the
following. The general hope is that the $E_8$ gauge bundle will give
rise to instanton corrections generating a non-vanishing
superpotential which is sufficiently complicated to stabilize
moduli~\cite{Lima:2001nh, Lima:2001jc, Buchbinder:2002ic,
  Buchbinder:2002pr, Buchbinder:2002wz, Buchbinder:2003pi,
  Buchbinder:2002ji}. However, this is far from obvious, especially in
view of unexpected cancellations between instantons in the same
homology class found in~\cite{Distler:1986wm, Distler:1987ee,
  Silverstein:1995re, Basu:2003bq, Beasley:2003fx}.  Now in our
case~\cite{Braun:2005nv, Braun:2005ux, Braun:2005bw, Braun:2005zv} the
Calabi-Yau threefold is not a toric complete intersection and the
vector bundle does not come from the ambient space, so the above
arguments do not apply. Still, it is not, a priori, clear that the
instanton contributions do not cancel for some other reason.  However,
as we are going to show in the following, the simplest smooth rigid
rational curves in $X$ are alone in their homology class, and no such
cancellation can occur. In fact, they contribute to the vector bundle
superpotential as will be explained elsewhere.

Another independent motivation is the following. Any (real
$2$-dimensional) surface in a torsion homology class cannot be
contracted by definition. Yet integrating any closed $2$-form over
this surface must give zero, since a multiple of the surface is
contractible. So whatever minimal volume surface there is in a torsion
homology class, its volume is not the integral over the \Kahler{}
form. In particular, the curve cannot be holomorphic and a D-brane
carrying the corresponding K-theory\footnote{We remind the reader that
  on a Calabi-Yau threefold $H^\even(Y,\Z) \simeq K^0(Y)$ and
  $H^\odd(Y,\Z) \simeq K^1(Y)$, so in particular the torsion parts are
  identical~\cite{doran-2006}.} charge cannot preserve any
supersymmetry (assuming no background fluxes).

As a final motivation, we note that, on general grounds,
$H_2(X,\Z)_\tors = H^3(X,\Z)_\tors$. Hence, if there is torsion then
there is a possibility for fractional Chern-Simons invariants. It was
argued in~\cite{Gukov:2003cy} that under favorable circumstances this
can generate a potential for complex structure moduli, \Kahler{}
moduli, and dilaton.

Given these motivations, we will only complete the first step and
count rational curves on $X$. Really, this means finding the instanton
correction $\FprepotXNP$ to the prepotential of the topological string.
This is usually written as a (convergent) power series in $h^{11}$
variables $q_a=e^{2\pi\iunit t^a}$. The novel feature of the
$3$-torsion curves on $X$ is that for each $3$-torsion generator we
need an additional variable $b_j$ such that $b_j^3=1$. The Fourier
series of the prepotential on $X$ becomes
\begin{equation}
  \FprepotXNP(q_1,q_2,q_3,\, b_1,b_2)
  = 
  \sum_{
    \begin{smallmatrix}
      n_1,n_2,n_3\in \Z \\ 
      m_1,m_2\in \Z_3
    \end{smallmatrix}  
  }
  n_{(n_1,n_2,n_3,m_1,m_2)} \Li_3
  \big( q_1^{n_1} q_2^{n_2} q_3^{n_3} b_1^{m_1} b_2^{m_2} \big)
  ,
\end{equation}
where $N_{(n_1,n_2,n_3,m_1,m_2)}$ is the instanton number in the curve
class $(n_1,n_2,n_3,m_1,m_2)$. Realizing this, we will investigate a
number of complementary ways to determine this prepotential:
\begin{itemize}
\item Part of the prepotential of the universal cover $\Xt$ was
  computed directly in~\cite{Hosono:1997hp}, and by carefully
  descending to the quotient $X=\Xt/(\ZZZ)$ we can compute the
  corresponding part of the prepotential of $X$.
\item The same part of the prepotential of $X$ can also be computed by
  directly counting curves on $X$.
\end{itemize}
These two A-model calculations will be carried out in this paper,
which we therefore entitle \partA. By construction, these computations
only yield a part of the prepotential, although an important one. To
overcome this limitation, we will use the B-model and mirror symmetry
in \partB, the companion paper~\cite{PartB}.  More precisely, we will
do the following:
\begin{itemize}
\item Mirror symmetry for the toric complete intersection $\Xt$
  provides an algorithm to compute instanton numbers. Unfortunately,
  there are many non-toric divisors which cannot be treated this way.
  It turns out that, after descending to $X$, precisely the torsion
  information is lost. In this approach, one can only compute
  $\FprepotXNP(q_1,q_2,q_3,\, 1,1)$.
\item As a pleasant surprise we find strong evidence that the manifold
  $X$ of principal interest is self-mirror. In particular, we attempt
  to compute the instanton numbers on the mirror $X^\ast$ by
  descending from the covering space $\Xt^\ast$. The toric embedding
  of $\Xt^\ast$ is such that all $19$ divisors are toric. A complete
  analysis including the full $\Z_3\oplus\Z_3$ torsion information
  would be feasible after some straightforward efficiency improvement
  of existing software \cite{Kreuzer:2002uu}.
\item Although the full quotient $X=\Xt/(\ZZZ)$ is not toric, it turns
  out that a certain partial quotient $\Xt/\Z_3$ can be realized as a
  toric variety. That way, one only has to deal with
  $h^{11}(\Xt/\Z_3)=7$ parameters, which is manageable on a computer.
  On the mirror $(\Xt/\Z_3)^\ast$, all divisors are toric and we can
  compute the expansion $\FprepotXNP(q_1,q_2,q_3,\, 1,b_2)$ to any
  desired degree. A symmetry argument allows one to recover the $b_1$
  dependence as well.
\end{itemize}
The result of these calculations is the complete prepotential
$\FprepotXNP(q_1,q_2,q_3,\, b_1,b_2)$. The instanton numbers can be
numerically computed for any integral homology class, limited only by
computing power. We preview these results in the conclusion of this
paper. A complete discussion is presented in~\cite{PartB}. 

To prepare the ground, we first have to compute the torsion curves on
$X$. We will do this in \autoref{part:topology} of the present paper.
In Sections~\ref{sec:CY} and~\ref{sec:groupaction} we define the
manifold $X=\tilde X/G$ as a free quotient and introduce appropriate
bases for the homology and cohomology of the cover.  In
\autoref{sec:Gprop} we compute the group homology and cohomology of
$\IZ_3$ and $\IZ_3\times\IZ_3$ with coefficients in the appropriate
(co)homology groups. These results are used in
\autoref{sec:CoHomology} to compute the integral homology groups of
the full and of the partial quotient with appropriate spectral
sequences. \autoref{sec:CoHomResult} contains a non-technical summary
of the torsion curves.

In \autoref{part:instantons} of the present paper, we proceed to do
the A-model analysis of the instanton numbers. As a simpler example
without torsion curves, we first recapitulate certain free quotients
of the quintic threefold in \autoref{sec:quintic}. Subsequently, in
Sections~\ref{sec:AmodelXt} and~\ref{sec:AmodelX} we investigate $X$
using the aforementioned A-model techniques. Finally, we present our
conclusions in \autoref{sec:conclusion}. An easily readable overview
over these results can be found in~\cite{Braun:2007tp}.


\newpage
\part{Torsion Curves}
\label{part:topology}

\section{The \CY{} Threefold}
\label{sec:CY}

\subsection{Covering Space}

The \CYm{} $X$ we are going to investigate is constructed as a free
$G\eqdef \ZZZ$ quotient of its universal covering space $\Xt$. As
usual, instead of working with a non-simply connected manifold it is
technically easier to analyze the group action on its covering space.
The simply connected \CY{} threefold $\Xt$ is one of Schoen's
threefolds~\cite{MR923487}. It can be described in various ways,
including the fiber product of two \dP9 surfaces, resolution of a
certain $T^6$ orbifold~\cite{Lust:2006zh}, or a complete intersection.
For concreteness we adopt the latter viewpoint in this section. One
first introduces the ambient variety $\CPambient$
with homogeneous coordinates
\begin{equation}
  \Big( 
  [x_0:x_1:x_2],~ 
  [t_0:t_1],~
  [y_0:y_1:y_2]
  \Big)
  \in 
  \CPambient
  .
\end{equation}
A generic complete intersection of a degree $(0,1,3)$ and a degree
$(3,1,0)$ polynomial is a smooth \CY{} threefold, but does not admit a
non-trivial $\ZZZ$ group action. However, the polynomials
\begin{subequations}
  \begin{align}
    \label{eq:B1}
    t_0 \Big( x_0^3+x_1^3+x_2^3 \Big) + 
    t_1 \Big( x_0 x_1 x_2 \Big)
    &\eqdef F_1 \\
    \label{eq:B2}
    \big( \lambda_1 t_0 + t_1\big)
    \Big( y_0^3+y_1^3+y_2^3 \Big) 
    +
    \big( \lambda_2 t_0 + \lambda_3 t_1\big)
    \Big( y_0 y_1 y_2 \Big)
    &\eqdef F_2 
    ,
  \end{align}
\end{subequations}
where $\lambda_1$, $\lambda_2$, $\lambda_3$ are complex parameters,
are invariant under the $G= \ZZZ$ action generated by
($\zeta\eqdef e^{\frac{2\pi i}{3}}$)
\begin{subequations}
\begin{equation}
  \label{eq:g1action}
  g_1:  
  \begin{cases}
    [x_0:x_1:x_2] \mapsto
    [x_0:\zeta x_1:\zeta^2 x_2]
    \\
    [t_0:t_1] \mapsto
    [t_0:t_1] 
    ~\text{(no action)}
    \\
    [y_0:y_1:y_2] \mapsto
    [y_0:\zeta y_1:\zeta^2 y_2]
  \end{cases}
\end{equation}
and
\begin{equation}
  \label{eq:g2action}
  g_2:  
  \begin{cases}
    [x_0:x_1:x_2] \mapsto
    [x_1:x_2:x_0]
    \\
    [t_0:t_1] \mapsto
    [t_0:t_1] 
    ~\text{(no action)}
    \\
    [y_0:y_1:y_2] \mapsto
    [y_1:y_2:y_0]
  \end{cases}
\end{equation}
\end{subequations}
This group action has fixed points in the ambient variety \CPambient,
but these do not satisfy eqns.~\eqref{eq:B1} and~\eqref{eq:B2}.
Hence, this \ZZZ{} group action on the complete intersection \CY{}
threefold
\begin{equation}
  \Xt 
  ~\eqdef~
  \Big\{
    \big([x_0:x_1:x_2], [t_0:t_1], [y_0:y_1:y_2]\big)  
  \Big|
  F_1=0 
  ,\, 
  F_2=0
  \big\}
  ~\subset~
  \CPambient
\end{equation}
is free. 

We point out that this $\ZZZ$ action is slightly different from the
$\ZZZ$ action investigated within the context of an heterotic standard
model~\cite{Braun:2004xv}. The group action we discuss in this paper
``does not act on the base $\CP^1$'' and, hence, is not included in
the classification~\cite{Braun:2004xv}. The reason we are using the
$\ZZZ$ action defined above is that it is more amenable to toric
methods, which will be important for the B-model computation later in
this paper. However, the curve counting can be easily extended to the
MSSM manifold~\cite{Braun:2004xv}, which we will present elsewhere.

Finally, let us review some facts about the homology and cohomology of
the universal cover $\Xt$, see~\cite{MR923487, Ovrut:2002jk}.  The
Hodge diamond of the Calabi-Yau threefold $\Xt$ is self-mirror and
given by 
\begin{equation}
  \label{eq:Hodgecover}
  h^{p,q}\big(\Xt\big) = ~
  \vcenter{\xymatrix@!0@=7mm@ur{
    1 &  0  &  0  & 1 \\
    0 &  19 &  19 & 0 \\
    0 &  19 &  19 & 0 \\
    1 &  0  &  0  & 1 
  }}
  \qquad \Rightarrow \quad
  H^i_\text{de Rham}\big(\Xt, \R\big) 
  =
  \begin{cases}
    \R
    & i=6 \\
    0
    & i=5 \\
    \R^{19}
    & i=4 \\
    \R^{40}
    & i=3 \\
    \R^{19} 
    & i=2 \\
    0
    & i=1 \\
    \R
    & i=0
    . 
  \end{cases}
\end{equation}
In general the dimension of the $i$-th de Rham cohomology is the same
as the rank of the $i$-th integral cohomology group, but the latter
might also contain torsion information which is not captured by de
Rham cohomology. However, a smooth complete intersection in a smooth
toric variety does not have any torsion in its integral
cohomology~\cite{Braun:2000zm}. This determines the integral
cohomology, and Poincar\'e duality eq.~\eqref{eq:PD} then yields the
integral homology groups. We conclude that
\begin{equation}
  H_{6-i}\big(\Xt, \Z\big) 
  =
  H^i\big(\Xt, \Z\big) 
  =
  \begin{cases}
    \Z
    & i=6 \\
    0
    & i=5 \\
    \Z^{19}
    & i=4 \\
    \Z^{40}
    & i=3 \\
    \Z^{19} 
    & i=2 \\
    0
    & i=1 \\
    \Z
    & i=0
    . 
  \end{cases}
\end{equation}

\subsection{The Quotient}

Having constructed $\Xt$ with a free $\ZZZ$ group action, we define
\begin{equation}
  X \eqdef \Xt \Big/ \big(\ZZZ\big)
  .
\end{equation}
On general grounds, $X$ is again a smooth \CY{} threefold with
fundamental group $\pi_1(X)=\ZZZ$. Since the defining
equations~\eqref{eq:B1},~\eqref{eq:B2} allow for three independent
coefficients up to $PGL(3)\times PGL(2)\times PGL(3)$ coordinate
changes if one wants to preserve the $\ZZZ$ symmetry, we expect that
there are $h^{21}(X)=3$ complex structure parameters. This turns out
to be true, as will be shown in more detail in
\autoref{sec:Cohinvariant}.

Moreover, we know the Euler numbers\footnote{Note that $\Xt$ will turn
  out to be self-mirror. Nevertheless, instanton corrections are
  present, part of which were been computed in~\cite{Hosono:1997hp,
    Hosono:1999qc, Klemm:1996hh}. There is a common misconception
  based on the free $K3\times T^2\big/\Z_2$ orbifold investigated
  in~\cite{Ferrara:1995yx, Aspinwall:1995mh} that self-mirror
  threefolds do not receive quantum corrections to the classical
  moduli space. Indeed, in that case, all rational curves come in
  families which happen not to contribute~\cite{Maulik:2006cf}, that
  is, their Gromov-Witten invariants vanish. However, this is not due
  to $K3\times T^2\big/\Z_2$ being self-mirror.
}
vanish,
\begin{equation}
  \chi\big(\Xt\big) = 
  2 h^{11}\big(\Xt\big) - 2 h^{21}\big(\Xt\big)
  = 0 = 9 \chi\big(X\big)
  .
\end{equation}
This fixes the Hodge numbers of the quotient $X=\Xt/(\ZZZ)$ to be
\begin{equation}
  h^{p,q}\big(X\big) = ~
  \vcenter{\xymatrix@!0@=7mm@ur{
    1 &  0  &  0  & 1 \\
    0 &  3  &  3  & 0 \\
    0 &  3  &  3  & 0 \\
    1 &  0  &  0  & 1 
  }}
\end{equation}
However, knowing the Betti numbers does not tell us everything about
the homology classes of curves. The integral homology groups
potentially contain \textdef{torsion}, that is, a finite subgroup.
For example, as we will show in \autoref{sec:CoHomology}
\begin{equation}
  H_2\big(X,\R\big)
  =
  \R\oplus\R\oplus\R=\R^3
  ,\quad 
  H_2\big( X,\Z \big) 
  = 
  \Z^3 \oplus \Big( \Z_3 \oplus \Z_3 \Big)
  .
\end{equation}
The subgroup $\Z_3 \oplus \Z_3$ consisting of $9$ elements is such a
torsion subgroup. Clearly, explicit knowledge of \emph{all} curve
homology classes is important when counting curves on~$X$.

\section{Group Action}
\label{sec:groupaction}

\subsection{Projections}
\label{sec:dP9}

As usual, instead of analyzing the quotient $X=\Xt/G$ directly we will
look at the $G= \ZZZ$ action on the covering space. In this section,
we find it particularly useful to exploit the property that $\Xt$ has
two projections to \dP9 surfaces. To see this, note that a degree
$(3,1)$ hypersurface in $\CP^2\times\CP^1$ is such a \dP9 surface,
also called a rational elliptic surface. Moreover, the defining
equations~\eqref{eq:B1} and~\eqref{eq:B2} do not depend on
$[y_0:y_1:y_2]$ and $[x_0:x_1:x_2]$, respectively.  Hence,
eq.~\eqref{eq:B1} and eq.~\eqref{eq:B2} define \dP9 surfaces with
natural projections $\pi_1:\Xt\to B_1$, $\pi_2:\Xt\to B_2$.  Finally,
each $B_1, B_2$ projects to the common $\CP^1$, yielding a commutative
diagram
\begin{equation}
  \label{eq:projections}
  \vcenter{\xymatrix@!0@=12mm{
      \dim_\C=3: && & \Xt \ar[dr]^{\pi_2} \ar[dl]_{\pi_1} \\
      \dim_\C=2: && B_1 \ar[dr] & & 
        B_2 \ar[dl] \\
      \dim_\C=1: && & \CP^1
      .
  }}
\end{equation}
By definition, this means that $\Xt$ is the fiber product of two
$dP_9$ surfaces, that is, $\Xt=B_1\times_{\CP^1} B_2$. In other words,
$\Xt$ is elliptically fibered over each $B_i$, $i=1,2$, and each $B_i$
is again elliptically fibered over the same $\CP^1$. In the remainder
of this section, we are going to detail the properties of these
\dP9 surfaces.

The $\ZZZ$ group action descends to $B_1$, $B_2$. Moreover, since the
action is trivial on the $\CP^1$, it must be the
translation\footnote{A point $z_0$ on an elliptic surface $\C/\Lambda$
  defines a group action $z\mapsto z+z_0$. A section of the elliptic
  fibration $B_i$ consists of a point in each fiber. Hence, a section
  $s$ defines a group action $t_s:B_i\to B_i$ by translation along
  each fiber.} by two independent sections. The existence of two
sections of order three determines the Kodaira fibers and \MWgrp{}
uniquely~\cite{MR1069483} to be
\begin{equation}
  \begin{gathered}
    \Sing(B_1) = 
    \Sing(B_2) = 
    4 I_3
    ,
    \\
    MW(B_1) = MW(B_2) = \Z_3 \oplus \Z_3
    .
  \end{gathered}  
\end{equation}
Recall that the \MWgrp{} is the set of all sections (which depends on
the moduli of the \dP9 surface) together with a group law
``$\boxplus$''. The \MW{} sum\footnote{We point out that the \MW{} sum
  ``$\boxplus$'' must be distinguished from the sum of homology
  classes, which we write as ``+''. For example, $\alpha\boxplus\beta$
  is again a section whereas $\alpha+\beta$ is a two-section.}
$\alpha\boxplus\beta$ of two sections $\alpha$, $\beta$ is the
fiberwise sum. In other words,
\begin{equation}
  \alpha\boxplus\beta=t_\alpha(\beta) = t_\beta(\alpha)
  .
\end{equation}
Let us label\footnote{In the following, it will always be clear from
  the context whether we are referring to $B_1$ or $B_2$. Hence we use
  the same symbol for divisors in $B_1$ and $B_2$.} the generating
sections $\mu$ and $\nu$ on $B_i$, $i=1,2$ such that
\begin{equation}
  \label{eq:B1MWgrp}
  MW(B_i) =
  \Big\{
  \sigma,\,
  \mu,\, 
  \mu \boxplus \mu,\,
  \nu,\,
  \nu\boxplus\mu,\,
  \nu\boxplus\mu\boxplus\mu,\,
  \nu\boxplus\nu,\, 
  \nu\boxplus\nu\boxplus\mu,\,
  \nu\boxplus\nu\boxplus\mu\boxplus\mu
  \Big\}
  ,
\end{equation}
with $\sigma$ being the zero section. Furthermore, note that each
vertical $I_3$ fiber is composed of three irreducible components,
intersecting in a triangle. We denote the $i$-th component of the
$j$-th $I_3$ Kodaira fiber by $\theta_{ji}$. Up to re-indexing the
divisors, there is only one possible intersection pattern between the
horizontal and vertical divisors, namely
\begin{equation}
  \label{eq:hvintersect}
  \begin{array}{c|cccccccccccc}
    (-)\cdot (-) & 
    \theta_{10} &
    \theta_{11} &
    \theta_{12} &
    \theta_{20} &
    \theta_{21} &
    \theta_{22} &
    \theta_{30} &
    \theta_{31} &
    \theta_{32} &
    \theta_{40} &
    \theta_{41} &
    \theta_{42} \\
    \hline
    \sigma &
    1&0&0 &
    1&0&0 &
    1&0&0 &
    1&0&0 
    \\
    \mu &
    1&0&0 &
    0&1&0 &
    0&1&0 &
    0&1&0 
    \\
    \nu &
    0&1&0 &
    1&0&0 &
    0&0&1 &
    0&1&0 
  \end{array}
\end{equation}
Finally, denote the class of an elliptic fiber by $f$. 

Recall the Hodge diamond, homology, and cohomology of \dP9 surfaces,
\begin{equation}
  h^{p,q}\big(B_i\big) = 
  \vcenter{\xymatrix@!0@=7mm@ur{
    0 &  0  &  1 \\
    0 &  10 &  0 \\
    1 &  0  &  0,
  }}
  \qquad 
  H_{4-i}(B_i, \Z\big) 
  =
  H^i(B_i, \Z\big) 
  =
  \begin{cases}
    \Z
    & i=4 \\
    0
    & i=3 \\
    \Z^{10} 
    & i=2 \\
    0
    & i=1 \\
    \Z
    & i=0
    . 
  \end{cases}
\end{equation}
Therefore, although the above $9+3\cdot 4+1$ divisors generate
$H_2(B_i,\Z)=\Z^{10}$, they cannot be linearly independent.  It is a
straightforward task to identify all relations, which we will do in
\autoref{sec:relations}. One possible integral basis~\cite{MR1104782,
  MR1081832} is
\begin{equation}
  \label{eq:Bbasis}
  H_2(B_i,\Z) = 
  \Span_\Z\Big\{
  \sigma,\,
  f,\,
  \theta_{11},\,
  \theta_{21},\,
  \theta_{31},\,
  \theta_{32},\,
  \theta_{41},\,
  \theta_{42},\,
  \mu,\,
  \nu
  \Big\}
  ,
\end{equation}
and we will use this integral basis in the following.

\subsection
[The $E_8$ Lattice]
[The E8 Lattice]
{The $\mathbf{E_8}$ Lattice}
\label{sec:E8}

There is another special basis for the homology of the \dP9 surfaces
in addition to eq.~\eqref{eq:Bbasis}. This other basis is the natural
basis choice for a generic \dP9 surface $B$, that is, one with $12
I_0$ singular fibers. In that case the \MWgrp{} is $E_8$.
\begin{figure}[htbp]
  \centering \input{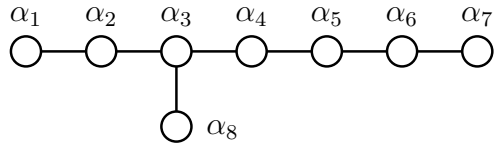}
  \caption{The $E_8$ Dynkin diagram.}
  \label{fig:E8Dynkin}  
\end{figure}
This means that the quotient
\begin{equation}
  H_2\big(B,\Z\big) \Big/ \Span_\Z\big\{\sigma, f \big\}
  = MW(B) = \Lambda_{E_8}
\end{equation}
is the $E_8$ root lattice with respect to the height pairing
\begin{equation}
  \label{eq:E8intersect}
  \left<s_1, s_2\right> = 
  1+s_1\cdot \sigma + s_2 \cdot \sigma - s_1\cdot s_2
  .
\end{equation}
Therefore, one obvious integral basis choice is to pick $8$ simple
roots together with $\sigma$ and $f$,
\begin{equation}
  \label{eq:BE8basis}
  H_2(B_i,\Z) = 
  \Span_\Z\Big\{
  \sigma,\, 
  f,\,
  \alpha_1,\,  \alpha_2,\,  \alpha_3,\,  \alpha_4,\,
  \alpha_5,\,  \alpha_6,\,  \alpha_7,\,  \alpha_8
  \Big\}
  .
\end{equation}
Of course, the generic \dP9 does not have the $\ZZZ$ group action
which we are interested in. For example, the \MW{} lattice in our case
needs to be $\Z_3\oplus\Z_3$ instead of $\Lambda_{E_8}$. However, the
homology groups do not know about the choice of complex structure.
Hence, although, in our case, the homology classes $\alpha_i$ cannot
be represented by sections, we can still use the same basis for
homology.  The $240$ roots of $E_8$ are readily identified as
\begin{equation}
  \Phi_{E_8} \eqdef
  \Big\{
    \alpha \in H_2(B,\Z)
    ~\Big|~
    \alpha\cdot f = 1
    ,\,
    \alpha\cdot \sigma = 0
    ,\,
    \alpha\cdot\alpha=-1
  \Big\}
  .
\end{equation}
The choice of simple roots is not unique. For convenience, we will
make the same choice as in~\cite{Hosono:1997hp}:
\begin{equation}
  \begin{aligned}
    \alpha_1 =&\; 
    2\sigma+2f-\mu, 
    \\
    \alpha_2 =&\; 
    2\sigma+2f-\theta_{21}-\theta_{31}-\theta_{41}-\mu,
    \\
    \alpha_3 =&\;
    \theta_{21}+\theta_{31}+\theta_{41}+2\mu-\nu, 
    \\
    \alpha_4 =&\;
    2\sigma+2f-\theta_{31}-\theta_{32}-\theta_{41}-\mu,
    \\
    \alpha_5 =&\;
    2\sigma+2f-\theta_{21}-\theta_{41}-\theta_{42}-\mu, 
    \\
    \alpha_6 =&\;
    -\theta_{11}+\theta_{21}+\theta_{31}+\theta_{41}+\theta_{42}+2\mu-\nu,
    \\
    \alpha_7 =&\;
    2\sigma+2f-\theta_{31}-\theta_{41}-\theta_{42}-\mu, 
    \\
    \alpha_8 =&\;
    -2\sigma-2f+\theta_{11}+\theta_{31}+2\theta_{32}+
    2\theta_{41}+\theta_{42}+3\nu
    .
  \end{aligned}
\end{equation}
To clarify, on a generic \dP9 surface $B$ the sections $\alpha_i$ can
be added by the usual \MW{} sum ``$\boxplus$'' defined previously.
However, the definition of ``$\boxplus$'' as fiberwise sum of points
on a torus depends on having actual sections, and not just the
homology classes. However, while on the special \dP9 surfaces $B_1$,
$B_2$ the homology classes $\alpha_i$ are still well-defined, they
need not contain a section anymore. Nevertheless, we can still define
the lattice sum
\begin{equation}
  \boxplus:\;
  \Lambda_{E_8} \times 
  \Lambda_{E_8} \to
  \Lambda_{E_8}
\end{equation}
on $\Lambda_{E_8}\subset B_1, B_2$ by taking it to the same as for the
generic \dP9 surface $B$.

\subsection{Action on the Base}
\label{sec:Baction}

We start by analyzing the base \dP9 surfaces $B_1$, $B_2$ which, as
discussed above, are again elliptically fibered over $\CP^1$. The
$G=\ZZZ$ group action\footnote{By abuse of notation we use
  $G=\left\{\Id,g_1,g_1^2,g_2,g_1 g_2, g_1^2 g_2, g_2^2, g_1 g_2^2,
    g_1^2 g_2^2 \right\}$ for the group action on $\Xt$ and for the
  induced action on $B_1$, $B_2$.} is fiberwise translation
\begin{equation}
  g_1 = t_\mu
  ,\quad
  g_2 = t_\nu
\end{equation}
by the two sections $\mu$, $\nu$ of order $3$ described previously.
Obviously, this maps the fiber to itself, $g_1(f) = g_2(f) = f$. On
any section, that is, any element of $MW(B_i)$, the group also acts in
the obvious way
\begin{equation}
  \begin{gathered}
    MW(B_i) = 
    \Span_\boxplus \big\{ \mu, \nu \big\}
    ~\Big/~
    \big( \boxplus_3 \mu = \boxplus_3 \nu = \sigma \big)
    ,
    \notag \\
    g_1(s) = s \boxplus \mu
    ,\quad
    g_2(s) = s \boxplus \nu
    .
  \end{gathered}
\end{equation}
Finally, the action on each $I_3$ Kodaira fiber either maps each
irreducible component to itself or cyclically permutes the irreducible
components, as explained in~\cite{Braun:2004xv}. From
eq.~\eqref{eq:hvintersect} we can read off that
\begin{equation}
  \label{eq:thetaaction}
  \begin{array}{c||cccccccccccc}
    D & 
    \theta_{10} &
    \theta_{11} &
    \theta_{12} &
    \theta_{20} &
    \theta_{21} &
    \theta_{22} &
    \theta_{30} &
    \theta_{31} &
    \theta_{32} &
    \theta_{40} &
    \theta_{41} &
    \theta_{42} \\
    \hline
    g_1(D) &
    \theta_{10} &
    \theta_{11} &
    \theta_{12} &
    \theta_{21} &
    \theta_{22} &
    \theta_{20} &
    \theta_{31} &
    \theta_{32} &
    \theta_{30} &
    \theta_{41} &
    \theta_{42} &
    \theta_{40} \\
    g_2(D) &
    \theta_{11} &
    \theta_{12} &
    \theta_{10} &
    \theta_{20} &
    \theta_{21} &
    \theta_{22} &
    \theta_{32} &
    \theta_{30} &
    \theta_{31} &
    \theta_{41} &
    \theta_{42} &
    \theta_{40} 
  \end{array}
\end{equation}
Using the relations from \autoref{sec:relations} we can now express
the $G$ action on $H_2(B_i,\Z)$ as $10\times 10$ matrices in the basis
eq.~\eqref{eq:Bbasis}. One obtains
\begin{equation}
  \label{eq:Bg1g2matrix}
  g_1 = 
  \left(
    \begin{smallmatrix}
      0&0&0&3&0&0&0&0&-1&-1\\
      0&1&0&3&0&1&0&1&-1&-1\\
      0&0&1&0&0&0&0&0&0&0\\
      0&0&0&-2&0&0&0&0&1&0\\
      0&0&0&-2&0&-1&0&0&1&1\\
      0&0&0&-1&1&-1&0&0&0&1\\
      0&0&0&-2&0&0&0&-1&1&1\\
      0&0&0&-1&0&0&1&-1&0&0\\
      1&0&0&-3&0&0&0&0&2&1\\
      0&0&0&0&0&0&0&0&0&1
    \end{smallmatrix} 
  \right)
  ,\quad
  g_2 =
  \left(
    \begin{smallmatrix}
      0&0&3&0&0&0&0&0&-1&-1\\
      0&1&3&0&1&0&0&1&-1&-1\\
      0&0&-2&0&0&0&0&0&0&1\\
      0&0&0&1&0&0&0&0&0&0\\
      0&0&-1&0&-1&1&0&0&1&0\\
      0&0&-2&0&-1&0&0&0&1&1\\
      0&0&-2&0&0&0&0&-1&1&1\\
      0&0&-1&0&0&0&1&-1&0&0\\
      0&0&0&0&0&0&0&0&1&0\\
      1&0&-3&0&0&0&0&0&1&2
    \end{smallmatrix}
  \right)
  .
\end{equation}

\subsection{Line Bundles}
\label{sec:divisoraction}

Having determined the group action on the base \dP9 surfaces, we now
investigate the action on $\Xt$. First, recall that by a happy
coincidence $h^{2,0}(\Xt)=0$ and therefore
\begin{equation}
  \label{eq:linebundleequiv}
  \begin{split}    
    \Pic\big(\Xt\big)
    \eqdef&\;
    \big\{
    \text{Algebraic line bundles on }\Xt
    \big\}
    \\
    =&\;
    \big\{
    \text{Topological line bundles on }\Xt
    \big\}
    = 
    H^2\big(\Xt,\Z\big) 
    = 
    H_4\big(\Xt,\Z\big) 
    .
  \end{split}
\end{equation}
In other words, 
\begin{itemize}
\item Each line bundle has a unique complex structure.
\item A line bundle is uniquely determined by its first Chern class.
\item Every line bundle $\Lsheaf$ can be written as
  $\Lsheaf=O_\Xt(D)$, and depends only on the homology class of the
  divisor $D\in H_4\big(\Xt,\Z\big)$.
\end{itemize}
Note that the identification $H_2=H^4$ does not involve any duality
(see \autoref{sec:dualpoincare}), which will be important later on. To
lift from $B_i$, $i=1,2$ to $\Xt$, we can use
\begin{itemize}
\item Pull back of line bundles: $\pi_i^\ast: \Pic(B_i)\to\Pic(\Xt)$.
\item Pull back in cohomology: $\pi_i^\ast: H^2\big(B_i,\Z\big)\to
  H^2\big(\Xt,\Z\big)$.
\item Preimage of divisors: $\pi_i^{-1}: H_2\big(B_i,\Z\big)\to
  H_4\big(\Xt,\Z\big)$.
\end{itemize}
All these notions commute with the identifications
eq.~\eqref{eq:linebundleequiv}. However, the pull backs of the $\dim
H_2(B_1,\Z)+\dim H_2(B_2,\Z)=20$ line bundles on the bases cannot be
independent in $H_4\big(\Xt,\Z\big)\simeq \Z^{19}$. As was shown
in~\cite{Donagi:2003tb, Ovrut:2003zj, Curio:1997rn, Ovrut:2002jk}, the
line bundles on $\Xt$ have a particularly nice description, that is,
the pullback of the line bundles to $\Xt$ yields a generating set of
$20$ line bundles, which must satisfy one relation. This relation is
that $\pi_1^{-1}(f)=\pi_2^{-1}(f)$, both being the Abelian surface
fiber of the fibration $\Xt\to\CP^1$.  Hence,
\begin{equation}
  \begin{split}
    H_4\big(\Xt, \Z\big) 
    =&\;
    \Big[
    \pi_1^{-1}H_2\big(B_1,\Z\big)
    \oplus
    \pi_2^{-1}H_2\big(B_2,\Z\big)
    \Big]
    \bigg/
    \left< \pi_1^{-1}(f) =\pi_2^{-1}(f) \right> 
    \\
    =&\;
    \Span_\Z \Big\{ 
    \pi_1^{-1}(f) =\pi_2^{-1} (f),\,
    \\&\;\phantom{\Span_\Z \Big\{}
    \pi_1^{-1}(\sigma),\,
    \pi_1^{-1}(\theta_{11}),\,
    \pi_1^{-1}(\theta_{21}),\,
    \pi_1^{-1}(\theta_{31}),\,
    \pi_1^{-1}(\theta_{32}),
    \\&\;\phantom{\Span_\Z \Big\{}    
    \pi_1^{-1}(\theta_{41}),\,
    \pi_1^{-1}(\theta_{42}),\,
    \pi_1^{-1}(\mu),\,
    \pi_1^{-1}(\nu),
    \\&\;\phantom{\Span_\Z \Big\{}
    \pi_2^{-1}(\sigma),\,
    \pi_2^{-1}(\theta_{11}),\, 
    \pi_2^{-1}(\theta_{21}),\,
    \pi_2^{-1}(\theta_{31}),\,
    \pi_2^{-1}(\theta_{32}),\,
    \\&\;\phantom{\Span_\Z \Big\{}
    \pi_2^{-1}(\theta_{41}),\,
    \pi_2^{-1}(\theta_{42}),\,
    \pi_2^{-1}(\mu),\,
    \pi_2^{-1}(\nu)
    \Big\}
    \simeq
    \Z^{19} 
    .        
  \end{split}
\end{equation}
Having determined the geometric action on the divisors of the surfaces
$B_i$ in \autoref{sec:Baction}, one can now easily determine the
$G=\ZZZ$ representation on $H_4(\Xt,\Z)$ in terms of $19\times
19$ integer matrices. Other than to note that we use them in the
following for some linear algebra computations, it is not particularly
enlightening to present the explicit matrices here. We denote this
representation as
\begin{equation}
  \label{eq:Rdualdef}
  R^\dual \eqdef H_4\big(\Xt, \Z\big)
  .
\end{equation}

\subsection{Curves}
\label{sec:curveaction}

Abstractly, the previous subsection boils down to the short exact
sequence
\begin{equation}
  0
  \longrightarrow
  \Z
  \longrightarrow
  H^2\big(B_1,\Z\big)
  \oplus
  H^2\big(B_2,\Z\big)
  \xrightarrow{\pi_1^\ast+\pi_2^\ast}
  H^2\big(\Xt,\Z\big)
  \longrightarrow
  0
  .
\end{equation}
Recall that the fiber product $\Xt=B_1\times_{\CP^1}B_2$ is a
hypersurface in $B_1\times B_2$.  The Poincar\'e dual (see
\autoref{sec:dualpoincare}) sequence
\begin{equation}
  0
  \longrightarrow
  H_2\big(\Xt,\Z\big)
  \xrightarrow{\pi_{1\ast}\oplus\pi_{2\ast}}
  \underbrace{
    H_2\big(B_1,\Z\big)
    \oplus
    H_2\big(B_2,\Z\big)
  }_{= H_2\big(B_1\times B_2,\Z\big)}
  \longrightarrow  
  \Z
  \longrightarrow
  0
\end{equation}
assures us that we can study the curves in $\Xt$ completely by looking
at their image in $B_1\times B_2$. All we have to do is determine the
curves in $B_1\times B_2$ that lie on the hypersurface $\Xt$. 

Let us introduce the notation 
\begin{equation}
  C_1\FPtimes C_2 
  ~=~
  \big( C_1\times C_2 \big) \cap \Xt
  ~\subset~
  \Xt
  ~\subset~
  B_1\times B_2
\end{equation}
for two curves $C_1\subset B_1$ and $C_2\subset B_2$. For example,
\begin{equation}
  \sigma \FPtimes \theta_{ij} = \ptset \times \theta_{ij}
  ,\quad
  \theta_{ij} \FPtimes \sigma = \theta_{ij} \times \ptset 
  .
\end{equation}
Also note that, for example, $\sigma \FPtimes \sigma$ is a section of
the Abelian surface fibration $\Xt\to\CP^1$. Using this notation, a
basis for $H_2(\Xt,\Z)$ is
\begin{equation}
  \label{eq:Xthomologybasis}
  \begin{split}
  H_2\big(\Xt,\Z\big) = 
  \Span_\Z \big\{
  &
  \sigma \FPtimes f ,\,
  f \FPtimes \sigma ,\,
  \\ &
  \sigma \FPtimes \theta_{11} ,\,
  \sigma \FPtimes \theta_{21} ,\,
  \sigma \FPtimes \theta_{31} ,\,
  \sigma \FPtimes \theta_{32} ,\,
  \sigma \FPtimes \theta_{41} ,\,
  \sigma \FPtimes \theta_{42} ,\,
  \\ &
  \theta_{11} \FPtimes \sigma ,\,
  \theta_{21} \FPtimes \sigma ,\,
  \theta_{31} \FPtimes \sigma ,\,
  \theta_{32} \FPtimes \sigma ,\,
  \theta_{41} \FPtimes \sigma ,\,
  \theta_{42} \FPtimes \sigma ,\,
  \\ &
  \sigma \FPtimes \sigma ,\,
  \mu \FPtimes \sigma ,\,
  \nu \FPtimes \sigma ,\,
  \sigma \FPtimes \mu ,\,
  \sigma \FPtimes \nu
  \big\}
  \simeq \Z^{19}
  .
  \end{split}
\end{equation}
The group action can now easily be determined from the group action on
the base, \autoref{sec:Baction}, and explicitly written in terms of
$19\times 19$ matrices. Again, we will use these matrices
computationally in the following, but find it unenlightening to
actually write them down here. We denote this representation
suggestively as
\begin{equation}
  \label{eq:Rdef}
  R \eqdef H_2\big(\Xt, \Z\big)
  .
\end{equation}
As we will now show, it is dual to the representation $H_4(\Xt,\Z)$.

\subsection{Poincar\'e Duality}
\label{sec:dual}

We now have defined a priori independent bases on
$H_4\big(\Xt,\Z\big)$ and $H_2\big(\Xt,\Z\big)$. But they are related
through the intersection pairing
\begin{equation}
  H_4\big(\Xt,\Z\big) \times H_2\big(\Xt,\Z\big) \to
  \Z=H_0\big(\Xt,\Z\big)
  ,
\end{equation}
which is one version of Poincar\'e duality (see
\autoref{sec:dualpoincare}). We can explicitly determine the
intersection numbers for our two bases in terms of elementary
intersection numbers on $B_1$ and $B_2$: For any two basis curves
$C_1,C_2\in\{\sigma,f,\theta_{11},\dots,\theta_{42},\mu,\nu\}$ and
section $s\in \{\sigma,\mu,\nu\}$
\begin{equation}
  \begin{gathered}
    \big( C_1 \FPtimes \sigma \big) \cdot \big( \pi_1^{-1} C_2 \big) 
    = C_1 \cdot C_2  =
    \big( \sigma \FPtimes C_1 \big) \cdot \big( \pi_2^{-1} C_2 \big) 
    ,\\
    \big( \sigma \FPtimes s \big) \cdot \big( \pi_1^{-1} C_2 \big) 
    = s \cdot C_2  =  
    \big( s \FPtimes \sigma \big) \cdot \big( \pi_2^{-1} C_2 \big) 
    ,\\
    \big( \sigma \FPtimes C_1 \big) \cdot \big( \pi_1^{-1} s \big) 
    = C_1 \cdot s  =  
    \big( C_1 \FPtimes \sigma \big) \cdot \big( \pi_2^{-1} s \big) 
    ,
  \end{gathered}
\end{equation}
and $0$ in the remaining cases. For example,
$(\theta_{11}\FPtimes\sigma)\cdot(\pi_2^{-1} \theta_{11})=0$.

This makes it easy to write down the explicit $19\times 19$
intersection matrix. One can check that its determinant is $1$, as it
should be. The inverse matrix is again integral and defines the
Poincar\'e dual of any curve or divisor. In particularly, it follows
that $R$ and $R^\dual$ , eqns.~\eqref{eq:Rdef}
and~\eqref{eq:Rdualdef}, are mutually dual representations, as we
already implied by the notation.

\subsection{Middle Dimension}
\label{sec:middle}

For completeness, let us also discuss the $G=\ZZZ$-action on the
middle dimensional homology group $H_3(\Xt, \Z)\simeq \Z^{40}$.  By
Poincar\'e duality, this representation must be self-dual.
Unfortunately, there seems to be no simple way to write down an
integral basis of three-cycles. We did construct a $G$-CW complex of
the $4$-skeleton of $\Xt$, that is, a cell complex on which $G$ acts
by permutation of cells. Given this, finding the action on homology
boils down to a lengthy linear algebra exercise on the corresponding
chain complex. With the help of a computer we found the explicit
$40\times 40$ representation matrices for $H3$. As above, we do not
write out the explicit matrices but simply define this $\ZZZ$
representation to be 
\begin{equation}
  H3 \eqdef H_3\big(\Xt, \Z\big) 
  .
\end{equation}
Note that we will only need information about $H3$ in
\autoref{sec:X3specseq}, where it could be replaced by some
independent toric computation.

\section{Properties of the Group Action}
\label{sec:Gprop}

\subsection{Describing Integer Representations}
\label{sec:modularrep}

Summarizing the results of \autoref{sec:groupaction}, the $G=\ZZZ$
group action on the homology and cohomology of $\Xt$ is
\begin{equation}
  \label{eq:RH3def}
  H_{6-i}\big(\Xt, \Z\big) 
  =
  H^i\big(\Xt, \Z\big) 
  =
  \begin{cases}
    \Z
    & i=6 \\
    0
    & i=5 \\
    R \simeq \Z^{19}
    & i=4 \\
    H3 \simeq \Z^{40}
    & i=3 \\
    R^\dual \simeq \Z^{19}
    & i=2 \\
    0
    & i=1 \\
    \Z
    & i=0 
    , \\
  \end{cases}
\end{equation}
where we used Poincar\'e duality as well,
see~\autoref{sec:dualpoincare}. Of course, we are really interested in
the quotient $X$ and not in the covering space $\Xt$. However, as we
will show is \autoref{sec:CoHomology}, the homology of the quotient
$X$ can be calculated from the $G$-action on the homology of $\Xt$.
More precisely, certain invariants, called \textdef{group homology},
of the group action on $H_\ast(\Xt,\Z)$ are the starting point for the
\CLss{}, which in turn computes $H_\ast(X,\Z)$. Dually, the \LSss{}
computes the cohomology on $X$ from the \textdef{group cohomology
  groups} of the group action on $H^\ast(\Xt,\Z)$. The purpose of this
section is to find the group homology groups of the
$G$-representations $H_q(\Xt,\Z)$ and group cohomology groups of the
$G$-representations $H^q(\Xt,\Z)$. These are denoted by 
\begin{equation}
  H_p\Big( G, H_q\big(\Xt,\Z\big) \Big)
  ,\quad
  H^p\Big( G, H^q\big(\Xt,\Z\big) \Big)  
  .
\end{equation}

An important point is that we are considering representations on
integer lattices. Many of the nice features of representation theory
on vector spaces no longer hold. In particular, there is no longer any
unique decomposition into a sum of irreducible representations. Since
the actual integer representations are so complicated, a nice way to
classify them is via their group homology and group cohomology.  This
is entirely analogous to the study of manifolds using their homology
and cohomology groups:
\begin{center}
  \renewcommand{\arraystretch}{2}
  \begin{tabular}{c||c}
    \parbox{5cm}{\centering
      Homology and cohomology\\ in topology} & 
    \parbox{5cm}{\centering
      Group homology and\\ group cohomology} 
    \\ \hline
    Manifold $X$ & Group $G$ 
    \\ \hline
    \parbox{5cm}{\centering      
      Coefficients $C=\Z$, $\R$,\\
      $\C$, twisted coefficients, $\dots$}
    & 
    Group representation $M$
    \\ \hline
    $H_\ast(X,C),~ H^\ast(X,C)$
    &
    $H_\ast(G,M),~ H^\ast(G,M)$    
  \end{tabular}
\end{center}
An inevitably confusing part of the computation below is that it
involves both the ``topological homology'' and the group homology.
Specifically, we need to consider the case where the
$G$-representation is one of the topological homology groups of $X$.
Then, for this representation, we must determine the group homology.

Let us start by defining the group homology and group cohomology. Take
any representation $M$ of a finite group $G$ on an integer
lattice\footnote{$M$ could also have $\Z$-torsion, that is, be of the
  form $\Z^n\oplus\Z_{r_1}\oplus\cdots\oplus \Z_{r_k}$. However the
  representations we are interested in will be of the form $\Z^n$
  only.}. In particular, we are interested in the cases where $M$ is
either $\Z$ (the trivial representation), $R$, $R^\dual$, or $H3$. The
representation defines a bundle $\widetilde{M}$ of lattices over the
classifying space $BG$ through its holonomy around $\pi_1(BG)=G$.  The
group (co)homology is defined to be the sheaf (co)homology,
\begin{equation}
  H_\ast\big( G, M \big) = H_\ast\big( BG, \widetilde{M} \big)
  ,\quad
  H^\ast\big( G, M \big) = H^\ast\big( BG, \widetilde{M} \big)
  .
\end{equation}
This is a formal, but rather unhelpful definition of group homology
and cohomology. However, although defined abstractly via classifying
spaces, the actual group homology groups are very computable.  All one
has to do is compute the cohomology (kernel modulo image) of a certain
complex, see~\cite{MR1324339, MR2035696}.  The boundary maps are given
explicitly in terms of the $G$-representation matrices.  Computing
kernel modulo image then boils down to finding the Smith normal form
of the boundary maps, which we calculate using Maple. Basic properties
include
\begin{itemize}
\item $H^0(G,M)=M^G$, the invariant subspace.
\item $H_0(G,M)=M_G$, the coinvariants (See
  \autoref{sec:Homcoinvariant})
\item $H^i(G,M)=0=H_i(G,M)$ for $i<0$.
\item $H^i(G,M)$ and $H_i(G,M)$ are \emph{finite} Abelian groups
  for $i>0$.
\end{itemize}

Finally, note that any $\ZZZ$ representation restricts to a $\Z_3$
representations for each choice of $\Z_3\subset \ZZZ$. We are going to
need these in the following. Let us write
\begin{equation}
  G = \ZZZ = G_1\times G_2 =
  \{g_1, g_1^2, g_1^3=1\} \times 
  \{g_2, g_2^2, g_2^3=1\}
  .
\end{equation}
Of course, there is also a third (diagonal) $\Z_3$ subgroup of $\ZZZ$,
which we denote by $G_{12}=\{1,g_1g_2, g_1^2g_2^2\}$. For example,
restriction of the $\ZZZ$-representation $R$, see eq.~\eqref{eq:Rdef},
then defines three $\Z_3$-representations
\begin{equation}
  R_1 \eqdef
  R|_{G_1}
  ,\quad
  R_2 \eqdef
  R|_{G_2}
  ,\quad
  R_{12} \eqdef
  R|_{G_{12}}
  ~\in~
  \Z_3\text{--Rep}
\end{equation}
corresponding to these three $\Z_3$ subgroups. There are the analogous
restrictions of $R^\dual$ and $H3$.

\subsection{Invariant Cohomology}
\label{sec:Cohinvariant}

We start by computing the invariant cohomology of $\Xt$. This is also
the degree zero group cohomology of the topological cohomology of
$\Xt$,
\begin{equation}
  H^i\big(\Xt,\Z\big)^G
  = 
  H^0\Big(    H^i\big(\Xt,\Z\big)   \Big)
  .
\end{equation}
In particular, let us discuss the case $i=2$. The invariants of a
$G=\ZZZ$ group representation are simple to compute. All one has to do
is find the kernel of $\Id-g_1$ and $\Id-g_2$, which is a
straightforward linear algebra exercise. For the \dP9 base surfaces,
one obtains\footnote{The middle dimensional homology is self-dual. On
  $B_1$, $B_2$ this is in degree $2$. This is why we are not careful
  in distinguishing the curves on $B_i$ and their Poincar\'e duals
  here.}
\begin{equation}
  \label{eq:H2BiInv}
  H^2\big(B_i,\Z\big)^G 
  \simeq
  H_2\big(B_i,\Z\big)^G =
  \Span\big\{ 
  f ,\, t
  \big\}
\end{equation}
where we defined\footnote{Geometrically, $t$ is the pull-back of the
  hyperplane divisor via the blow-up map $B_i\to\CP^2$.}
\begin{equation}
  \label{eq:tdef}
  \begin{split}
    t \eqdef&\;
    -3\sigma-3 f+
    \theta_{11}+
    \theta_{21}+
    2\theta_{31}+2\theta_{32}+
    3\theta_{41}+\theta_{42}+
    3\mu+3\nu
    \\ =&\;
    5f+5\sigma -2 \alpha_1 -\alpha_2 +\alpha_8
    .   
  \end{split}  
\end{equation}
On the Calabi-Yau threefold $\Xt$, the degree-$2$ invariant cohomology
group is then (see~\cite{Braun:2004xv})
\begin{equation}
  \label{eq:H2invcohomology}
  H^2\big(\Xt,\Z\big)^G
  \simeq 
  H_4\big(\Xt,\Z\big)^G
  =
  \Span\big\{
  \pi_1^{-1}(f)=\pi_2^{-1}(f)
  ,\, \pi_1^{-1}(t)
  ,\, \pi_2^{-1}(t)
  \big\}
  .
\end{equation}
Let us define the invariant cohomology generators to
be\footnote{\label{fn:divisors}Again, we explicitly write the
  identification $H^2\simeq H_4$ as $c_1\big(\Osheaf(-)\big)$. This
  identification will be implicit in the future.}
\begin{equation}
  \label{eq:phitaudef}
\begin{split}
  \phi 
  \eqdef&\;
  c_1\Big( \Osheaf\big(\pi_1^{-1}(f)\big)\Big)
  =
  c_1\Big( \Osheaf\big(\pi_2^{-1}(f)\big)\Big)
  ,\\
  \tau_1
  \eqdef&\;
  c_1\Big( \Osheaf\big(\pi_1^{-1}(t)\big)\Big)
  ,\quad
  \tau_2
  \eqdef 
  c_1\Big( \Osheaf\big(\pi_2^{-1}(t)\big)\Big)  
  \quad \in H^2\big(\Xt,\Z\big)
  , 
\end{split}
\end{equation}
so that
\begin{equation}
  \label{eq:H2invcohomologyphitau}
  H^2\big(\Xt,\Z\big)^G
  \simeq 
  H_4\big(\Xt,\Z\big)^G
  =
  \Span_\Z \big\{
  \phi,\, \tau_1,\, \tau_2
  \big\}
  .
\end{equation}
The triple intersection numbers are encoded in the products of $\phi$,
$\tau_1$, $\tau_2$. One finds that
\begin{equation}
  H^\even\big(\Xt,\Z\big)^G = 
  \Z[\tau_1,\tau_2,\phi]
  \Big/
  \left<
    \phi^2
    ,\, 
    \tau_1^3
    ,\, 
    \tau_2^3
    ,\,
    \tau_1 \phi = 3 \tau_1^2
    ,\,
    \tau_2 \phi = 3 \tau_2^2
  \right>
  .
\end{equation}
Similarly, one can compute the invariant part under the $G=\ZZZ$
action of all cohomology groups of $\Xt$. We find that
\begin{equation}
  \label{eq:XtInvCohomology}
  H^0\Big( H^i\big(\Xt, \Z\big) \Big)
  =
  H^i\big(\Xt, \Z\big)^G
  =
  \begin{cases}
    \Z
    & i=6 \\
    0
    & i=5 \\
    \Z^3
    & i=4 \\
    \Z^8
    & i=3 \\
    \Z^3
    & i=2 \\
    0
    & i=1 \\
    \Z
    & i=0 
    .
  \end{cases}
\end{equation}

As far as cohomology with real (or complex) coefficients is concerned,
the cohomology of the quotient is simply the invariant cohomology on
the covering space. That is, for example,
\begin{equation}
  H^2\big( \Xt, \R \big)^G 
  = 
  \Span_\R \big\{
  \phi,\, \tau_1,\, \tau_2
  \big\}
  =
  \R^3
  \quad \Rightarrow \quad
  H^2\big( X, \R \big)
  = 
  \R^3
  ,
\end{equation}
and, in particular, $h^{11}(X)=3$. However, determining the cohomology
with integral coefficients on $X$ is far more difficult and will be
the subject of \autoref{sec:CoHomology}.

\subsection{Coinvariant Homology}
\label{sec:Homcoinvariant}

The dual notion to invariant cohomology is coinvariant homology, also
known as the degree zero group homology group of the homology groups
of $\Xt$,
\begin{equation}
  H_i\big(\Xt,\Z\big)_G
  = 
  H_0\Big(    H_i\big(\Xt,\Z\big)   \Big)
  .  
\end{equation}
Since we are mainly interested in curves, we are going to consider the
$i=2$ case in detail. It turns out that there is a clear reason why
the coinvariant curves are of particular interest. To see this,
consider the $G=\ZZZ$-quotient map
\begin{equation}
  q: \Xt \to X
  .
\end{equation}
This map of manifolds determines the push-forward $q_\ast$ of homology
groups as follows. Pick any $2$-cycle $\Ct\subset \Xt$, and let us
denote its image by $C \eqdef q\big(\Ct\big)\, \subset X$.
\begin{itemize}
\item If $\dim_\R C <2$, then $q_\ast\big(\Ct\big) = 0$.
\item If $q|_\Ct: \Ct\to C$ is one-to-one, then $q_\ast\big(\Ct\big) = C$.
\item If $q|_\Ct: \Ct\to C$ is $n$-to-one, then $q_\ast\big(\Ct\big) =
  n C$.
\end{itemize}
One tautological property of the push-forward is that 
\begin{equation}
  q_\ast\big(\Ct\big) 
  = 
  q_\ast\Big(g(\Ct)\Big) 
  \quad
  \forall g\in G,\, \Ct \in H_2\big(\Xt,\Z\big)
  .
\end{equation}
In other words,
\begin{equation}
  q_\ast\Big(\Ct - g(\Ct) \Big) 
  = 0
  \quad
  \forall g\in G,\, \Ct \in H_2\big(\Xt,\Z\big)
  .
\end{equation}
Put yet differently, there are obvious relations
\begin{equation}
  \label{eq:Idef}
  I \eqdef
  \Span_\Z\Big\{
  \Ct - g(\Ct) \;\Big|\;
  g\in G,\, \Ct\in H_2(\Xt,\Z) \Big\}
  \subset H_2\big(\Xt,\Z\big)
\end{equation}
that push forward to zero. The quotient by these relations is called
the \textdef{coinvariant} homology,
\begin{equation}
  H_2\big(\Xt,\Z\big)_G
  \eqdef 
  H_2\big(\Xt,\Z\big) \Big/ I
  .
\end{equation}
The push-forward map obviously factorizes
\begin{equation}
  \vcenter{\xymatrix@!0@R=20mm@C=16mm{
      H_2\big(\Xt,\Z\big) 
      \ar[rr]^{q_\ast} 
      \ar[rd]_{\text{mod }I} 
      &&
      H_2\big(X,\Z\big) \\
      &
      H_2\big(\Xt,\Z\big)_G
      \ar[ru]_{\hat{q}_\ast}
    }}  
\end{equation}
One nice set of generators for the relations $I$ using the notation
of eq.~\eqref{eq:Xthomologybasis} is
\begin{equation}
  \label{eq:coinvH2rel}
  \begin{gathered}
    \sigma \FPtimes \theta_{ij} = \sigma \FPtimes \theta_{11}
    \quad \forall i=1,2,3,4;\, j=0,1,2;
    \\
    \theta_{ij} \FPtimes \sigma = \theta_{11} \FPtimes \sigma
    \quad \forall i=1,2,3,4;\, j=0,1,2;
    \\
    \sigma \FPtimes f = 3\, \sigma \FPtimes \theta_{11}
    ,
    \qquad
    f \FPtimes \sigma = 3\, \theta_{11} \FPtimes \sigma
    ,
    \\
    2\, \sigma \FPtimes \sigma = 
    \mu \FPtimes \sigma + \sigma \FPtimes \mu
    , 
    \qquad 
    \sigma \FPtimes \sigma + \nu \FPtimes \sigma = 
    2\, \sigma \FPtimes \nu    
    ,
    \\
    3 \big( \sigma \FPtimes \mu - \sigma\FPtimes \sigma \big) = 0
    ,
    \qquad
    3 \big( \sigma \FPtimes \nu - \sigma\FPtimes \sigma \big) = 0
    .
  \end{gathered}
\end{equation}
Interestingly, the last two relations can only be obtained with an
overall factor of $3$, but not without! For example, take
\begin{equation}
  \begin{split}
  \Ct_1 \eqdef&\;
  2\, \sigma \FPtimes \theta_{31} 
  -2\, \sigma \FPtimes \theta_{41}
  + \theta_{21} \FPtimes \sigma
  + \theta_{31} \FPtimes \sigma
  + 3\, \mu\FPtimes\sigma 
  - 3\, \nu\FPtimes\sigma
  ,\\
  \Ct_2 \eqdef&\;
   2\, \sigma \FPtimes \theta_{32} 
  +2\, \sigma \FPtimes \theta_{41}
  -2\, \theta_{31} \FPtimes \sigma
  - \theta_{32} \FPtimes \sigma
  - \theta_{41} \FPtimes \sigma
  - \theta_{42} \FPtimes \sigma
  ,
  \end{split}
\end{equation}
then
\begin{equation}
  \Ct_1 - g_1\big(\Ct_1\big) + \Ct_2 - g_2\big(\Ct_2\big)
  = 
  3 \big( \sigma \FPtimes \mu - \sigma\FPtimes \sigma \big)
  .
\end{equation}
We conclude that the coinvariant homology of $\Xt$ can be written as
\begin{equation}
  \label{eq:H2GcoinvGenerators}
  \begin{split}
    H_2\big(\Xt,\Z\big)_G =&\;
    \big(\sigma\FPtimes\theta_{11}\big) \Z
    \oplus
    \big(\theta_{11}\FPtimes\sigma\big) \Z
    \oplus
    \big(\sigma\FPtimes\sigma\big) \Z
    \\
    &\;
    \oplus
    \big( \sigma \FPtimes \mu - \sigma\FPtimes \sigma \big) \Z_3
    \oplus
    \big( \sigma \FPtimes \nu - \sigma\FPtimes \sigma \big) \Z_3
    \\
    \simeq &\; \Z^3 \oplus \Z_3 \oplus \Z_3    
    .
  \end{split}
\end{equation}
Moreover, the push-downs of the generating curves have clear geometric
interpretations:
\begin{itemize}
\item X is again elliptically fibered over $B_1/G_1$ and $B_2/G_2$.
  The homology class of the fiber is
  $q_\ast(\theta_{11}\FPtimes\sigma)$ and
  $q_\ast(\sigma\FPtimes\theta_{11})$, respectively.
\item Due to the two independent elliptic fibrations, X is also
  fibered by Abelian surfaces $X\to\CP^1$. Note that, since the $G$
  action on $\Xt$ is by translation along fibers, it does not act on
  the base $\CP^1$. The zero section is
  $q_\ast(\sigma\FPtimes\sigma)$.
\item The torsion curves $q_\ast(\sigma \FPtimes \mu - \sigma\FPtimes
  \sigma)$ and $q_\ast(\sigma \FPtimes \nu - \sigma\FPtimes \sigma)$
  are differences of sections of the Abelian surface fibration.
\end{itemize}

Similarly to the above, we have computed all of the coinvariant
homology groups of $\Xt$ with respect to $G=\ZZZ$, and found
\begin{equation}
  \label{eq:XtCoinvHomology}
  H_0\Big( H_i\big(\Xt, \Z\big) \Big)
  =
  H_i\big(\Xt, \Z\big)_G
  =
  \begin{cases}
    \Z
    & i=6 \\
    0
    & i=5 \\
    \Z^3 \oplus \Z_3
    & i=4 \\
    \Z^8 \oplus \big(\Z_3\big)^4
    & i=3 \\
    \Z^3 \oplus \Z_3 \oplus \Z_3
    & i=2 \\
    0
    & i=1 \\
    \Z
    & i=0 
    .
  \end{cases}
\end{equation}

Recall that, modulo torsion, the invariant (co)homology of $\Xt$ is
the (co)homology of $X$. Is the coinvariant homology of $\Xt$ exactly
equal to the homology of the quotient $X$, including the torsion
subgroups? In general, this is not an easy question, and one needs
extra generators and extra relations. However, as we will show in
\autoref{sec:CoHomology}, in degree $2$ the coinvariant homology does
capture the whole torsion information, that is
\begin{equation}
  \hat{q}_\ast\Big[  
  \underbrace{H_2\big(\Xt,\Z\big)_{G,~\tors}} 
  _{= \Z_3\oplus\Z_3}
  \Big]
  = 
  H_2\big(X,\Z\big)_\tors
  =
  \Z_3\oplus\Z_3
  .
\end{equation}
On the other hand, the free part $H_2(\Xt,\Z)_{G,~\free}\simeq\Z^3$
does not push down to the whole $H_2(X,\Z)$, as we will discuss later
in detail.

\subsection{Group (Co)homology Groups}
\label{sec:groups}

So far, we have only computed the degree $0$ group homology and group
cohomology groups of the representations $R$, $R^\dual$, $H3$ in
eq.~\eqref{eq:RH3def}. However, in order to compute the homology of
the quotient $X$, which will be done in the next section, we also need
the higher group homology and group cohomology groups.

Because the case of a cyclic group ($\Z_3$) is simpler, let us first
consider the restriction of $R$, $R^\dual$, $H3$ to different $Z_3$
subgroups of $G=\ZZZ$.  Since we have the group action given in terms
of explicit integer matrices, finding any particular group
(co)homology group is just a linear algebra exercise, see
\autoref{sec:modularrep}.  Combined with the fact that the positive
degree cohomology groups of a cyclic group are $2$-periodic, this
determines all $\Z_3$ group (co)homology groups. We have computed all
of these group (co)homology groups, and found that they are
\begin{equation}
  \label{eq:HRresult}
  \begin{aligned}
    H^j\big(\Z_3, R_i\big) = 
    H^j\big(\Z_3, R_i^\dual\big) 
    &\simeq
    \begin{cases}
      \Z_3\oplus\Z_3 & j=2k \\
      \Z_3 & j=2k+1 \\
      \Z^7 & j = 0
    \end{cases}
    \\
    H_j\big(\Z_3, R_i\big) = 
    H_j\big(\Z_3, R_i^\dual\big) 
    &\simeq
    \begin{cases}
      \Z_3 & j=2k \\
      \Z_3\oplus\Z_3 & j=2k+1 \\
      \Z^7\oplus\Z_3 & j = 0
    \end{cases}
  \end{aligned}
\end{equation}
and
\begin{equation}
  \label{eq:HH3result}
  \begin{aligned}
    H^j\big(\Z_3,H3_i\big)
    = 
    H^j\big(\Z_3,H3_i^\dual\big)
    &\simeq
    \begin{cases}
      \big(\Z_3\big)^6 & j=2k \\
      \big(\Z_3\big)^2 & j=2k+1 \\
      \Z^{16} & j = 0
    \end{cases}
    \\
    H_j\big(\Z_3,H3_i\big)
    = 
    H_j\big(\Z_3,H3_i^\dual\big)
    &\simeq
    \begin{cases}
      \big(\Z_3\big)^2 & j=2k \\
      \big(\Z_3\big)^6 & j=2k+1 \\
      \Z^{16}\oplus\Z_3\oplus \Z_3 & j = 0
    \end{cases}
  \end{aligned}
\end{equation}
independently of whether $i=1$, $2$, or $12$. 

Finally, we will need the group homology and group cohomology of
$\ZZZ$. We have already determined the degree zero part in
Subsections~\ref{sec:Cohinvariant} and~\ref{sec:Homcoinvariant}, but
will need some of the higher degrees in the following. They turn out
to be
\begin{equation}
  \renewcommand{\arraystretch}{1.5}
  \label{eq:HZ3Z3result}
  \begin{array}{c|ccccccc}
    i & 0 & 1 & 2 & 3 & 4 & 6 & \cdots \\ \hline
    H_i\big(G,R\big) & 
    \Z^3 \oplus (\Z_3)^2
    & (\Z_3)^5 & (\Z_3)^5 & (\Z_3)^8 & (\Z_3)^8 & (\Z_3)^{11}
    &\cdots
    \\
    H_i\big(G,R^\dual\big) & 
    \Z^3 \oplus \Z_3
    & (\Z_3)^4 & (\Z_3)^4 & (\Z_3)^7 & (\Z_3)^7 & (\Z_3)^{10}
    &\cdots
    \\
    H^i\big(G,R\big) & 
    \Z^3 &
    \Z_3 & (\Z_3)^4 & (\Z_3)^4 & (\Z_3)^7 & (\Z_3)^7 
    &\cdots
    \\
    H^i\big(G,R^\dual\big) &     
    \Z^3 & 
    (\Z_3)^2 & (\Z_3)^5 & (\Z_3)^5 & (\Z_3)^8 & (\Z_3)^8
    &\cdots
    \\
    H_i\big(G,H3\big) &
    \Z^8 \oplus (\Z_3)^4 & 
    (\Z_3)^{12} & (\Z_3)^9 & (\Z_3)^{17} & (\Z_3)^{14} & (\Z_3)^{22}
    &\cdots
    \\
    H^i\big(G,H3\big) & 
    \Z^8 &
    (\Z_3)^4 & (\Z_3)^{12} & (\Z_3)^9 & (\Z_3)^{17} & (\Z_3)^{14}
    &\cdots
  \end{array}
\end{equation}
Interestingly, this proves that the representation $R$ is not
isomorphic to its dual.


\section{Homology and Cohomology}
\label{sec:CoHomology}

\subsection{General Form}
\label{sec:CoHomResult}

We now have all the information necessary to compute the homology and
cohomology groups with integer coefficients on $\Xt/\Z_3$ and
$\Xt/(\ZZZ)$. However, since this involves many mathematical details,
we first preview the results. The non-mathematically oriented reader
is advised to peruse this subsection only, skipping the remainder of
\autoref{sec:CoHomology}.

We begin by considering the integral homology groups. As we have
already mentioned, the rank of the integral homology of the quotient
is determined by the rank of the coinvariant homology of $\Xt$.  For
$X/\Z_3$, this can be read off from the degree-$0$ group homology
groups ($j=0$) in eqns.~\eqref{eq:HRresult} and~\eqref{eq:HH3result}.
Similarly, the $i=0$ column in eq.~\eqref{eq:HZ3Z3result} provides
this information for $X=\Xt/(\ZZZ)$. However, this only determines the
free part of the homology of $X$ and gives us no information on the
torsion part, which must be computed in another way. Note that,
although there are in principle seven non-vanishing homology groups on
a $6$-dimensional manifold, only four of them can contain a torsion
subgroup. Moreover, using Poincar\'e duality and the Universal
Coefficient Theorem, there are only two distinct torsion subgroups,
each occurring twice in the homology of the $6$-dimensional
manifold~\cite{MR0321934}. In our case, one of the torsion subgroups
is simply determined from the group action and the ensuing fundamental
groups $\pi_1(\Xt/\Z_3)=\Z_3$ and $\pi_1(X)=\Z_3\oplus\Z_3$. We denote
the remaining unknown finite subgroup by $T_3$ and $T_{33}$,
respectively. Putting all of this information together, the integral
homology of the quotients must be of the form
\begin{equation}
  \label{eq:X3X33cohomology}
  H_i\big(\Xt\big/\Z_3, \Z\big) 
  \simeq
  \begin{cases}
    \Z
    \\
    0
    \\
    \Z^{7} \oplus \Z_3
    \\
    \Z^{16} \oplus T_3
    \\
    \Z^{7} \oplus T_3
    \\
    \Z_3
    \\
    \Z
    , \\
  \end{cases}
  \quad
  H_i\big(\Xt\big/(\ZZZ), \Z\big) 
  \simeq
  \begin{cases}
    \Z
    & i=6 \\
    0
    & i=5 \\
    \Z^{3} \oplus \big(\Z_3\big)^2
    & i=4 \\
    \Z^{8} \oplus T_{33}
    & i=3 \\
    \Z^{3} \oplus T_{33}
    & i=2 \\
    \big(\Z_3\big)^2
    & i=1 \\
    \Z
    & i=0
    . \\
  \end{cases}
\end{equation}
In the remainder of this section, we are going to compute $T_3$ and
$T_{33}$. The result will be that
\begin{equation}
  T_3 \simeq \Z_3
  ,\quad
  T_{33} \simeq \Z_3\oplus\Z_3
  .
\end{equation}

In fact, we can be more precise and identify the geometry of the
torsion curves. We will see that the torsion curves are images of
curves on the covering space $\Xt$, something that is not automatic.
Explicitly, the push-forward by the quotient maps $\Hat{q}:\Xt\to
\Xt/\Z_3$ and $q:\Xt\to X$ is an isomorphism
\begin{equation}
  \label{eq:H2GisSigmaTors}
  \begin{split}
    \hat{q}_\ast:&~
    H_2\big(\Xt,\Z\big)_{\Z_3,\tors}
    ~\isolongrightarrow~
    H_2\big(\Xt/\Z_3,\Z\big)_\tors
    ,\\
    q_\ast:&~
    H_2\big(\Xt,\Z\big)_{G,\tors}
    ~\isolongrightarrow~
    H_2\big(X,\Z\big)_\tors
  \end{split}
\end{equation}
between the torsion parts of coinvariant homology on $\Xt$ and the
homology on the quotient. Note that the free parts of the respective
homology groups are equal as well, raising the obvious question
whether the push-forward is an isomorphism for the whole integral
homology. For the intermediate quotient, $\Xt/\Z_3$, this is indeed so
and
\begin{equation}
  \hat{q}_\ast:~
  H_2\big(\Xt,\Z\big)_{\Z_3}
  ~\isolongrightarrow~
  H_2\big(\Xt/\Z_3,\Z\big)  
  .
\end{equation}
However, on $X$ there is the following subtlety. The degree-$2$
homology classes on any simply connected manifold, for example $\Xt$,
can always be represented by spheres and, therefore, the image of
$q_\ast$ is a linear combination of spheres. But on $X$ not every
degree-$2$ homology class can be represented by spheres. To make this
more precise, we denote the spherical homology classes by
$\Sigma_2(X,\Z)$. A convenient definition is to start with $\pi_2(X)$,
the second homotopy group of $X$, and look at its image in homology,
that is,
\begin{equation}
  \label{eq:Sigmadef}
  \Sigma_2(X,\Z) \eqdef 
  \img \big[ \pi_2(X) \big] \subset H_2(X,\Z)
  .
\end{equation}
In our case, it turns out that
\begin{equation}
  \begin{gathered}
    \Sigma_2\big(\Xt/\Z_3,\Z\big) = H_2\big(\Xt/\Z_3,\Z\big)
    ,
    \\
    \Sigma_2\big(X,\Z\big)_\tors = H_2\big(X,\Z\big)_\tors
    ,    
  \end{gathered}
\end{equation}
while
\begin{equation}
  \Sigma_2\big(X,\Z\big)_\free \subsetneq H_2\big(X,\Z\big)_\free
\end{equation}
is a sublattice of index $3$. To summarize, the push-forward by the
quotient maps actually is an isomorphism
\begin{equation}
  \label{eq:H2GisSigma}
  \hat{q}_\ast:~
  H_2\big(\Xt,\Z\big)_{\Z_3}
  ~\isolongrightarrow~
  \Sigma_2\big(\Xt/\Z_3,\Z\big)
  ,\qquad 
  q_\ast:~
  H_2\big(\Xt,\Z\big)_G
  ~\isolongrightarrow~
  \Sigma_2\big(X,\Z\big)
  ,
\end{equation}
between the coinvariant homology and the homology classes that are
representable by linear combinations of spheres. Since we are only
interested in the genus $0$ worldsheet instantons for the purposes of
this paper, we actually only need $\Sigma_2$ and not $H_2$.

As a final remark, note that $X$ is a non-toric example where the
mirror symmetry conjecture of~\cite{Batyrev:2005jc} holds: Let $Y$ and
$Y^\ast$ be a pair of mirror Calabi-Yau threefolds. Then it is
conjectured\footnote{This mirror conjecture can be written in terms of
  integral cohomology as well. The equivalent statement then is $H^2(
  Y, \Z)_\tors = H^3( Y^\ast, \Z)_\tors$.} that
\begin{equation}
  H_1\big( Y, \Z\big)_\tors = H_2\big( Y^\ast, \Z\big)_\tors
  .
\end{equation}
Previously~\cite{Batyrev:2005jc}, this has been checked for the $16$
toric hypersurfaces with non-trivial fundamental group. In those $16$
cases $H_1\big( Y, \Z\big)_\tors=\pi_1(Y)$ is non-trivial while
$H_2\big( Y, \Z\big)_\tors=0$, and their mirror manifolds satisfy the
above relation. In our case, $X$ is ,presumably, self-mirror and, in
contrast to the toric hypersurface case, its mirror is again a free
quotient. The homology of $X$ again satisfies the above mirror
relation $H_1\big( X, \Z\big)_\tors=T_{33}=H_2\big( X,\Z\big)_\tors$.

\subsection{Spectral Sequences}
\label{sec:specseq}

We are now going to compute the remaining unknown torsion subgroups
$T_3$, $T_{33}$ in eq.~\eqref{eq:X3X33cohomology}. To do so, we will
rely on two spectral sequences which we will review below. Applying
one of these spectral sequences in \autoref{sec:X3specseq}, we will
compute the integral cohomology of $\Xt/\Z_3$. Using the other
spectral sequence, we will then attempt to compute $H_2(X,\Z)$ in
\autoref{sec:Xspecseq} and find that there are two possible answers.
Finally, in \autoref{sec:Xresult}, we resolve this ambiguity and
determine the integral homology and cohomology of $X$.

The cohomology version of the aforementioned spectral sequences
is~\cite{MR0037505, JPSerre}
\begin{theorem}[\LSss]
  For any manifold $Y$ with free\footnote{More generally, this
    spectral sequence computes the $G$-equivariant cohomology. For
    free group actions, this is the same as the cohomology of the
    quotient.} $G$ action, there is a cohomology spectral sequence
  \begin{equation}
    \label{eq:LSSS}
    E_2^{p,q} = 
    H^p\Big( G, H^q\big(Y,\Z\big) \Big)
    \quad\Longrightarrow\quad
    H^{p+q}\Big( Y/G, \Z \Big)
    .
  \end{equation}
  In particular, $E_2^{0,q} = H^q(Y,\Z)^G$ is the invariant
  cohomology. 
\end{theorem}
The analogous sequence for homology groups is~\cite{MR0023524}
\begin{theorem}[\CLss]
  For any manifold $Y$ with free $G$ action, there is a homology
  spectral sequence
  \begin{equation}
    \label{eq:CLSS}
    E^2_{p,q} = 
    H_p\Big( G, H_q\big(Y,\Z\big) \Big)
    \quad\Longrightarrow\quad
    H_{p+q}\Big( Y/G, \Z \Big)
    .
  \end{equation}
  In particular, $E^2_{0,q} = H_q(Y,\Z)_G$ is the coinvariant
  homology.
\end{theorem}
Hence, the \CLss{} describes the precise relationship between
coinvariant homology and the homology of the quotient. Dually, the
\LSss{} describes the precise relationship between invariant
cohomology and the cohomology of the quotient.

\subsection{The Partial Quotient}
\label{sec:X3specseq}

As a warm-up exercise, and since we are going to need some of these
results in the following, we begin with the computation of the
cohomology of the partial quotient $\Xt/G_i$, where $G_i\simeq \Z_3$
(see \autoref{sec:modularrep}). It turns out that nothing depends on
whether we consider $G_1$, $G_2$, or $G_{12}$, so we need not make any
distinction between them in this subsection. Note that, while the
$\ZZZ$ group action is not toric, any single $\Z_3$ subgroup can be
chosen to act only by phase multiplications. For example, in the
coordinates used in eqns.~\eqref{eq:B1} and \eqref{eq:B2}, the $g_1$
action, eq.~\eqref{eq:g1action}, is toric.  Hence, the partial
quotient can also be treated using toric methods, see
\xautoref{sec:Bquotient} in \partB~\cite{PartB}. In particular, its
integral homology groups could be computed as
in~\cite{Batyrev:2005jc}.

We use the \LSss{} to compute the cohomology of $X/G_i$ starting from
the $G_1$ group action on the cohomology of $\Xt$. The $E_2$ tableau
consists of the group cohomology groups computed in
eqns.~\eqref{eq:HRresult} and~\eqref{eq:HH3result},
\begin{equation}
  \label{eq:GiLSssE2}
  E_2^{p,q}\big(\Xt/G_i\big) = 
  \vcenter{\xymatrix@=1mm{
      {\scriptstyle q=6}\hspace{1.6mm} &
      \Z & 0 & \Z_3 & 0 & \Z_3 & 0 & \Z_3 & 0
      & \cdots \\
      {\scriptstyle q=5}\hspace{1.6mm} &
      0 & 0 & 0 & 0 & 0 & 0 & 0 & 0
      & \cdots \\
      {\scriptstyle q=4}\hspace{1.6mm} &
      \Z^7 & \Z_3 & 
      \Z_3^2 \ar[drr] & 
      \Z_3 \ar[drr] & 
      \Z_3^2 & 
      \Z_3 & \Z_3^2 & \Z_3
      & \cdots \\
      {\scriptstyle q=3}\hspace{1.6mm} &
      \Z^{16} & \Z_3^2 & \Z_3^6 & 
      \Z_3^2 & 
      \Z_3^6 \ar[drr] & 
      \Z_3^2 \ar[drr] & 
      \Z_3^6 & \Z_3^2 
      & \cdots \\
      {\scriptstyle q=2}\hspace{1.6mm} &
      \Z^7 
      & \Z_3 & \Z_3^2 & \Z_3 & \Z_3^2 & \Z_3 & \Z_3^2 & \Z_3
      & \cdots \\
      {\scriptstyle q=1}\hspace{1.6mm} &
      0 & 0 & 0 & 0 & 0 & 0 & 0 & 0
      & \cdots \\
      {\scriptstyle q=0}\hspace{1.6mm} &
      \Z & 0 & \Z_3 & 0 & \Z_3 & 0 & \Z_3 & 0
      & \cdots \\
      \ar[]+/r 3.4mm/+/u 1.7mm/;[rrrrrrrrr]+/r 3mm/+/r 3.4mm/+/u 1.7mm/
      \ar[]+/r 3.4mm/+/u 1.7mm/;[uuuuuuu]+/u  2mm/+/r 3.4mm/+/u 1.7mm/
      & 
      {\vbox{\vspace{3.5mm}}\scriptstyle p=0} & 
      {\vbox{\vspace{3.5mm}}\scriptstyle p=1} & 
      {\vbox{\vspace{3.5mm}}\scriptstyle p=2} & 
      {\vbox{\vspace{3.5mm}}\scriptstyle p=3} & 
      {\vbox{\vspace{3.5mm}}\scriptstyle p=4} & 
      {\vbox{\vspace{3.5mm}}\scriptstyle p=5} & 
      {\vbox{\vspace{3.5mm}}\scriptstyle p=6} & 
      {\vbox{\vspace{3.5mm}}\scriptstyle p=7} & 
    }}\,. 
\end{equation}
The $E_2$ tableau is obviously not bounded to the right. However, in
the $E_\infty$ tableau all entries with $p+q>6$ have to vanish since
$H^{p+q}\big(\Xt/\Z_3,\Z\big)=0$ if $p+q>6$. Hence, the superfluous
entries must be removed by higher differentials. Since the $E_2$
tableau is $2$-periodic for sufficiently large $p$, we first consider
the case where every differential starts or ends in the periodic
range. Counting the ranks of possible differentials, the entries can
only be completely removed if every non-zero differential either
starts or ends in the $q=3$ row. And, moreover, each such differential
starting or ending at $q=3$ must have maximal rank.

This argument determines all differentials for sufficiently large $p$,
but we also need the differentials for small $p$. Note that the
cohomology Leray-Serre spectral sequence is actually a spectral
sequence of $H^\ast(\Z_3,\Z)$-algebras. Therefore, the differentials
\begin{equation}
  d_r^{p,q}: E_r^{p,q}\longrightarrow E_r^{p+r,q-r+1}
\end{equation}
for $p\gg 0$ are all induced from $d_r^{0,q}$, $d_r^{1,q}$, and
multiplication with the generator in $E_r^{2,0}$. Hence we know all
$d_2$ differentials, not only the ones with $p\gg 0$. Therefore, we
determine the next tableau to be
\begin{equation}
  \label{eq:GiLSssE3}
  E_3^{p,q} = 
  \vcenter{\xymatrix@=1mm{
      {\scriptstyle q=6}\hspace{1.6mm} &
      \Z & 0 & \Z_3 & 0 & \Z_3 & 0 & \Z_3 & 0
      & \cdots \\
      {\scriptstyle q=5}\hspace{1.6mm} &
      0 & 0 & 0 & 0 & 0 & 0 & 0 & 0
      & \cdots \\
      {\scriptstyle q=4}\hspace{1.6mm} &
      \Z^7 
      & 0 & 0 & 0 & 0 & 0 & 0 & 0
      & \cdots \\
      {\scriptstyle q=3}\hspace{1.6mm} &
      \Z^{16} & \Z_3 & \Z_3^2 & 0 & \Z_3^2 & 0 & \Z_3^2 & 0 
      & \cdots \\
      {\scriptstyle q=2}\hspace{1.6mm} &
      \Z^7 & 
      \Z_3 \ar[ddrrr]|(0.5){d_3} & 
      0 & 0 & 0 & 0 & 0 & 0
      & \cdots \\
      {\scriptstyle q=1}\hspace{1.6mm} &
      0 & 0 & 0 & 0 & 0 & 0 & 0 & 0
      & \cdots \\
      {\scriptstyle q=0}\hspace{1.6mm} &
      \Z & 0 & \Z_3 & 0 & \Z_3 & 0 & \Z_3 & 0
      & \cdots \\
      \ar[]+/r 3.4mm/+/u 1.7mm/;[rrrrrrrrr]+/r 3mm/+/r 3.4mm/+/u 1.7mm/
      \ar[]+/r 3.4mm/+/u 1.7mm/;[uuuuuuu]+/u  2mm/+/r 3.4mm/+/u 1.7mm/
      & 
      {\vbox{\vspace{3.5mm}}\scriptstyle p=0} & 
      {\vbox{\vspace{3.5mm}}\scriptstyle p=1} & 
      {\vbox{\vspace{3.5mm}}\scriptstyle p=2} & 
      {\vbox{\vspace{3.5mm}}\scriptstyle p=3} & 
      {\vbox{\vspace{3.5mm}}\scriptstyle p=4} & 
      {\vbox{\vspace{3.5mm}}\scriptstyle p=5} & 
      {\vbox{\vspace{3.5mm}}\scriptstyle p=6} & 
      {\vbox{\vspace{3.5mm}}\scriptstyle p=7} & 
    }}\,. 
\end{equation}
The $d_3$ drawn above must vanish, since the range has to survive
until $d_4^{0,3}:\Z^{16}\to \Z_3$. Hence, $E_3^{p,q}=E_4^{p,q}$ and
the $d_4$-cohomology is 
\begin{equation}
  \label{eq:GiLSssEinfty}
  E_5^{p,q} = 
  E_\infty^{p,q} = 
  \vcenter{\xymatrix@=1mm{
      {\scriptstyle q=6}\hspace{1.6mm} &
      \Z & 0 & 0 & 0 & 0 & 0 & 0 & 0
      & \cdots \\
      {\scriptstyle q=5}\hspace{1.6mm} &
      0 & 0 & 0 & 0 & 0 & 0 & 0 & 0
      & \cdots \\
      {\scriptstyle q=4}\hspace{1.6mm} &
      \Z^7 
      & 0 & 0 & 0 & 0 & 0 & 0 & 0
      & \cdots \\
      {\scriptstyle q=3}\hspace{1.6mm} &
      \Z^{16} & \Z_3 & \Z_3 & 0 & 0 & 0 & 0 & 0 
      & \cdots \\
      {\scriptstyle q=2}\hspace{1.6mm} &
      \Z^7 & 
      \Z_3 & 0 & 0 & 0 & 0 & 0 & 0
      & \cdots \\
      {\scriptstyle q=1}\hspace{1.6mm} &
      0 & 0 & 0 & 0 & 0 & 0 & 0 & 0
      & \cdots \\
      {\scriptstyle q=0}\hspace{1.6mm} &
      \Z & 0 & \Z_3 & 0 & 0 & 0 & 0 & 0
      & \cdots \\
      \ar[]+/r 3.4mm/+/u 1.7mm/;[rrrrrrrrr]+/r 3mm/+/r 3.4mm/+/u 1.7mm/
      \ar[]+/r 3.4mm/+/u 1.7mm/;[uuuuuuu]+/u  2mm/+/r 3.4mm/+/u 1.7mm/
      & 
      {\vbox{\vspace{3.5mm}}\scriptstyle p=0} & 
      {\vbox{\vspace{3.5mm}}\scriptstyle p=1} & 
      {\vbox{\vspace{3.5mm}}\scriptstyle p=2} & 
      {\vbox{\vspace{3.5mm}}\scriptstyle p=3} & 
      {\vbox{\vspace{3.5mm}}\scriptstyle p=4} & 
      {\vbox{\vspace{3.5mm}}\scriptstyle p=5} & 
      {\vbox{\vspace{3.5mm}}\scriptstyle p=6} & 
      {\vbox{\vspace{3.5mm}}\scriptstyle p=7} & 
    }}\,.   
\end{equation}
Looking at the diagonals, there are no extension ambiguities and 
we can read off the cohomology. The Universal Coefficient Theorem then
fixes the homology. The result is
\begin{equation}
  \label{eq:HomCohXtZ3}
  H^i\big(\Xt\big/\Z_3, \Z\big) 
  \simeq
  \begin{cases}
    \Z      \\
    \Z_3    \\
    \Z^7\oplus \Z_3      \\
    \Z^{16} \oplus \Z_3  \\
    \Z^7 \oplus \Z_3     \\
    0  \\
    \Z
  \end{cases}
  \qquad \Rightarrow \quad  
  H_i\big(\Xt\big/\Z_3, \Z\big) 
  \simeq
  \begin{cases}
    \Z
    & i=6 \\
    0
    & i=5 \\
    \Z^7\oplus \Z_3
    & i=4 \\
    \Z^{16} \oplus \Z_3
    & i=3 \\
    \Z^7 \oplus \Z_3
    & i=2 \\
    \Z_3
    & i=1 \\
    \Z
    & i=0
    .
  \end{cases}
\end{equation}
Hence, we have determined $T_3$ in eq.~\eqref{eq:X3X33cohomology} to
be
\begin{equation}
  T_3\simeq \Z_3
  .
\end{equation}

Now that we know the result, let us return to the corresponding \CLss.
The bottom part of the $E^3$ tableau is
\begin{equation}
  \label{eq:GiCLssE3}
  E^3_{p,q}\big(\Xt/G_i\big) =
  \vcenter{\xymatrix@=2mm{
      {\scriptstyle q=2}\hspace{1.6mm} &
      \ar@{<-}[ddrrr]|(0.58){d^3_{(i)}}
      \Z^7\oplus \mathemph{\Z_3} & 
      \vdots & \vdots &\vdots & \vdots  & \adots \\
      {\scriptstyle q=1}\hspace{1.6mm} &
      0 & 0 & 0 & 0 & 0 
      & \cdots \\
      {\scriptstyle q=0}\hspace{1.6mm} &
      \Z & \Z_3 & 0 & \Z_3 & 0
      & \cdots \\
      \ar[]+/r 3.4mm/+/u 1.7mm/;[rrrrrr]+/r 3mm/+/r 3.4mm/+/u 1.7mm/
      \ar[]+/r 3.4mm/+/u 1.7mm/;[uuu]+/u  2mm/+/r 3.4mm/+/u 1.7mm/
      & 
      {\vbox{\vspace{3.5mm}}\scriptstyle p=0} & 
      {\vbox{\vspace{3.5mm}}\scriptstyle p=1} & 
      {\vbox{\vspace{3.5mm}}\scriptstyle p=2} & 
      {\vbox{\vspace{3.5mm}}\scriptstyle p=3} & 
      {\vbox{\vspace{3.5mm}}\scriptstyle p=4} & 
      {\vbox{\vspace{3.5mm}}\scriptstyle \cdots }&
    }}
  .
\end{equation}
From the cohomology computation, we know that the torsion curve
$\mathemph{\Z_3}$ has to survive\footnote{That is, must not be removed
  by differentials or extensions.} to
\begin{equation}
  H_2\big(\Xt/G_i,\Z\big) 
  =
  H_2\big(\Xt,\Z\big)_{G_i} 
  \simeq
  \Z^7 \oplus \mathemph{\Z_3}
  .
\end{equation}
Hence, the above differential 
\begin{equation}
  d^3_{(i)} :~
  E^3_{3,0}\big(\Xt/G_i\big) 
  \stackrel{0}{\longrightarrow}
  E^3_{0,2}\big(\Xt/G_i\big)
\end{equation}
must vanish. We will need this result in the following.

\subsection{The Full Quotient}
\label{sec:Xspecseq}

We now compute the degree-$2$ homology groups of $X=\Xt/G$ with
$G=\ZZZ$ using the Cartan-Leray spectral sequence. The bottom part,
which does not depend on $d_2$, is
\begin{equation}
  \label{eq:GCLssE3}
  \lrstack[b]{E^2_{p,q}\big(\Xt/G\big) =}{\quad E^3_{p,q}\big(\Xt/G\big) =}
  \vcenter{
    \xymatrix@=2mm{
      {\scriptstyle q=2}\hspace{1.6mm} &
      \ar@/_/@{<-}[ddrrr]|(0.35){d^3}
      \Z^3\oplus \Z_3\oplus \Z_3 
      & \vdots & \vdots &\vdots & \vdots  & \adots \\
      {\scriptstyle q=1}\hspace{1.6mm} &
      0 & 0 & 0 & 0 & 0 
      & \cdots \\
      {\scriptstyle q=0}\hspace{1.6mm} &
      \Z & \big(\Z_3\big)^2 & \Z_3 & \big(\Z_3\big)^3 & \big(\Z_3\big)^2
      & \cdots \\
      \ar[]+/r 3.4mm/+/u 1.7mm/;[rrrrrr]+/r 3mm/+/r 3.4mm/+/u 1.7mm/
      \ar[]+/r 3.4mm/+/u 1.7mm/;[uuu]+/u  2mm/+/r 3.4mm/+/u 1.7mm/
      & 
      {\vbox{\vspace{3.5mm}}\scriptstyle p=0} & 
      {\vbox{\vspace{3.5mm}}\scriptstyle p=1} & 
      {\vbox{\vspace{3.5mm}}\scriptstyle p=2} & 
      {\vbox{\vspace{3.5mm}}\scriptstyle p=3} & 
      {\vbox{\vspace{3.5mm}}\scriptstyle p=4} & 
      {\vbox{\vspace{3.5mm}}\scriptstyle \cdots }&
    }}
  .
\end{equation}
Knowing the differential in the $\Xt/G_i$ spectral sequence above, we
can determine the differential $d^3$ in the $\Xt/G$ spectral sequence
as follows. The quotient map
\begin{equation}
  q_i: \Xt/G_i \longrightarrow \Xt/G
\end{equation}
induces a morphism of spectral sequences 
\begin{equation}
  q_{i\ast}:
  \Big\{ E_{\bullet,\bullet}^r\big(\Xt/G_i\big), d^r_{(i)} \Big\}
  \longrightarrow
  \Big\{ E_{\bullet,\bullet}^r\big(\Xt/G\big), d^r \Big\}
  \,.
\end{equation}
In particular, for $r=3$ there is a commutative diagram
\begin{equation}
  \vcenter{\xymatrix{
      \Z_3
      \simeq
      E^3_{3,0}\big(\Xt/G_i\big) 
      \ar[r]^{d^3_{(i)}=0}
      \ar[d]^{G_i\subset G}
      & 
      E^3_{0,2}\big(\Xt/G_i\big) 
      \simeq
      \Z_3 \oplus \Z^7
      \ar@{->>}[d]^{q_{i\ast}}
      \\
      \Z_3\oplus\Z_3\oplus\Z_3
      \simeq
      E^3_{3,0}\big(\Xt/G\big) 
      \ar[r]^{d^3}
      & 
      E^3_{0,2}\big(\Xt/G\big) 
      \simeq
      \Z_3 \oplus \Z_3 \oplus \Z^3 
      \,.
    }}
\end{equation}
The $E^3_{p,0}$ terms are just group homology, and only depend on the
group. It is fairly clear that the inclusion $G_1\subset G$ and
$G_2\subset G$ map onto two of the three $\Z_3$ summands in
$H_3(G;\Z)$. A bit of homological algebra, see \autoref{sec:image},
shows that the inclusion of the diagonal $G_{12}\subset G$ then maps
onto the third summand. So we can find $3$ generators of
$E^3_{3,0}\big(\Xt/G\big)=H_3(G,\Z)$ which are induced from some
$E^3_{3,0}\big(\Xt/G_i\big)$. Moreover,
\begin{equation}
  q_{i\ast}: 
  \underbrace{H_2\big(\Xt,\Z\big)_{G_i}}_{=E^3_{0,2}\big(\Xt/G_i\big)}
  \longrightarrow
  \underbrace{H_2\big(\Xt,\Z\big)_{G}}_{=E^3_{0,2}\big(\Xt/G\big)}
\end{equation}
is surjective, since enlarging the group only adds more relations to
the coinvariant homology. Therefore, commutativity forces
\begin{equation}
  d^3 = 0
  \,.
\end{equation}
To summarize, we found that the following entries in the tableau
eq.~\eqref{eq:GCLssE3} survive to $r=\infty$,
\begin{equation}
  \label{eq:GCLssGEinfty}
  E^\infty_{p,q}\big(\Xt/G\big) =
  \vcenter{\xymatrix@=1mm{
      {\scriptstyle q=2}\hspace{1.6mm} &
      \Z^3\oplus \Z_3\oplus \Z_3 
      & \vdots & \vdots & \adots \\
      {\scriptstyle q=1}\hspace{1.6mm} &
      0 & 0 & 0
      & \cdots \\
      {\scriptstyle q=0}\hspace{1.6mm} &
      \Z & \big(\Z_3\big)^2 & \Z_3 
      & \cdots \\
      \ar[]+/r 3.4mm/+/u 1.7mm/;[rrrr]+/r 3mm/+/r 3.4mm/+/u 1.7mm/
      \ar[]+/r 3.4mm/+/u 1.7mm/;[uuu]+/u  2mm/+/r 3.4mm/+/u 1.7mm/
      & 
      {\vbox{\vspace{3.5mm}}\scriptstyle p=0} & 
      {\vbox{\vspace{3.5mm}}\scriptstyle p=1} & 
      {\vbox{\vspace{3.5mm}}\scriptstyle p=2} & 
      {\vbox{\vspace{3.5mm}}\scriptstyle \cdots }&
    }}
  .
\end{equation}

Having determined the endpoint of the Cartan-Leray spectral sequence
for $\Xt/G$, we still do not quite know its homology. We have to solve
one extension ambiguity, which takes the form of the short exact
sequence
\begin{equation}
  \label{eq:HomGses}
  0
  \longrightarrow
  \underbrace{
    H_2\big(\Xt,\Z\big)_G
  }_{\simeq \Z^3 \oplus \Z_3 \oplus \Z_3}
  \stackrel{q_\ast}{\longrightarrow}
  H_2\big(\Xt\big/G,\Z\big)
  \longrightarrow
  \underbrace{
    H_2\big(G,\Z\big)
  }_{\simeq \Z_3}
  \longrightarrow
  0
  \,,
\end{equation}
where the first map $q_\ast$ is just the pushforward by the $\ZZZ$
quotient map
\begin{equation}
  q: \Xt \longrightarrow \Xt/G = X
  .
\end{equation}
Depending on which extension is realized, the homology group could
either be
\begin{equation}
  \label{eq:H2ambiguity}
  H_2\big(X,\Z\big) 
  \simeq
  \Z^3 \oplus \big(\Z_3\big)^2
  \quad\text{or}\quad
  \Z^3 \oplus \big(\Z_3\big)^3
  .
\end{equation}
This leaves two possibilities, either $T_{33}=(\Z_3)^2$ or
$T_{33}=(\Z_3)^3$, for the torsion group in
eq.~\eqref{eq:X3X33cohomology}. In the next subsection, we will fix
this ambiguity.

\subsection{A Higher Differential and Final Result}
\label{sec:Xresult}

Recall that there is also a \LSss{} for the cohomology of the quotient
$X=\Xt/G$. Its $E_2$ tableau reads
\begin{equation}
  \label{eq:GLSssE2}
  E_2^{p,q}\big(\Xt/G\big) =
  \vcenter{
    \xymatrix@=2mm{
      {\scriptstyle \vdots}\hspace{1.6mm} &
      \vdots & \vdots & \vdots &\vdots & \vdots & \vdots & \adots \\
      {\scriptstyle q=3}\hspace{1.6mm} &
      \Z^8 & \Z_3^4 & \Z_3^{12} & 
      \Z_3^9 & \Z_3^{17} & \Z_3^{14} 
      & \cdots \\
      {\scriptstyle q=2}\hspace{1.6mm} &
      \Z^3 \ar[ddrrr]|(0.51){d_3}   & 
      \Z_3^2 & \Z_3^5 & \Z_3^5 & \Z_3^8 & \Z_3^8 
      & \cdots \\
      {\scriptstyle q=1}\hspace{1.6mm} &
      0 & 0 & 0 & 0 & 0 & 0
      & \cdots \\
      {\scriptstyle q=0}\hspace{1.6mm} &
      \Z & 0 & \Z_3^2 & \Z_3 & 
      \Z_3^3 & \Z_3^2
      & \cdots \\
      \ar[]+/r 3.4mm/+/u 1.7mm/;[rrrrrrr]+/r 3mm/+/r 3.4mm/+/u 1.7mm/
      \ar[]+/r 3.4mm/+/u 1.7mm/;[uuuuu]+/u  2mm/+/r 3.4mm/+/u 1.7mm/
      & 
      {\vbox{\vspace{3.5mm}}\scriptstyle p=0} & 
      {\vbox{\vspace{3.5mm}}\scriptstyle p=1} & 
      {\vbox{\vspace{3.5mm}}\scriptstyle p=2} & 
      {\vbox{\vspace{3.5mm}}\scriptstyle p=3} & 
      {\vbox{\vspace{3.5mm}}\scriptstyle p=4} & 
      {\vbox{\vspace{3.5mm}}\scriptstyle p=5} & 
      {\vbox{\vspace{3.5mm}}\scriptstyle \cdots }&
    }}
  .
\end{equation}
With this in mind, there are two dual ways of fixing the ambiguity
encountered in the previous subsection:
\begin{enumerate}
\item Identify the short exact sequence eq.~\eqref{eq:HomGses} with
  the sequence~\cite{MR0013312, Aspinwall:1994uj}
  \begin{equation}
    0
    \longrightarrow
    \Sigma_2\big(\Xt/G,\Z\big)
    \hooklongrightarrow
    H_2\big(\Xt\big/G,\Z\big)
    \longrightarrow
    H_2\big(G,\Z\big)
    \longrightarrow
    0    
    ,
  \end{equation}
  where $\Sigma_2$ are the homology classes of degree $2$ which are
  representable by spheres, see eq.~\eqref{eq:Sigmadef}. If one can
  find a higher genus holomorphic curve in $\Xt/G$ whose homology
  class is not representable by spheres, then the short exact sequence
  does not split. This way to fix the ambiguity was used
  in~\cite{Aspinwall:1994uj} for a certain quotient of the quintic.
\item If the differential $d_3:E_3^{0,2}\to E_3^{3,0}$ in
  eq.~\eqref{eq:GLSssE2} is non-trivial, then $E_\infty^{3,0}=0$ and
  the torsion part $H^3\big(\Xt/G,\Z\big)_\tors$ is at most
  $E_2^{1,2}=(\Z_3)^2$. Hence the second possibility in
  eq.~\eqref{eq:H2ambiguity} would be ruled out, fixing the
  ambiguity. 
\end{enumerate}
We will follow the latter route and compute
\begin{equation}
  d_3:~
  \underbrace{
    H^2\big(\Xt,\Z\big)^G
  }_{\simeq \Z^3}
  \longrightarrow
  \underbrace{
    H^3\big(G,\Z\big)
  }_{\simeq \Z_3}  
  .
\end{equation}
Note that we can identify two key objects with certain line bundles on
$\Xt$. Recall the correspondence between $H^2\big(\Xt,\Z\big)$ and
line bundles via the first Chern class, \autoref{sec:divisoraction}:
\begin{itemize}
\item $H^2\big(\Xt,\Z\big)^G$ are the $G$-invariant line bundles.
\item Evaluating the \LSss{}, eq.~\eqref{eq:GLSssE2}, yields
  \begin{equation}
    \ker\big(d_3\big)\oplus \big(\Z_3\big)^2 = 
    \left[ \bigoplus_{p+q=2} E_\infty^{p,q} \right] =
    H^2\big( X, \Z\big)
    .
  \end{equation}
  Pulling back to $\Xt$ via the quotient map kills the torsion part
  $(\Z_3)^2$, and we obtain
  \begin{equation}
    q^\ast\Big[ H^2(X,\Z) \Big]
    = \ker\big(d_3\big)
    \subset H^2\big(\Xt,\Z\big)^G
    \subset H^2\big(\Xt,\Z\big)
  \end{equation}
  But the pull-backs of line bundles on the quotient $X=\Xt/G$ are
  precisely the $G$-equivariant line bundles on $\Xt$. Hence,
  $\ker\big(d_3\big)$ are the $G$-equivariant line bundles.
\end{itemize}
The differential $d_3$ is either zero or surjective. Therefore,
$\ker(d_3)$ is either all of $H^2(\Xt,\Z)^G$ or an index-$3$
sublattice, respectively. In fact, the latter is true:
\begin{example}
  Consider the line bundle
  \begin{equation}
    \OsheafXt\big( \tau_i \big) 
    =
    \OsheafXt\Big( \pi_i^{-1} (t) \Big) 
    = 
    \pi_i^\ast \Big( \Osheaf_{B_i}(t) \Big)
  \end{equation}
  on $\Xt$, which is pulled back from one of the base \dP9{} surfaces
  $B_i$. This line bundle is $G$-invariant but not $G$-equivariant.
\end{example}
\begin{proof}
  The line bundle is invariant because $\pi_i^{-1} (t)$ is an
  invariant divisor class, see eq.~\eqref{eq:H2invcohomology}.  It
  remains to show that the line bundle is not equivariant. Assume, on
  the contrary, that $\pi_i^\ast \big( \Osheaf_{B_i}(t) \big)$ were
  equivariant. Then
  \begin{equation}
    \pi_{i\ast}\left[
      \pi_i^\ast \Big( \Osheaf_{B_i}(t) \Big)
    \right]
    = 
    \Osheaf_{B_i}(t)    
  \end{equation}
  would be equivariant, and hence $\Osheaf_{B_i}(t)|_f =
  \Osheaf_f(t\cdot f) = \Osheaf_f\big(3 \ptset\big)$ would be
  $G$-equivariant.  But $G\simeq\ZZZ$ acts on $f\simeq T^2$ by two
  independent order-$3$ translations, so any equivariant bundle must
  have degree divisible by $9$. Hence the degree $3$ line bundle
  $\Osheaf_f(t\cdot f)$ cannot be equivariant, contradicting our
  assumption.
\end{proof}
To summarize, the differential $d_3$ had to remove the
invariant-but-not-equivariant line bundles when descending to $X$ and
,hence, had to be nontrivial. Therefore, the torsion part
$H^3\big(\Xt,\Z\big)_\tors \simeq H_2\big(\Xt,\Z\big)_\tors$ in
eq.~\eqref{eq:H2ambiguity} can be at most $(\Z_3)^2$ and, therefore,
\begin{equation}
  H_2\big(X,\Z\big) 
  \simeq 
  \Z^3 \oplus \big(\Z_3\big)^2
  ,\quad
  H^3\big(X,\Z\big) 
  \simeq
  \Z^8 \oplus \big(\Z_3\big)^2
  \,.
\end{equation}
It follows that we have determined $T_{33}$ in
eq.~\eqref{eq:X3X33cohomology} to be
\begin{equation}
  T_{33}\simeq \Z_3\oplus\Z_3
  .
\end{equation}
This fixes the last ambiguity in the
integral homology and cohomology of $X$. The final result is
\begin{equation}
  \label{eq:HomCohXtZ3Z3}
  H^i\big(X, \Z\big) 
  =
  H_{6-i}\big(X, \Z\big) 
  \simeq
  \begin{cases}
    \Z
    & i=6 \\
    \Z_3\oplus\Z_3
    & i=5 \\
    \Z^3\oplus \Z_3\oplus\Z_3
    & i=4 \\
    \Z^8 \oplus \Z_3\oplus\Z_3
    & i=3 \\
    \Z^3 \oplus \Z_3\oplus\Z_3
    & i=2 \\
    0
    & i=1 \\
    \Z
    & i=0
    .
  \end{cases}
\end{equation}


\newpage
\part{Instantons}
\label{part:instantons}

\section{Quotients of the Quintic}
\label{sec:quintic}

\subsection{Curves and \Kahler{} Classes}
\label{sec:quinticcurves}

Having found the complete integral homology and cohomology groups
including torsion, we turn to the second topic of this paper, that is,
computing the Gromov-Witten invariants, or instanton numbers, on
$X=\Xt/(\ZZZ)$. We begin by reviewing the simpler and well-studied
case of the quintic Calabi-Yau threefold and its $\Z_5$ and
$\Z_5\times\Z_5$ quotients. Although the quintic and its quotients do
not have torsion curves, we will encounter some subtleties associated
with the group quotients that are also relevant to our case.

In particular, consider the one-parameter family
\begin{equation}
  Q \eqdef
  \left\{ 
    z_0^5 + z_1^5 + z_2^5 + z_3^5 + z_4^5 + 
    \psi^5 z_0 z_1 z_2 z_3 z_4
    =0
  \right\}
  \subset \CP^4
\end{equation}
of quintic threefolds. The defining equation is invariant under the
$\Z_5\times\Z_5\subset PGL(5,\C)$ group action
\begin{equation}
  \begin{split}
    [z_0:z_1:z_2:z_3:z_4]
    \mapsto&\;
    [z_1:z_2:z_3:z_4:z_0]
    \\
    [z_0:z_1:z_2:z_3:z_4]
    \mapsto&\;
    [z_0:
    e^{  \frac{2\pi i}{5}} z_1:
    e^{  \frac{4\pi i}{5}} z_2:
    e^{  \frac{6\pi i}{5}} z_3:
    e^{  \frac{8\pi i}{5}} z_4]
    .
  \end{split}
\end{equation}
The group action has fixed points in $\CP^4$, but they do not lie on
the hypersurface $Q$. Hence, the quotients\footnote{Of course, there
  are $6$ different $\Z_5$ subgroups in $\Z_5\times \Z_5$. However,
  that distinction will not be relevant in the following.} $Q/\Z_5$
and $Q/\big(\Z_5\times\Z_5\big)$ are smooth Calabi-Yau threefolds. Let
us put a bar over quantities on the $\Z_5$ quotient and use a double
bar for the $\Z_5\times\Z_5$ quotient,
\begin{equation}
  \Bar{Q} \eqdef Q/\Z_5
  ,\qquad
  \Bar{\Bar{Q}} \eqdef Q\Big/\big(\Z_5\times\Z_5\big) = 
  \Bar{Q} / \Z_5
  .
\end{equation}
The rational cohomology is always one-dimensional in each even degree,
generated by the hyperplane class of the ambient $\CP^4$. However, if
one keeps track of the proper normalization, things are slightly more
complicated. Moreover, there are torsion $1$-cycles corresponding to
the discrete Wilson lines on the quotients.

Recall that $h^{11}(Q)=1$ and $h^{21}(Q)=101$. Note specifically that
there is only a single \Kahler{} modulus. Thus, while the odd degree
cohomology groups are fairly large, the even degree cohomology, that
is $H^\even=H^0\oplus H^2\oplus H^4 \oplus H^6$, is very manageable.
For the quintic and its quotients they are
\begin{subequations}
  \begin{align}
    \label{eq:Qcohring}
    H^\even\big(Q,\Z\big) =&\;
    \Z[\xi_2, \xi_4] \Big/ 
    \big< 
    \xi_2^2 = 5 \xi_4
    ,\,
    (\dim\!>\!6)\;
    \big>
    \\
    H^\even\left(\Bar{Q},\Z\right) =&\;
    \Z\left[
      \Bar{\xi}_2, \Bar{\tau}_2
    \right] \Big/ 
    \big< 
    5 \Bar{\tau}_2
    ,\,
    \Bar{\tau}_2^2
    ,\,
    \Bar{\tau}_2 \Bar{\xi}_2 
    ,\,
    (\dim\!>\!6)\;
    \big>
    \\
    H^\even\left(\Bar{\Bar{Q}},\Z\right) =&
    \begin{array}[t]{r@{}l}
      \Z\left[
        \Bar{\Bar{\xi}}_2, \Bar{\Bar{\tau}}_2, 
        \Bar{\Bar{\rho}}_2, \Bar{\Bar{\xi}}_4, \Bar{\Bar{\xi}}_6
      \right] \Big/ 
      \big<&\strut
      5 \Bar{\Bar{\tau}}_2
      ,\,
      5 \Bar{\Bar{\rho}}_2
      ,\,
      \Bar{\Bar{\tau}}_2^2
      ,\,
      \Bar{\Bar{\tau}}_2 \Bar{\Bar{\rho}}_2
      ,\,
      \Bar{\Bar{\rho}}_2^2      
      ,\,
      \\ &\strut
      \Bar{\Bar{\tau}}_2 \Bar{\Bar{\xi}}_2 
      ,\,
      \Bar{\Bar{\tau}}_2 \Bar{\Bar{\xi}}_4
      ,\,
      \Bar{\Bar{\rho}}_2 \Bar{\Bar{\xi}}_2 
      ,\,
      \Bar{\Bar{\rho}}_2 \Bar{\Bar{\xi}}_4
      ,\,
      \\[1ex] &\strut
      \Bar{\Bar{\xi}}_2^2 = 5 \Bar{\Bar{\xi}}_4
      ,\,
      \Bar{\Bar{\xi}}_2 \Bar{\Bar{\xi}}_4 = 5 \Bar{\Bar{\xi}}_6
      ,\,
      (\dim\!>\!6)\;
      \big>            
      ,
    \end{array}
  \end{align}  
\end{subequations}
where the subscripts on the generators are their dimension and we do
not explicitly write the relations imposed by dimension $>6$ terms.
Note the appearance of torsion classes $\Bar{\tau}_2$,
$\Bar{\Bar{\tau}}_2$, and $\Bar{\Bar{\rho}}_2$. These are the first
Chern classes of flat line bundles (the Wilson lines).

The pull backs under the successive quotients can be determined by
computing the higher differentials in the \LSss. This is tedious but
straightforward, and we will not present the details. One finds that
\begin{equation}
  \vcenter{\xymatrix{
      \strut& H^0 \strut& 
      \strut& H^2 \strut& 
      \strut& H^4 
      \strut& H^6 \strut\\
      H^\even(Q,\Z) \ar@{}|-{\displaystyle=}[r] & 
      \Z 
      \ar@{}|{\displaystyle\oplus}[r] & 
      \xi_2 \Z 
      \ar@{}|{\displaystyle\oplus}[r] & 
      0 
      \ar@{}|{\displaystyle\oplus}[rr] &   & 
      \xi_4 \Z 
      \ar@{}|{\displaystyle\oplus}[r] & 
      \xi_2\xi_4\Z     
      \\
      H^\even\left(\Bar{Q},\Z\right) \ar@{}|-{\displaystyle=}[r] & 
      \Z
      \ar[u]_{\times 1} \ar@{}|{\displaystyle\oplus}[r] & 
      \Bar{\xi}_2 \Z
      \ar[u]_{\times 1} \ar@{}|{\displaystyle\oplus}[r]& 
      \Bar{\tau}_2 \Z_5
      \ar[u] \ar@{}|{\displaystyle\oplus}[r] & 
      0 
      \ar@{}|{\displaystyle\oplus}[r] & 
      \Bar{\xi}_2^2 \Z
      \ar[u]_{\times 5} \ar@{}|{\displaystyle\oplus}[r] & 
      \Bar{\xi}_2^3 \Z
      \ar[u]_{\times 5} 
      \\
      H^\even\left(\Bar{\Bar{Q}},\Z\right) \ar@{}|-{\displaystyle=}[r] & 
      \Z
      \ar[u]_{\times 1} \ar@{}|{\displaystyle\oplus}[r] & 
      \Bar{\Bar{\xi}}_2 \Z
      \ar[u]_{\times 5} \ar@{}|{\displaystyle\oplus}[r] & 
      \Bar{\Bar{\tau}}_2 \Z_5
      \ar[u]_{\times 1} \ar@{}|{\displaystyle\oplus}[r] & 
      \Bar{\Bar{\rho}}_2 \Z_5
      \ar[u] \ar@{}|{\displaystyle\oplus}[r] & 
      \Bar{\Bar{\xi}}_4 \Z
      \ar[u]_{\times 5} \ar@{}|{\displaystyle\oplus}[r] & 
      \Bar{\Bar{\xi}}_6 \Z
      \ar[u]_{\times 5}
      \save
      "4,2"."2,2"*+<5mm,7mm>\frm{^\}}
      \restore
      \save
      "4,4"."2,3"."4,5"*+<5mm,7mm>\frm{^\}}
      \restore
      \save
      "4,6"."2,6"*+<5mm,7mm>\frm{^\}}
      \restore
      \save
      "4,7"."2,7"*+<5mm,7mm>\frm{^\}}
      \restore
      ,
    }}
\end{equation}
where we picked integral generators in each even cohomology group. By
separating the different degrees, one can easily read off any even
cohomology group. For example, $H^2(\Bar{Q},\Z)=\Z\oplus\Z_5$ and it
is generated by $\Bar{\xi}_2$ and $\Bar{\tau}_2$. We observe that
there is only a single \Kahler{} modulus on $Q$, $\Bar{Q}$, and
$\Bar{\Bar{Q}}$. However, when comparing them there is a subtlety
involving the correct integral normalization.  The integral generator
$\Bar{\xi}_2$ pulls back to the integral generator $\xi_2$, while the
integral generator $\Bar{\Bar{\xi}}_2$ pulls back to five times the
integral generator $\Bar{\xi}_2$.

The corresponding Poincar\'e dual push downs in homology are
\begin{equation}
  \vcenter{\xymatrix{
      \strut& H_0 
      \strut& H_2 \strut& 
      \strut& H_4 \strut& 
      \strut& H_6 \strut\\
      H_\even(Q,\Z) \ar@{}|-{\displaystyle=}[r] & 
      \ptset\Z 
      \ar[d]_{\times 1} \ar@{}|{\displaystyle\oplus}[r] & 
      C \Z 
      \ar[d]_{\times 1} \ar@{}|{\displaystyle\oplus}[r] & 
      0 
      \ar[d] \ar@{}|{\displaystyle\oplus}[rr] &   & 
      D \Z 
      \ar[d]_{\times 5} \ar@{}|{\displaystyle\oplus}[r] & 
      Q Z     
      \ar[d]_{\times 5} 
      \\
      H_\even\left(\Bar{Q},\Z\right) \ar@{}|-{\displaystyle=}[r] & 
      \ptset\Z
      \ar[d]_{\times 1} \ar@{}|{\displaystyle\oplus}[r] & 
      \Bar{C} \Z
      \ar[d]_{\times 5}\ar@{}|{\displaystyle\oplus}[r]& 
      \Bar{\tau}_4 \Z_5
      \ar[d]_{\times 1} \ar@{}|{\displaystyle\oplus}[r] & 
      0 
      \ar[d] \ar@{}|{\displaystyle\oplus}[r] & 
      \Bar{D} \Z
      \ar[d]_{\times 5} \ar@{}|{\displaystyle\oplus}[r] & 
      \Bar{Q} \Z
      \ar[d]_{\times 5}
      \\
      H_\even\left(\Bar{\Bar{Q}},\Z\right) \ar@{}|-{\displaystyle=}[r] & 
      \ptset\Z
      \ar@{}|{\displaystyle\oplus}[r] & 
      \Bar{\Bar{C}} \Z
      \ar@{}|{\displaystyle\oplus}[r] & 
      \Bar{\Bar{\tau}}_4 \Z_5
      \ar@{}|{\displaystyle\oplus}[r] & 
      \Bar{\Bar{\rho}}_4 \Z_5
      \ar@{}|{\displaystyle\oplus}[r] & 
      \Bar{\Bar{D}} \Z
      \ar@{}|{\displaystyle\oplus}[r] & 
      \Bar{\Bar{Q}} \Z
      \save
      "4,2"."2,2"*+<5mm,7mm>\frm{^\}}
      \restore
      \save
      "4,3"."2,3"*+<5mm,7mm>\frm{^\}}
      \restore
      \save
      "4,5"."2,4"."4,6"*+<5mm,7mm>\frm{^\}}
      \restore
      \save
      "4,7"."2,7"*+<5mm,7mm>\frm{^\}}
      \restore
      ,
    }}
\end{equation}
where $C$, $\Bar{C}$, $\Bar{\Bar{C}}$ and $D$, $\Bar{D}$,
$\Bar{\Bar{D}}$ are generating curves\footnote{$C$ and $\Bar{C}$ can
  be taken to be rational curves, whereas the homology class of
  $\Bar{\Bar{C}}$ can \emph{not} be represented by a rational
  curve~\cite{Aspinwall:1994uj}. $\Bar{\Bar{C}}$ can be represented by
  a genus 1 curve.} and divisors, respectively.  Furthermore, we
denote by $\Bar{\Bar{\tau}}_4$, $\Bar{\Bar{\rho}}_4$ and
$\Bar{\tau}_4$ the torsion generators in $H_4(\Bar{Q},\Z)$ and
$H_4(\Bar{\Bar{Q}},\Z)$.  We observe again that, while the curve
classes are abstractly the same $1$-dimensional lattice
\begin{equation}
  H_2\big(Q,\Z\big)
  \simeq
  H_2\big(\Bar{Q},\Z\big)
  \simeq
  H_2\big(\Bar{\Bar{Q}},\Z\big)
  \simeq \Z
  ,
\end{equation}
the normalization of the curves is subtle. The $\Z_5$-quotient of the
generator $C$ is again a generator, but the $\Z_5$-quotient of the
generator $\Bar{C}$ is five times a generator in
$H_2(\Bar{\Bar{Q}},\Z)$.

\subsection{Instantons on the Quintic}
\label{sec:quinticinstanton}

We now turn to the worldsheet instanton corrections to certain Yukawa
couplings. To be more precise, we consider the $E_8\times E_8$
heterotic string on the quintic $Q$ (and, similarly, $\Bar{Q}$,
$\Bar{\Bar{Q}}$) with the standard embedding. This choice of gauge
bundle breaks $E_8\to E_6$.  Recall that the massless $E_6$ matter
fields correspond to the bundle-valued cohomology groups
\begin{equation}
  H^1\big(Q,TQ\big)
  ,\hspace{2cm}
  H^1\big(Q,TQ^\dual\big)  
  = 
  H^1\big(Q,\Omega_Q\big)  
  = 
  H^{1,1}(Q)
  = \xi_2 \C
\end{equation}
for the \Rep{27} and \barRep{27} representations, respectively.
Conveniently, there is a single \barRep{27} matter field corresponding
to $H^1(Q,TQ^\dual)$ and we will only consider its Yukawa couplings.
These can be computed by calculating a three-point function in the
A-model\footnote{Conversely, the Yukawa couplings of the fields coming
  from $H^1(Q,TQ)$ are a three-point function in the B-model.}
topological string. More precisely, the harmonic form associated with
the generator $\xi_2\in H^{1,1}(Q)$ corresponds to a chiral operator
$\mathscr{O}_{\xi_2}$ in the conformal field theory.  Classically, the
Yukawa coupling is just the triple overlap integral of $\xi_2$, or,
equivalently, the triple intersection number of the Poincar\'e dual
divisor. The result is that
\begin{equation}
  \big<\mathscr{O}_{\xi_2}^3\big>_\text{classical}
  = 
  \int_Q \xi_2 \wedge \xi_2 \wedge \xi_2 
  = 
  \int_Q 5 \xi_2 \wedge \xi_4
  = 
  5
  ,
\end{equation}
where we used the relation eq.~\eqref{eq:Qcohring} and the fact that
$\xi_2\wedge\xi_4$ is the properly normalized volume form. Due to a
non-renormalization theorem, there are no perturbative corrections.
However, genus $0$ worldsheet instantons can and do contribute. The
triumph of mirror symmetry was that this duality allows one to
actually calculate the instanton effects. For example, the correctly
normalized three-point function for the quintic turns out to
be~\cite{Candelas:1990rm}
\begin{subequations}
\begin{flalign}
  \label{eq:OQ}
  \hspace{25mm}
  \big<\mathscr{O}_{\xi_2}^3\big>
  =&\;
  5 + 2875 q + 4876875 q^2 + \cdots  
  , &
\end{flalign}
where $q=e^{2\pi\iunit t}$ is the minimal instanton action.
Similarly, the three-point function for the $\Z_5$ and
$\Z_5\times\Z_5$ quotient are given by~\cite{Aspinwall:1994uj}
\begin{flalign}
  \label{eq:OQbar}
  \hspace{25mm}
  \big<\Bar{\mathscr{O}}_{\Bar{\xi}_2}^3\big>
  =&\;
  1 + 575 q + 975375 q^2 + \cdots  
  ,
  &
  \\
  \label{eq:OQbarbar}
  \hspace{25mm}
  \big<\Bar{\Bar{\mathscr{O}}}_{\Bar{\Bar{\xi}}_2}^3\big>
  =&\;
  25 + 14375 q^5 + 24384375 q^{10} + \cdots  
  .
\end{flalign}
\end{subequations}

To count the number of instantons $n_d$ of volume $d$, one has to
compare these results with the formal $q$-series for the
instanton-corrected Yukawa coupling. This has the general
form~\cite{Candelas:1990rm}
\begin{equation}
  \label{eq:Oqseries}
  \big<\mathscr{O}^3\big> = 
  \kappa_{111}
  +
  \sum_{d=1}^\infty n_d d^3 \frac{q^d}{1-q^d}
  ,
\end{equation}
where $\kappa_{111}$ is the triple intersection number. Note that each
minimal curve can be wrapped multiply times, contributing at different
volumes. In the instanton expansion above, this is already taken into
account by the factor
\begin{equation}
   \frac{q^d}{1-q^d}
   = 
   q^d + 
   q^{2d} + 
   q^{3d} + 
   \cdots
   = 
   \sum_{i=1}^\infty 
   q^{id}
   .
\end{equation} 
Comparing the instanton-corrected three-point functions in
eqns.~\eqref{eq:OQ}, \eqref{eq:OQbar}, and \eqref{eq:OQbarbar} to the
general form of the instanton series eq.~\eqref{eq:Oqseries}, we can
read of the non-vanishing instanton numbers
\begin{equation}
  \renewcommand{\arraystretch}{1.5}
  \begin{array}{c|c|c}
    Q & \Bar{Q} & \Bar{\Bar{Q}} 
    \\ \hline\hline
    \kappa_{111} = 5 &
    \Bar{\kappa}_{111} = 1 &
    \Bar{\Bar{\kappa}}_{111} = 25 
    \\
    n_1 = 2875 &    
    \Bar{n}_1 = 575 = \frac{n_1}{5} &
    \Bar{\Bar{n}}_5 = 115 = \frac{n_1}{25}  
    \\
    n_2 = 609250  &    
    \Bar{n}_2 = 121850 = \frac{n_2}{5} &
    \Bar{\Bar{n}}_{10} = 24370 = \frac{n_2}{25}  
    \\
    n_3 = 317206375 &    
    \Bar{n}_3 = 63441275 = \frac{n_3}{5} &
    \Bar{\Bar{n}}_{15} = 12688255 = \frac{n_3}{25}  
    \\
    \vdots & \vdots & \vdots
    .
\end{array}
\end{equation}
We make two important observations, both of which apply to
$X=\Xt/(\ZZZ)$ as well:
\begin{itemize}
\item The number of rational curves on the quotient of some freely
  acting group $G$ is $\tfrac{1}{|G|}$ times the number of
  corresponding rational curves on the covering space.
\item Even if a curve class is primitive (not a multiple of another
  curve) on the covering space, its image on the quotient can still be
  non-primitive.
\end{itemize}
To summarize, we first computed the relations between the degree-$2$
homology and cohomology in the quintic $Q$ and its quotients
$\Bar{Q}$, $\Bar{\Bar{Q}}$. This allows one to compute the classical
$\barRep{27}^3$ Yukawa couplings. The classical result on the quintic
can be extended to the complete worldsheet instanton corrected
three-point functions using mirror symmetry. By comparing the
resulting instanton expansion with the formal $q$-series of the Yukawa
couplings, one can read off the instanton numbers on the covering
space $Q$. The corresponding instanton numbers on $\Bar{Q}$,
$\Bar{\Bar{Q}}$ are $\frac{1}{5}$ and $\frac{1}{25}$, respectively, of
the instanton numbers on $Q$. This last result is true for all free
quotients, and will be used in the following.

Having established these results, we now warn the reader that we will
\emph{not} continue to work with the Yukawa couplings.  Rather, we
will calculate the genus $0$ prepotential instead. For the quintic,
this amounts to the triple integral over the \Kahler{} modulus $t$,
\begin{equation}
  \Fprepot{Q}(q) 
  = 
  \iiint
  \big<\mathscr{O}_{\xi_2}^3\big> 
  \diff t^3
  =
  \frac{1}{3!}
  \kappa_{111} t^3 + p_2(t) +
  \frac{1}{(2\pi \iunit)^3}
  \underbrace{
    \sum_{d=1}^\infty n_d \Li_3(q^d)
  }_{
    \eqdef\FprepotNP{Q}(q) 
  }
  ,
\end{equation}
where $p_2(t)$ is a quadratic polynomial and
$\Li_3(q)=\sum_{n=1}^\infty \tfrac{q^n}{n^3}$ takes care of
multi-covers of the same curve. Clearly, the non-perturbative part
$\FprepotNP{Q}(q)$ of the prepotential contains the same
information about the instanton numbers as the three-point functions.
The real advantage of this formulation is that there is always only
one prepotential, whereas, for example on the $19$-parameter
Calabi-Yau $\Xt$, there would be $\binom{19+3-1}{3}=1330$ three-point
functions. On a general Calabi-Yau threefold, $Y$, with $r=h^{11}(Y)$
\Kahler{} moduli $t^1,\dots,t^r$, the prepotential is of the form
\begin{multline}
  \label{eq:PrepotExpansion}
  \Fprepot{Y}(q_1,\dots,q_r) 
  =\;
  \frac{1}{3!}
  \sum_{1\leq a \leq b \leq c \leq r}
  \kappa_{abc} t^a t^b t^c + p_2(t^1,\dots,t^r) 
  \\
  +
  \frac{1}{(2\pi \iunit)^3}
  \underbrace{
    \sum_{d_1,\dots,d_r} n_{(d_1,\dots,d_r)} 
    \Li_3
    \Bigg(\prod_{i=1}^r q_i^{d_i} \Bigg)
  }_{
    \eqdef\FprepotNP{Y}(q_1,\dots,q_r) 
  }
  ,  
\end{multline}
where $q_i=e^{2\pi\iunit t^i}$. The three-point functions can be
recovered as
\begin{equation}
  \left< 
    \mathscr{O}_i \mathscr{O}_j \mathscr{O}_\ell
  \right> = 
  \partial_{t^i}
  \partial_{t^j}
  \partial_{t^\ell}
  \Fprepot{Y}(q_1,\dots,q_r)
  .
\end{equation}


\section
[A-Model on the Covering Space $\mathbf{\Xt}$]
[A-Model on the Covering Space]
{A-Model on the Covering Space $\mathbf{\Xt}$}
\label{sec:AmodelXt}

\subsection{Curves}
\label{sec:Amodcurves}

We now return to the main objective of this paper, which is to compute
the instanton numbers (Gromov-Witten invariants) for the Calabi-Yau
threefold $X$ defined in \autoref{sec:CY}. However, before graduating
to the non-simply connected $X$, we first have to understand the
universal cover $\Xt$. Fortunately, a generic Schoen \CY{} threefold,
that is, the fiber product of two generic \dP9 surfaces, was studied
in~\cite{Hosono:1997hp}. Using the $E_8$ \MWgrp{} of a generic \dP9,
they expressed the prepotential in terms of $E_8$ theta functions, see
also~\cite{Donagi:1996yf}. Our covering space $\Xt$ is such a Schoen
\CY{} threefold, although one with a special $\ZZZ$ symmetry. In our
case, the \MWgrp{}s are just $MW(B_i)=\Z_3\oplus\Z_3$.  However,
although the actual curves change\footnote{This phenomenon is already
  familiar from the quintic, for which there are $375$ isolated curves
  and $50$ one-parameter families at the Fermat point, while
  generically all $2875=5\cdot 375+20 \cdot 50$ are isolated.}  as we
move to a $\ZZZ$ symmetric point in the complex structure moduli
space, the instanton numbers do not jump. So we might just as well use
the instanton numbers computed for generic complex structure moduli.

In the remainder of this subsection, we will review the above A-model
computation. Let $\Bhat_1$, $\Bhat_2$ be two generic \dP9 surfaces
($12I_0$ Kodaira fibers), and define the fiber product
\begin{equation}
  \Xhat = \Bhat_1 \times_{\CP^1} \Bhat_2
  .
\end{equation}
The surfaces $\Bhat_i$ now have infinitely many sections forming the
$E_8$ root lattice
\begin{equation}
  MW\big(\Bhat_i\big) \simeq \Lambda_{E_8}
  = \bigg(
  \Big\{ 
  \mathop{\boxplus}\limits_{i=1}^8 \, 
  \big(\boxplus_{n_i}\alpha_i\big) 
  ~\Big|~ n_i \in \Z
  \Big\}
  ,
  \, \big<-,-\big>
  \bigg)
  ,
\end{equation}
where we will use the notation of \autoref{sec:E8} for a choice of
simple roots. The \CY{} threefold $\Xhat\to\CP^1$ is fibered by
Abelian surfaces, so we again have a group law on the sections. This
defines the group
\begin{equation}
  MW\big(\Xhat\big) = 
  \Big\{ s_1 \FPtimes s_2
  ~\Big|~
  s_1 \in MW(\Bhat_1), s_2 \in MW(\Bhat_2) 
  \Big\}
  = MW\big(\Bhat_1\big) \oplus MW(\Bhat_2) 
  .
\end{equation}
Now we can describe part of the rational curves in $\Xhat$:
\begin{itemize}
\item Vertical curves\footnote{In other words, curves that project to
    a point in the base $\CP^1$. Put differently, curves $C$ such that
    $C\cdot \phi=0$, where $\phi$ is the $T^4$ fiber, see
    eq.~\eqref{eq:phitaudef}.} are precisely the components of
  singular fibers. The Abelian surface fibration $\Xhat\to\CP^1$ has
  $12$ singular fibers of type $I_0\times T^2$ and $12$ singular
  fibers of type $T^2\times I_0$, so there are $24$ families. The
  moduli space $\mathscr{M}_\text{Vert}$ of each family is a $T^2$, so
  $\chi\big(\mathscr{M}_\text{Vert}\big)=0$ and they do not contribute
  to the instanton numbers.
\item The sections in $MW(\Xhat)$ are the only \emph{smooth} rational
  curves $s$ with $s\cdot\phi=1$. 
\item Each (smooth) section $s$ passes through the singular fibers of
  $\Xhat\to\CP^1$. Pick, for example, one such $I_0\times T^2$.
  Amongst the one-parameter family of $I_0$, there is precisely one
  $I_0^s$ which intersects $s$. Therefore, $s \cup I_0^s$ is an
  isolated (reducible) rational curve. Those curves are called
  \textdef{pseudo-sections} in~\cite{Hosono:1997hp}, and all curves
  $C$ with $C\cdot \phi=1$ are either sections or of this form.
\item Multi-sections, that is, curves $C$ with $C\cdot \phi\geq 2$,
  are not yet understood.
\end{itemize}
These curves contribute to the instanton numbers with some (integral)
multiplicity. Roughly, the multiplicity is the Euler characteristic of
the moduli space of the curve (this needs to be refined if the moduli
space is singular). Hence,
\begin{itemize}
\item The moduli space $\mathscr{M}_\text{Vert}$ of each vertical
  curve is a $T^2$, so $\chi\big(\mathscr{M}_\text{Vert}\big)=0$ and
  they do not contribute to the instanton numbers.
\item Sections do not have infinitesimal deformations, $N_{s|\Xhat}
  = \Osheaf_s(-1)\oplus\Osheaf_s(-1)$. Hence, they contribute to the
  instanton numbers with multiplicity $1$. The volume of such a
  section is
  \begin{equation}
    V_s = \int_s J = s\cdot J
    ,
  \end{equation}
  where $J\in H^2\big(\Xhat,\R\big)$ is the \Kahler{} form.
\item Consider a pseudo-section $P$ consisting of a section $s$ and
  covering the $i$-th Kodaira fiber $m_i$ times. Then it contributes
  to the instanton numbers with a pre-factor (see~\cite{Hosono:1997hp,
    MR1622601})
  \begin{equation}
    \label{eq:pseudomult}
    n(P) = \prod_{i=1}^{24} p\big(m_i\big)
    ,
  \end{equation}
  where $p(k)$ is the number of partitions of $k\in\Z_\geq$. By
  definition, the homology class of a pseudo-section is
  \begin{equation}
    P = s + 
    \sum_{i=1}^{12} m_i \big( f\FPtimes\sigma  \big) +
    \sum_{i=13}^{24} m_i \big( \sigma\FPtimes f \big)
    ,
  \end{equation}
  where we labeled the Kodaira fibers such that the first $12$ are in
  the first fiber direction and the remaining $12$ are in the other
  fiber direction. Hence, the volume of a general pseudo-section is
  \begin{equation}
    V_P = \int_P J = 
    \int_{P_s} J ~+~ 
    \sum_{i=1}^{12} m_i
    \int_{f\FPtimes\sigma} J ~+~ 
    \sum_{i=13}^{24} m_i
    \int_{\sigma\FPtimes f} J
    .
  \end{equation}  
\end{itemize}

\subsection{Prepotential}
\label{sec:Amodelprepot}

Using the above knowledge about the curves, one can directly write
down their non-perturbative contribution to the
prepotential~\cite{Hosono:1997hp}. One obtains
\begin{multline}
  \label{eq:FXtNPinitial}
  \FprepotXtNP
  =
  \sum_{
    \begin{smallmatrix}
      s_1\FPtimes s_2\\ \in MW(\Xhat)
    \end{smallmatrix}}
  e^{2\pi\iunit \int_{s_1\FPtimes s_2} \omega}
  \left(
    \sum_{m=0}^\infty p(m) 
    e^{2\pi \iunit m \int_{f\FPtimes\sigma} \omega}
  \right)^{12}
  \left(
    \sum_{n=0}^\infty p(n) 
    e^{2\pi \iunit n \int_{\sigma\FPtimes f} \omega}
  \right)^{12}
  \\
  +
  (\text{contribution of curves with $C\cdot\phi\geq 2$})
\end{multline}
for the genus zero contribution to the prepotential on $\Xt$, where
$\omega = B + \iunit J$ is the complexified \Kahler{} form.  Note
that multi-covers of a pseudo-section contribute at the same order as
multi-sections, which is why we did not need to include the $\Li_3$
accounting for multi-covers at order $p$.

Let us define coordinates $t^a$ on the $19$-dimensional \Kahler{}
moduli space as
\begin{equation}
  \label{eq:KahlerModuliCoords}
  \omega = 
  t^1 \phi
  + t^2 \left( \pi_1^{-1} \sigma \right)
  + \sum_{i=1}^{8} t^{i+2} \big(\pi_1^{-1} \alpha_{i}\big)
  + t^{11} \left( \pi_2^{-1} \sigma \right)
  + \sum_{i=1}^{8} t^{i+11} \big(\pi_2^{-1} \alpha_{i}\big)
  ,
\end{equation}
where we used the basis for the cohomology adapted to the $E_8$
lattice given in eq.~\eqref{eq:BE8basis}. In addition, define the
Fourier-transformed coordinates
\begin{equation}
  \label{eq:p0q8r8def}
  \begin{gathered}
    p_0 \eqdef e^{2\pi\iunit t^1}
    = e^{2\pi\iunit \int_{PD(\phi)} \omega}
    ,
    \\
    q_0 \eqdef e^{2\pi\iunit t^2}
    ,\quad
    q_1 \eqdef e^{2\pi\iunit t^3}
    ,\dots,~
    q_8 \eqdef e^{2\pi\iunit t^{10}}
    ,
    \\
    r_0 \eqdef e^{2\pi\iunit t^{11}}
    ,\quad
    r_1 \eqdef e^{2\pi\iunit t^{12}}
    ,\dots,~
    r_8 \eqdef e^{2\pi\iunit t^{19}}
    .
  \end{gathered}  
\end{equation}
It follows that
\begin{equation}
  \begin{gathered}
    e^{2\pi \iunit \int_{f\FPtimes\sigma} \omega}
    =
    \prod_{i=0}^8 q_i
    ,\qquad
    e^{2\pi \iunit \int_{\sigma\FPtimes f} \omega}  
    =
    \prod_{i=0}^8 r_i
    ,\\
    e^{2\pi\iunit \int_{s_1\FPtimes s_2} \omega}
    =
    p_0 
    q_0^{s_1\cdot\sigma}
    \prod_{i=1}^8 q_i^{s_1\cdot \alpha_i} 
    r_0^{s_1\cdot\sigma}
    \prod_{i=1}^8 r_i^{s_2\cdot \alpha_i}
    ,    
  \end{gathered}
\end{equation}
and, hence,
\begin{multline}
  \FprepotXtNP
  =
  p_0
  \left(
    \sum_{
      \begin{smallmatrix}
        s_1\in \\ MW(\Bhat_1)
      \end{smallmatrix}}
    q_0^{s_1\cdot\sigma}
    \prod_{i=0}^8 q_i^{s_1\cdot \alpha_i} 
  \right)
  \left(
    \sum_{
      \begin{smallmatrix}
        s_2\in \\ MW(\Bhat_2)
      \end{smallmatrix}}
    r_0^{s_2\cdot\sigma}
    \prod_{i=0}^8 r_i^{s_2\cdot \alpha_i}
  \right)
  \times \\ \times
  \left( 
    \sum_{m=0}^\infty p(m) 
    \prod_{i=0}^8 q_i^m
  \right)^{12}
  \left(
    \sum_{n=0}^\infty p(n) 
    \prod_{i=0}^8 r_i^n
  \right)^{12}
  +
  O(p_0^2)
  .
\end{multline}
Finally, we note the appearance of the generating function for
partitions,
\begin{equation}
  \label{eq:GenfnPartitions}
  P(q) \eqdef
  \sum_{i=0}^\infty p(i) q^i =
  \frac{q^\frac{1}{24}}{\eta(\tfrac{1}{2\pi\iunit}\ln q)}
  ,
\end{equation}
and the $E_8$ theta function\footnote{Usually, the theta function is
  written as $\ThetaEeight(\tau_0;\tau_1,\dots,\tau_8)$ with
  $q_i=e^{2\pi\iunit\tau_i}$. However, we will use our notation since
  we are going to work with the Fourier-transformed variables
  everywhere.} (using eq.~\eqref{eq:E8intersect})
\begin{equation}
  \ThetaEeight(q_0; q_1,\dots, q_8) \eqdef 
  \sum_{\gamma \in \Lambda_{E_8}} 
  q_0^{\frac{1}{2} \left<\gamma,\gamma\right>}
  \prod_{i=1}^8 
  q_i^{\left<\gamma,\alpha_i\right>}
  = 
  \sum_{s \in MW(\Bhat)} 
  q_0^{\sigma\cdot s+1}
  \prod_{i=1}^8 
  q_i^{1+s\cdot\sigma-s\cdot\alpha_i}
  .
\end{equation}
Therefore,
\begin{equation}
  \label{eq:Aprepot}
  \FprepotXtNP
  (p_0,q_0,\dots,q_8,r_0,\dots,r_8)
  =
  \frac{p_0}{q_0 r_0}~
  \At(q_0,\dots,q_8)
  \At(r_0,\dots,r_8)
  +
  O(p_0^2)
  ,
\end{equation}
where we defined the auxiliary function
\begin{equation}
  \label{eq:Aq08def}
  \At(q_0,\dots,q_8)
  \eqdef
  \ThetaEeight\bigg( 
    \prod_{i=0}^8 q_i; q_1^{-1},\dots, q_8^{-1} \bigg)
  P\bigg(\prod_{i=0}^8 q_i\bigg)^{12}
\end{equation}
and the analogous expression for $\At(r_0,\dots,r_8)$. Note the
occurrence of negative powers of $q_0,\dots,q_8,r_0,\dots,r_8$. This
is simply an artifact of working in a basis that is adapted to the
$E_8$ lattice structure. In a basis adapted to the Mori cone and the
\Kahler{} cone, only positive powers will appear. Nevertheless, by
expanding the expression for the prepotential as a series in the $19$
variables $p_0$, $q_0,\dots,q_8$, $r_0,\dots,r_8$ and comparing this
with the general form eq.~\eqref{eq:PrepotExpansion}, one can read of
the instanton numbers on $\Xt$. Clearly, the instanton numbers will be
indexed by $19$ different degrees, making this expansion very
cumbersome.  Hence, we will refrain from presenting them explicitly.


\section{A-Model for Quotients}
\label{sec:AmodelX}

\subsection{Instantons and the Path Integral}
\label{sec:InstGeneral}

Before delving into the actual computation of the prepotential and
instanton numbers on the quotients of $\Xt$, we need to understand the
effect of torsion homology classes on the instanton sum. The
worldsheet instantons in question for an arbitrary Calabi-Yau
threefold $Y$ are holomorphic maps $\gamma:\Sigma\to Y$ from the
string worldsheet $\Sigma$ to the target space $Y$. The path integral
sums over all such curves. If we ignore torsion in the homology for a
moment, then the effect of an instanton is to add a factor
\begin{equation}
  e^{\iunit S} 
  \big[\gamma:\Sigma\to Y\big] = e^{2\pi \iunit \int_\Sigma \gamma^\ast \omega}
\end{equation}
to the path integral, where $S$ is the instanton action and
\begin{equation}
  \label{eq:omegarational}
  \omega = B + \iunit J  
  = \sum_a \big( B + \iunit J \big)^a \; e_a
  \quad \in H^2\big(Y,\C\big)  
\end{equation}
is the complexified \Kahler{} class\footnote{Since we are really using
  topological strings on a Calabi-Yau threefold, there cannot be any
  flux. That is, we require that $\diff B=0$ for the purposes of this
  paper.} expanded in some suitable basis $\{e_a\}$ of harmonic forms.
Changing variables to
\begin{equation}
  \label{eq:qrational}
  q_a = e^{2\pi\iunit ( B + \iunit J )^a}
  ,
\end{equation}
the instanton factor can be written as
\begin{equation}
  e^{\iunit S} [\gamma] = \prod_a q_a^{d_a}
\end{equation}
with exponents
\begin{equation}
  d_a = \int_\Sigma e_a
  \quad \in \Z_\geq
  .
\end{equation}
Here and everywhere else we assume that the chosen basis $\{e_a\}$ is
suitably normalized and, therefore, the exponents $d_a$ are integers.

Now, let us assume that $H_2(Y,\Z)$ contains some non-zero torsion
part. Since everything said so far only depends only on the integral
$\int_\Sigma$, one might at first think that the torsion part of the
homology class $\Sigma\in H_2(Y,\Z)$ does not enter the path integral
at all.  However, there is one fallacy in the above reasoning, namely,
that the $B$-field need not be globally defined. So, strictly
speaking, the integral $\int_\Sigma B$ is not defined. The correct way
is to think about the instanton factor for a flat $B$-field, $\diff
B=0$, as a map assigning to each worldsheet a non-zero complex
number\footnote{By definition, $\Cunits\eqdef\C-\{0\}$ as a
  multiplicative group.}
\begin{equation}
  e^{\iunit S}  :~ H_2\big(Y,\Z\big) \to \Cunits
  ,
\end{equation}
which can only be written in terms of an integral if one is willing to
ignore a subtlety. This subtlety~\cite{Aspinwall:1995rb} is that the
homology classes can have torsion, that is,
\begin{equation}
  H_2\big(Y,\Z\big) = 
  H_2\big(Y,\Z\big)_\free \oplus H_2\big(Y,\Z\big)_\tors
  =
  \Z^r \oplus \Big( \Z_{m_1} \oplus \cdots \oplus \Z_{m_k} \Big)
  ,
\end{equation}
where $r$ is the rank and the $m_i$, $i=1,\dots,k$ are the torsion
coefficients. If there is no torsion, that is, $k=0$, then the above
description is perfectly valid. However, in general one needs in
addition to the free generators
\begin{flalign}
  \hspace{30mm}
  q_a ~&\in \Hom\Big[ H_2\big(Y,\Z\big)_\free, \Cunits \Big]
  ,
  &
  a =&\;  1,\dots,r 
  \hspace{30mm}
\end{flalign}
the torsion generators
\begin{flalign}
  \hspace{30mm}
  b_i ~&\in \Hom\Big[ H_2\big(Y,\Z\big)_\tors, \Cunits \Big]
  ,
  &
  i =&\;  1,\dots,k
  ,
  \hspace{30mm}
\end{flalign}
where
\begin{equation}
  b_i^{m_i} = 1
  .
\end{equation}
In terms of this basis, the instanton factor must be expanded to
\begin{equation}
  e^{\iunit S} [\gamma] = 
  \prod_{a=1}^r q_a^{d_a}  
  ~
  \prod_{i=1}^k b_i^{\delta_i}    
\end{equation}
with integral exponents 
\begin{equation}
  d_a \in \big\{0, 1, 2, \dots \big\}
  ,\quad
  \delta_i \in \big\{0,\dots,m_i-1\big\}
  ,
\end{equation}
provided that the basis $q_a$, $b_i$ is correctly normalized. This
describes the contribution of any given instanton to the path
integral. The non-perturbative correction to the prepotential, see
eq.~\eqref{eq:PrepotExpansion}, generalizes in the obvious way to
\begin{multline}
  \label{eq:TorsPrepotExpansion}
  \Fprepot{Y}(q_1,\dots,q_r,\, b_1, \dots, b_k ) 
  =\;
  \frac{1}{3!}
  \sum_{1\leq a \leq b \leq c \leq r}
  \kappa_{abc} t^a t^b t^c + p_2(t^1,\dots,t^r) 
  \\
  +
  \frac{1}{(2\pi \iunit)^3}
  \underbrace{
    \sum_{\substack{d_1,\dots,d_r \\ \delta_1,\dots,\delta_k}} 
    n_{(d_1,\dots,d_r,\, \delta_1,\dots,\delta_k)} 
    \Li_3
    \Bigg(\prod_{a=1}^r q_a^{d_a} \prod_{i=1}^k b_i^{\delta_i} \Bigg)
  }_{
    \eqdef\FprepotNP{Y}(q_1,\dots,q_r,\, b_1,\dots,b_k) 
  }
  ,  
\end{multline}

Finally, let us remark on the proper normalization. In principle, the
normalization of the $q_a$, $b_i$ has to be such that they form an
integral basis for $\Hom\big[ H_2(Y,\Z), \Cunits \big]$. However,
since we are only considering the genus $0$ instantons in the
following, one need only consider curve classes that are representable
by spheres. Therefore, we will use generators
\begin{equation}
  \label{eq:qabiGen}
  \begin{aligned}
    q_a ~&\in \Hom\Big[ \Sigma_2\big(Y,\Z\big)_\free, \Cunits \Big]
    ,
    &
    a =&\;  1,\dots,r 
    ,
    \\
    b_i ~&\in \Hom\Big[ \Sigma_2\big(Y,\Z\big)_\tors, \Cunits \Big]
    ,
    &
    i =&\;  1,\dots,k
    ,
  \end{aligned}
\end{equation}
see eq.~\eqref{eq:Sigmadef}. These are more practical for our
purposes, but keep in mind that they might have to be subdivided to
write the higher genus prepotential, as we saw in
\autoref{sec:quintic}.
\begin{table}
  \centering
  \renewcommand{\arraystretch}{1.5}
  \begin{tabular}{c|cccc}
    \parbox[c][12mm]{25mm}{\centering Calabi-Yau \\ threefold}
    & 
    $r$ & 
    \parbox{25mm}{\centering Free \\ generators} & 
    $\big\{m_1,\dots,m_k\big\}$ & 
    \parbox{25mm}{\centering Torsion \\ generators}  
    \\ \hline
    $\Xt$ &
    $19$ & $\big\{p_0,q_0,\dots,q_8,r_0,\dots,r_8\big\}$ &
    $\varnothing$ & $\varnothing$ 
    \\
    $\Xb=\Xt/G_1$ &
    $7$ & $\big\{P,Q_1,Q_2,Q_3,R_1,R_2,R_3\big\}$ &
    $\big\{3\big\}$ & $\big\{b_1\big\}$
    \\ 
    $X=\Xt/G$ &
    $3$ & $\big\{p,q,r\big\}$ &
    $\big\{3,3\big\}$ & $\big\{b_1,b_2\big\}$
  \end{tabular}
  \caption{Variables used in this paper to expand the prepotential for
    different Calabi-Yau threefolds.}
  \label{tab:variables}
\end{table}
Since we will be interested in the prepotential for $\Xt$ and two of
its quotients, we list the names for the generators
eq.~\eqref{eq:qabiGen} in \autoref{tab:variables}. We refer the reader
to the respective sections for detailed definitions.

\subsection
[Quotienting the A-Model on $\Xt$]
[Quotienting the A-Model]
{Quotienting the A-Model on $\mathbf{\Xt}$}
\label{sec:quotientA}

We finally have everything in place to compute the prepotential on the
quotient $X=\Xt/G$. On general grounds, the $G=\ZZZ$-orbits of a
$\CP^1\subset \Xt$ must be $|G|$=9 distinct rational curves since
there is no fixed-point free holomorphic map $\CP^1\to\CP^1$. Hence,
there is a one-to-one correspondence between one rational curve on $X$
and a set of $|G|$ rational curves on $\Xt$, permuted by $G$.

Therefore, to compute the genus $0$ prepotential on the quotient $X$,
we should 
\begin{enumerate}
\item Start with the prepotential on $\Xt$. For the purposes of this
  subsection, we consider only the terms linear in $p_0$. This part of
  the prepotential was computed in eq.~\eqref{eq:FXtNPinitial}.
\item Impose the relations
  \begin{equation}
    \label{eq:CgCidentify}
    e^{2\pi\iunit \int_{\Ct} \omega} = 
    e^{2\pi\iunit \int_{g(\Ct)} \omega}
  \end{equation}
  for all $g\in G$ and for all curves $\Ct \in H_2(\Xt,\Z) \simeq
  \Z^{19}$.
\item Divide by $|G|$.
\end{enumerate}
Note that setting $\Ct=g(\Ct)$ in $H_2(\Xt,\Z)$ yields by definition
the coinvariant homology $H_2(\Xt,\Z)_G$, see eq.~\eqref{eq:Idef}.
Now, in general, this might not be enough to describe $H_2(X,\Z)$
since there are potentially higher differentials in the \CLss,
eq.~\eqref{eq:CLSS}. However, as we discovered in
\autoref{sec:CoHomology}, there are no such subtleties in our case
and, according to eq.~\eqref{eq:H2GisSigma}, the homology classes of
rational curves on $X$ are identified with the coinvariant homology on
$\Xt$.

So all we have to do is to implement the relation
eq.~\eqref{eq:CgCidentify} in the expression for the prepotential on
$\Xt$, eq.~\eqref{eq:FXtNPinitial}. This can be done by restricting
the complexified \Kahler{} class $\omega$, only allowing classes that
yield the same result when integrated over $\Ct$ or $g(\Ct)$. Those
classes are precisely the $G$-invariant \Kahler{} classes, see
eq.~\eqref{eq:H2invcohomology}. Hence, we would like to
set\footnote{This particular choice of generators has the added
  advantage that its basis elements also span the $G$-invariant
  \Kahler{} cone~\cite{Gomez:2005ii}
  \begin{equation}
    \Kcone\big(\Xt\big)^G = 
    \Span_{\R_>} \big\{ \phi, \tau_1, \tau_2 \big\}
    .
  \end{equation}
  As a consequence, the Fourier series of the prepotential will only
  contain non-negative powers.
}
\begin{equation}
  \label{eq:omegainvariant}
  \begin{split}
  \omega \;&=
  t^1_R \phi + t^2_R \tau_1 + t^3_R \tau_2
  \\ &=
  (t^1_R +5 t^2_R +5 t^3_R) \phi 
  \\ &\qquad +
  t^2_R \pi_1^{-1}(5\sigma) + 
  t^2_R \pi_1^{-1}(-2\alpha_1) + 
  t^2_R \pi_1^{-1}(-\alpha_2) + 
  t^2_R \pi_1^{-1}(\alpha_8) 
  \\ &\qquad +
  t^3_R \pi_2^{-1}(5\sigma) + 
  t^3_R \pi_2^{-1}(-2\alpha_1) + 
  t^3_R \pi_2^{-1}(-\alpha_2) + 
  t^3_R \pi_2^{-1}(\alpha_8)
  ,
  \end{split}
\end{equation}
where we used eqns.~\eqref{eq:H2invcohomologyphitau}
and~\eqref{eq:tdef}.  Unfortunately, this is not yet the correct way
to implement the relations in eq.~\eqref{eq:CgCidentify}. In fact,
this restriction on $\omega$ is too strong. Recall that two of the
relations in the coinvariant homology, see eq.~\eqref{eq:coinvH2rel},
only have to hold with a certain multiplicity, namely
\begin{equation}
  \label{eq:I3tors}
  3 \big( \sigma \FPtimes \mu - \sigma\FPtimes \sigma \big) = 0  
  ,\quad
  3 \big( \sigma \FPtimes \nu - \sigma\FPtimes \sigma \big) = 0
  .
\end{equation}
However, demanding that $\omega$ be $G$-invariant enforces a stronger
relation, one without the multiplicity, and, hence, kills the torsion
information.

To capture the torsion information, we need to add two more \Kahler{}
classes which feel the torsion curves. We choose
\begin{equation}
  \label{eq:betadef}
  \begin{split}
      \beta_1 \eqdef&\;
      \pi_1^{-1}\big( 
        -6\sigma+3\theta_{21}+4\theta_{31}+2\theta_{32}
        +4\theta_{41}+2\theta_{42}+6\mu
      \big)
      \\&\qquad
      +
      \pi_2^{-1}\big( 
        6\sigma-3\theta_{21}-4\theta_{31}-2\theta_{32}
        -4\theta_{41}-2\theta_{42}-6\mu
      \big)
      \\
      =&\;
      \pi_1^{-1}\big( 
        -24 \sigma
        + \alpha_1+3\alpha_2+6\alpha_3+4\alpha_4
        +3\alpha_5+3\alpha_6+ \alpha_7+3\alpha_8 
      \big)
      \\&\qquad
      +
      \pi_2^{-1}\big( 
        24 \sigma
        - \alpha_1-3\alpha_2-6\alpha_3-4\alpha_4
        -3\alpha_5-3\alpha_6- \alpha_7-3\alpha_8
        \big)
      \\
      =&\;
      PD\big( \sigma\FPtimes\mu \big)-PD\big( \mu\FPtimes\sigma \big)
      ,
      \\
      \beta_2 \eqdef&\;
      -27\phi 
      +
      \pi_1^{-1}\big( 
        12\sigma-6\theta_{11}-4\theta_{31}-8\theta_{32}
        -8\theta_{41}-4\theta_{42}-12\nu
      \big)
      \\&\qquad
      +
      \pi_2^{-1}\big( 
        6\sigma-3\theta_{11}-2\theta_{31}-4\theta_{32}
        -4\theta_{41}-2\theta_{42}-6\nu
      \big)
      \\
      =&\;
      \pi_1^{-1}\big( 
        24\sigma
        -2\alpha_1-4\alpha_2-6\alpha_3-4\alpha_4
        -2\alpha_5                    -6\alpha_8
      \big)
      \\&\qquad
      +
      \pi_2^{-1}\big( 
        12\sigma
        - \alpha_1-2\alpha_2-3\alpha_3-2\alpha_4
        - \alpha_5                   -3\alpha_8
      \big)
      \\
      =&\;
      PD\big( \sigma\FPtimes\nu \big)+2PD\big( \nu\FPtimes\sigma \big)
      -45\phi
      .
  \end{split}
\end{equation}
These two additional \Kahler{} classes, $\beta_1$ and $\beta_2$, have
exactly the right property: They are perpendicular to all relations in
the coinvariant homology, eq.~\eqref{eq:coinvH2rel}, except for the
last two (reproduced in eq.~\eqref{eq:I3tors}) that only need to hold
with multiplicity three. That is,
\begin{equation}
  \begin{gathered}
    \big(
    \sigma \FPtimes \theta_{mn} - \sigma \FPtimes \theta_{11}
    \big)
    \cdot \beta_i = 0
    \quad \forall m=1,2,3,4;\, n=0,1,2;
    \\
    \big(
    \theta_{mn} \FPtimes \sigma - \theta_{11} \FPtimes \sigma
    \big)
    \cdot \beta_i = 0
    \quad \forall m=1,2,3,4;\, n=0,1,2;
    \\
    \big(
    \sigma \FPtimes f - 3\, \sigma \FPtimes \theta_{11}
    \big)
    \cdot \beta_i = 0
    ,
    \qquad
    \big(
    f \FPtimes \sigma - 3\, \theta_{11} \FPtimes \sigma
    \big)
    \cdot \beta_i = 0
    ,
    \\
    \big(
    2\, \sigma \FPtimes \sigma -
    \mu \FPtimes \sigma + \sigma \FPtimes \mu
    \big)
    \cdot \beta_i = 0
    , 
    \qquad 
    \big(
    \sigma \FPtimes \sigma + \nu \FPtimes \sigma -
    2\, \sigma \FPtimes \nu    
    \big)
    \cdot \beta_i = 0
  \end{gathered}
\end{equation}
for $i=1,2$. Moreover, with respect to the two curve classes on $\Xt$
that push-forward to the torsion curve generators, see
eq.~\eqref{eq:H2GcoinvGenerators}, they form a dual basis:
\begin{equation}
  \begin{aligned}
    \big( \sigma \FPtimes \mu - \sigma\FPtimes \sigma \big) 
    \cdot \beta_1 =&\; 1
    ,
    &\qquad
    \big( \sigma \FPtimes \mu - \sigma\FPtimes \sigma \big) 
    \cdot \beta_2 =&\; 0      
    ,
    \\
    \big( \sigma \FPtimes \nu - \sigma\FPtimes \sigma \big)
    \cdot \beta_1 =&\; 0  
    ,
    &\qquad
    \big( \sigma \FPtimes \nu - \sigma\FPtimes \sigma \big)
    \cdot \beta_2 =&\; 1
    .
  \end{aligned}
\end{equation}
Hence, instead of restricting $\omega$ to the $3$-dimensional
invariant space eq.~\eqref{eq:omegainvariant}, we now restrict
$\omega$ to lie in the $5$-dimensional subspace of \Kahler{} forms
\begin{equation}
  \omega = 
  t^1_R \phi + t^2_R \tau_1 + t^3_R \tau_2
  + t^4_R \beta_1 + t^5_R \beta_2
  .
\end{equation}
As usual, it is more convenient to work with the Fourier-transformed
variables
\begin{equation}
  \label{eq:pqrb1b2def}
  p \eqdef e^{2\pi\iunit t^1_R}
  ,\quad
  q \eqdef e^{2\pi\iunit t^2_R}
  ,\quad
  r \eqdef e^{2\pi\iunit t^3_R}
  ,\quad
  b_1 \eqdef e^{2\pi\iunit t^4_R}
  ,\quad
  b_2 \eqdef e^{2\pi\iunit t^5_R}
  ,
\end{equation}
where
\begin{equation}
  b_1^3 = 1
  ,\quad
  b_2^3 = 1  
\end{equation}
since they correspond to the torsion curve classes. The
$5$-dimensional subset of the \Kahler{} moduli space parametrized by
the $t^a_R$ can, of course, be expressed in terms of special linear
combinations of the $19$ \Kahler{} moduli $t^a$ defined in
eq.~\eqref{eq:KahlerModuliCoords}. Then, using the definitions
eqns.~\eqref{eq:p0q8r8def} and~\eqref{eq:pqrb1b2def}, we obtain the
relations
\begin{equation}
  \label{eq:Z3Z3quotientVariables}
  \begin{gathered}
    p_0 = p q^5 r^5 
    \\
    \begin{gathered}
      q_0 = q^5   
      \\
      \begin{aligned}
        q_1 =&\; q^{-2} b_1 b_2   & q_2 =&\; q^{-1} b_2^2     \\
        q_3 =&\; 1              & q_4 =&\; b_1 b_2^2       \\
        q_5 =&\; b_2            & q_6 =&\; 1              \\
        q_7 =&\; b_1            & q_8 =&\; q              
      \end{aligned}
    \end{gathered}
    \qquad
    \begin{gathered}
      r_0 = r^5        
      \\
      \begin{aligned}
        r_1 =&\; r^{-2} b_1^2 b_2^2 & r_2 =&\; r^{-1} b_2        \\
        r_3 =&\; 1               & r_4 =&\; b_1^2 b_2        \\
        r_5 =&\; b_2^2           & r_6 =&\; 1               \\
        r_7 =&\; b_1^2           & r_8 =&\; q              
        .
      \end{aligned}
    \end{gathered}
  \end{gathered}
\end{equation}

We now have everything in place to compute the genus $0$ prepotential
on $X=\Xt/G$. Imposing the curve relations eq.~\eqref{eq:CgCidentify}
on the instanton sum for the prepotential on $\Xt$,
eq.~\eqref{eq:FXtNPinitial}, is completely equivalent to substituting
eq.~\eqref{eq:Z3Z3quotientVariables} in the final expression for the
prepotential on $\Xt$, eq.\eqref{eq:Aprepot}. The non-perturbative
prepotential on the quotient is then $\frac{1}{|G|}$ times the
prepotential on the covering space after the replacement. The result
is
\begin{equation}
  \label{eq:PrepotXfromXt}
  \begin{split}
    \FprepotXNP(p,q,r, b_1,b_2) =&\;
    \frac{1}{|G|}~ \FprepotXtNP
    (p_0,q_0,\dots,q_8,r_0,\dots,r_8)
    \Big|_{p_0=pq^5r^5,\dots,r_8=q}
    \\ 
    =&\;
    \frac{1}{9}
    p
    A(q,b_1,b_2)
    A(r,b_1^{-1},b_2^{-1})
    +
    O(p^2)
  ,
  \end{split}
\end{equation}
where we defined the auxiliary function, see eq.~\eqref{eq:Aq08def},
\begin{equation}
  \label{eq:Aqb1b2def}
  \begin{split}
    A(q,b_1,b_2)
    \eqdef&\;
    \At\big(
    q^5, 
    q^{-2} b_1 b_2, q^{-1} b_2^2, 1, b_1 b_2^2,
    b_2, 1, b_1, q \big)
    \\
    =&\;
    \ThetaEeight\big( 
    q^3;\, 
    q^2 b_1^2 b_2^2,\, q b_2,\, 1,\, b_1^2 b_2,\, 
    b_2^2,\, 1,\, b_1^2,\, q^{-1} \big)
    P\big( q^3 \big)^{12}    
  \end{split}
\end{equation}
and an analogous expression for $A(r,b_1^{-1},b_2^{-1})$. Expanding
$A(q,b_1,b_2)$ as a power series, we find
\begin{equation}
  \label{eq:Amodform}
  \begin{split}
    A(q,b_1,b_2)
    =&\;
    \Big(1+4 q+14 q^2+28 q^3+57 q^4+84 q^5
         +148 q^6+196 q^7+\cdots\Big)    
    \\ &\quad\times
    (1+b_1+b_1^2)(1+b_2+b_2^2)
    P(q^3)^{12}
    \\
    =&\;
    \Big(
      1+4 q+14 q^2+40 q^3+105 q^4+252 q^5+574 q^6 +1240 q^7
      +\cdots\Big)    
    \\ &\quad\times
    (1+b_1+b_1^2)(1+b_2+b_2^2)
    \\ 
    \in&\; \Z[[q]] \otimes 
    \Z[b_1,b_2]\big/ \left<b_1^3=1,b_2^3=1\right>
    .
  \end{split}
\end{equation}
Since the series expansion is invariant under
$(b_1,b_2)\mapsto(b_1^{-1},b_2^{-1})=(b_1^2,b_2^2)$, we only have to
replace $q\mapsto r$ in eq.~\eqref{eq:Amodform} to obtain the series
expansion for $A(r,b_1^{-1},b_2^{-1})$. 

To conclude, we have computed an explicit closed form for the
prepotential on $X=\Xt/(\ZZZ)$ at linear order in $p$. This was done
by starting with the prepotential on $\Xt$ and suitably ``modding
out'' the $\ZZZ$ action. One can now expand the prepotential
eq.~\eqref{eq:PrepotXfromXt} as a power series and compare it with the
general form eq.~\eqref{eq:TorsPrepotExpansion}, thereby reading off
the instanton numbers. The impatient reader can find them in
\autoref{tab:1qrb1b2Inst} on page~\pageref{tab:1qrb1b2Inst}. However, before
we come to that, we will calculate the prepotential on $X$ directly in
the next subsection. In the course of this alternative computation, we
will find that the expression eq.~\eqref{eq:PrepotXfromXt} can be
significantly simplified.

\subsection
[Directly on the Quotient $X$]
[Directly on the Quotient]
{Directly on the Quotient $\mathbf{X}$}
\label{sec:Aquotient}

Instead of working with generic \dP9 surfaces, one can also work
directly with the special surfaces in eq.~\eqref{eq:B1} and
eq.~\eqref{eq:B2}. In order to admit a vertical $G=\ZZZ$ group action,
they have a special complex structure such that
\begin{itemize}
\item There are $9$ sections, $MW(B_i)=\Z_3\oplus\Z_3$.
\item The elliptic fibration $B_i\to\CP^1$ has $4 I_3$ Kodaira fibers.
\end{itemize}
The three irreducible components of each of the four $I_3$ fibers are
permuted by the four different $\Z_3$ subgroups of $G$.  Therefore,
the quotient $X=\Xt/G$ is still fibered by Abelian surfaces, having
$4$ singular fibers of the type $T^2\times I_0$ and $4$ singular
fibers of the type $I_0 \times T^2$. We can immediately identify the
following curves on the quotient $X$:
\begin{itemize}
\item $9$ sections $s_{ij}$ in $MW(X)=\Z_3\oplus\Z_3$, all
  distinguished by $H_2(X,\Z)_\tors=\Z_3\oplus\Z_3$.
\item The fiber classes $f_1$ and $f_2$ under the two different
  elliptic fibrations. 
\end{itemize}
Following exactly the same reasoning as in \autoref{sec:Amodcurves},
one can write down the instanton contribution from the pseudo-sections
to the genus $0$ prepotential \emph{directly on the quotient $X$}. The
result is
\begin{multline}
  \FprepotXNP
  =
  \sum_{
    \begin{smallmatrix}
      s_{ij}\in \\ MW(X)
    \end{smallmatrix}}
  e^{2\pi\iunit \int_{s_{ij}} \omega}
  \left(
    \sum_{m=0}^\infty p(m) 
    e^{2\pi \iunit m \int_{f_1} \omega}
  \right)^{4}
  \left(
    \sum_{n=0}^\infty p(n) 
    e^{2\pi \iunit n \int_{f_2} \omega}
  \right)^{4}
  \\
  +
  (\text{contribution of multi-sections})
  .
\end{multline}
We now pick variables for the complexified \Kahler{} moduli
space on $X$ such that
\begin{equation}
  e^{2\pi\iunit \int_{s_{ij}} \omega}
  = p b_1^i b_2^j
  ,\quad
  e^{2\pi \iunit \int_{f_1} \omega}
  = q
  ,\quad
  e^{2\pi \iunit \int_{f_2} \omega}
  = r  
  .
\end{equation}
Expanding the prepotential in these variables, we obtain 
\begin{equation}
  \label{eq:PrepotX}
  \begin{split}
    \FprepotXNP
    (p,q,r,b_1,b_2)
    =&\;
    \left( \sum_{i,j=0}^2 p b_1^i b_2^j \right)
    P(q)^4
    P(r)^4
    +O(p^2)
    \\
    =&\;
    p
    (1+b_1+b_1^2)
    (1+b_2+b_2^2)
    P(q)^4
    P(r)^4
    +O(p^2)
    .
  \end{split}
\end{equation}

Note that this expression appears to be distinct from
eq.~\eqref{eq:PrepotXfromXt}. However, although the two formulas look
very different, they must be identical functions of $p,q,r,b_1,b_2$.
Indeed, as we now show, this is the case. Note that the difficult part
in the first expression for the prepotential is the $E_8$ theta
function in the function $A$, see eq.~\eqref{eq:Aqb1b2def}. First, let
us ignore $b_1$ and $b_2$ for the moment, that is, set $b_1=b_2=1$,
and recall~\cite{MR1672085}
\begin{theorem}[Zagier]
  \begin{equation}
    \ThetaEeight\big( 
    q^3;\, 
    q^2,\, q,\, 1,\, 1,\, 
    1,\, 1,\, 1,\, q^{-1} \big)
    P\big( q^3 \big)^{12}
    = 
    9 P(q)^4
    \quad
    \in \Z[[q]]
    .
  \end{equation}
\end{theorem}
Using this identity, we can eliminate the $E_8$ theta function from
the function $A(q,1,1)$. A short computation then shows the equality
of the two expressions for the prepotential, eqns~\eqref{eq:PrepotX}
and~\eqref{eq:PrepotXfromXt}.

Putting $b_1$ and $b_2$ back into $A(q,b_1,b_2)$, it is very
suggestive that Zagier's identity ought to be generalized to
\begin{multline}
  \label{eq:ThetaGeneralized}
  \ThetaEeight\big( 
    q^3;\, 
    q^2 b_1^2 b_2^2,\, q b_2,\, 1,\, b_1^2 b_2,\, 
    b_2^2,\, 1,\, b_1^2,\, q^{-1} \big)
  P\big( q^3 \big)^{12}
  = \\ =
  (1+b_1+b_1^2)
  (1+b_2+b_2^2)
  P(q)^4
  \\
  \in
  \Z[[q]] \otimes \Z[b_1,b_2]\big/ 
  \left<b_1^3=1,b_2^3=1\right>      
  .
\end{multline}
Using a computer, we have expanded both sides of
eq.~\eqref{eq:ThetaGeneralized} up to degree $10$ and found agreement.
This generalized identity implies the equality of the two expressions
\begin{equation}
  \begin{split}
    \Big\{ 
    \text{$p$-linear part of eq.~\eqref{eq:PrepotXfromXt}} 
    \Big\}
    =&\;
    \frac{1}{9}
    p
    A(q,b_1,b_2)
    A(r,b_1^{-1},b_2^{-1})
    \\
    =&\;
    \frac{1}{9}
    p
    \big(1+b_1+b_1^2\big)^2
    \big(1+b_2+b_2^2\big)^2
    P(q)^4
    P(r)^4
    \\
    =&\;
    p
    (1+b_1+b_1^2)
    (1+b_2+b_2^2)
    P(q)^4
    P(r)^4  
    \\
    =&\;
    \Big\{ 
    \text{$p$-linear part of eq.~\eqref{eq:PrepotX}} 
    \Big\}
  \end{split}
\end{equation}
for the genus $0$ prepotential at linear order in $p$, where we used
that $b_1^3=1=b_2^3$. We conclude that the two expressions for the
prepotential on $X$ in eqns.~\eqref{eq:PrepotX}
and~\eqref{eq:PrepotXfromXt} are indeed the same function.

Expanding our formula for the instanton generated genus $0$
prepotential as a power series and comparing it with the general form
given in eq.~\eqref{eq:TorsPrepotExpansion}, one can finally read off
the instanton numbers computed using the A-model. We will do this in
the following subsection.

\subsection{Instanton Numbers}
\label{sec:A-numbers}

Recall from eq.~\eqref{eq:HomCohXtZ3Z3} that to correctly distinguish
all homology classes of curves, we need $5$ numbers
\begin{equation}
  (n_1,n_2,n_3,m_1,m_2) \in
  \Z\oplus\Z\oplus\Z\oplus\Z_3\oplus\Z_3
  \simeq H_2\big(X,\Z\big)
  .
\end{equation}
The effect of the torsion homology classes is that, for any curve on
$X$, we can assign quantum numbers $m_1,m_2\in\{0,1,2\}$ in addition
to the degrees $n_1,n_2,n_3\in\Z$. With this in mind, and using
eq.~\eqref{eq:pqrb1b2def}, the general form of the instanton expression
eq.~\eqref{eq:TorsPrepotExpansion} becomes
\begin{equation}
  \label{eq:XTorsPrepotExpansion}
  \FprepotXNP(p,q,r,\, b_1,b_2 ) 
  =
  \sum_{\substack{n_1,n_2,n_3\in\Z \\ m_1,m_2\in\Z_3}} 
  n_{(n_1,n_2,n_3,m_1,m_2)}
  \Li_3
  \Big( p^{n_1} q^{n_2} r^{n_3} b_1^{m_1} b_2^{m_2} \Big)
  .
\end{equation}
where $n_{(n_1,n_2,n_3,m_1,m_2)}$ is the number of instantons in the
given homology class. Comparing this with the series expansion of the
formula for the prepotential, either eq.~\eqref{eq:PrepotXfromXt}
or~\eqref{eq:PrepotX}, allows us to read off the instanton numbers.

As we explained previously, our A-model computation only yielded the
genus $0$ prepotential up to linear order in $p$, that is, for
$n_1\leq 1$. The constant part in $p$ vanishes, so all of these
instanton numbers are zero,
\begin{equation}
  n_{(0,n_2,n_3,m_1,m_2)} = 0 
  \quad \forall n_2,n_3\in\Z,\, m_1,m_2\in \Z_3
  .
\end{equation}
At linear order in $p$, that is, $n_1=1$, the instanton numbers do not
vanish. Interestingly, the instanton number does not depend on the
torsion part of the homology class. That is,
\begin{equation}
  n_{(1,n_2,n_3,m_1,m_2)}
  =
  n_{(1,n_2,n_3,0,0)}
  \quad 
  \forall m_1,m_2\in \{0,1,2\}
  .
\end{equation}
The underlying reason for this is another geometric $\ZZZ$ group
action. Unlike $G\simeq \ZZZ$, this additional group acts on $X$ and
has fixed points, see \partB~\cite{PartB}, \xautoref{sec:p3check}. On
the homology classes $(1,n_2,n_3,m_1,m_2)$ its action is generated by
$m_1\mapsto (m_1+1)\mod 3$ and $m_2\mapsto (m_2+1)\mod 3$. Since the
prepotential must respect this symmetry, the corresponding instanton
numbers are equal.

We list the instanton numbers for $n_2,n_3\leq9$ in
\autoref{tab:1qrb1b2Inst}.
\begin{table}[htpb]
  \centering
  \renewcommand{\arraystretch}{1.3}
  \newcommand{\s}{\scriptstyle}
  \begin{tabular}{c|cccccccccc}
    \backslashbox{$\mathrlap{n_2}$}{$\mathclap{n_3~}$}
    &
    $0$ & $1$ & $2$ & $3$ & $4$ & $5$ & $6$ & $7$ & $8$ & $9$ 
    \\ \hline
    $0$ &
    $1$&$4$&$14$&$40$&$105$&$252$&$574$&$\s1240$&$\s2580$&$\s5180$
    \\
    $1$ &
    $4$&$16$&$56$&$160$&$420$&$\s1008$&$\s2296$&$\s4960$&$\s10320$&$\s20720$
    \\
    $2$ &
    $14$&$56$&$196$&$560$&$\s1470$&$\s3528$&$\s8036$&$\s17360$&$\s36120$&$\s72520$
    \\
    $3$ &
    $40$&$160$&$560$&$\s1600$&$\s4200$&$\s10080$&$\s22960$&$\s49600$&$\s103200$&$\s207200$
    \\
    $4$ &
    $105$&$420$&$\s1470$&$\s4200$&$\s11025$&$\s26460$&$\s60270$&$\s130200$&$\s270900$&$\s543900$
    \\
    $5$ &
    $252$&$\s1008$&$\s3528$&$\s10080$&$\s26460$&$\s63504$&$\s144648$&$\s312480$&$\s650160$&$\s1305360$
    \\
    $6$ &
    $574$&$\s2296$&$\s8036$&$\s22960$&$\s60270$&$\s144648$&$\s329476$&$\s711760$&$\s1480920$&$\s2973320$
    \\
    $7$ &
    $\s1240$&$\s4960$&$\s17360$&$\s49600$&$\s130200$&$\s312480$&$\s711760$&$\s1537600$&$\s3199200$&$\s6423200$
    \\
    $8$ &
    $\s2580$&$\s10320$&$\s36120$&$\s103200$&$\s270900$&$\s650160$&$\s1480920$&$\s3199200$&$\s6656400$&$\s13364400$
    \\
    $9$ &
    $\s5180$&$\s20720$&$\s72520$&$\s207200$&$\s543900$&$\s1305360$&$\s2973320$&$\s6423200$&$\s13364400$&$\s26832400$
  \end{tabular}
  \caption{Instanton numbers $n_{(1,n_2,n_3,\ast,\ast)}$ 
    computable in the A-model. In this case (for $n_1=1$), 
    the instanton number is independent of the 
    torsion part of the homology class.}
  \label{tab:1qrb1b2Inst}
\end{table}
Note the symmetry under the exchange $n_2\leftrightarrow n_3$. This is
already visible in the expression for the prepotential, which is
invariant under the exchange $q\leftrightarrow r$,
\begin{equation}
  \FprepotXNP
  (p,r,q,b_1,b_2)
  =
  \bigg( \sum_{i,j=0}^2 p b_1^i b_2^j \bigg)
  P(q)^4
  P(r)^4
  +O(p^2)
  =
  \FprepotXNP
  (p,q,r,b_1,b_2)
  .
\end{equation}
The underlying geometric reason is that we can exchange the factors in
the fiber product
\begin{equation}
  \Xt = B_1 \times_{\CP^1} B_2 \simeq B_2 \times_{\CP^1} B_1
  .
\end{equation}
Unwinding the definitions, one can show that this geometric exchange
corresponds precisely to the exchange of $q$ and $r$.

The instanton numbers calculated using the A-model, and presented
above, have one glaring limitation. Namely, they are restricted to
$n_1\leq 1$. That is, we can only compute the prepotential to linear
order in $p$. Using mirror symmetry, we will be able to overcome this
restriction in \partB{}~\cite{PartB}.

\subsection
[The Partial Quotient $\Xb$]
[The Partial Quotient]
{The Partial Quotient $\mathbf{\Xb}$}

Since $G=G_1\times G_2=\ZZZ$ is generated by two independent $\Z_3$
actions, there are the obvious partial quotients
\begin{equation}
  \vcenter{\xymatrix@!0@=15mm{
      & \Xt 
      \ar[dr]^{\text{mod } G_2} 
      \ar[dl]_{\text{mod } G_1} 
      \ar[dd]|{\text{mod } G}
      \\
      \mathllap{\Xb \eqdef}
      \Xt\big/G_1 \ar[dr]_{\text{mod } G_2}  
      & & 
      \Xt\big/G_2 \ar[dl]^{\text{mod } G_1}
      \\
      & X 
    }}.
\end{equation}
Having just computed the prepotential on $X$, there is little
intrinsic interest in the simpler partial quotients. However, note
that the $G_1$ quotient $\Xb\eqdef\Xt/G_1$ is again a toric variety
since $G_1$ acts only by phase rotations on the coordinates, see
eq.~\eqref{eq:g1action}. This observation will enable us to compute
the instanton numbers using the B-model, as we will in
\partB~\cite{PartB}. To this end, we will need the correct variable
substitution analogous to eq.~\eqref{eq:Z3Z3quotientVariables} but for
the final $G_2$ quotient $X=\Xb/G_2$. This is why we will analyze the
partial quotient $\Xb$ in this subsection. In the same way as for the
full $G=\ZZZ$ quotient, we can compute part of its prepotential by
properly descending $\Xt\to \Xb$.

Because we will have to compare our basis for divisors with the basis
that is natural in toric geometry, let us first have a closer look at
the $G_1$ invariant cohomology of $\Xt$.  First, the $G_1$ invariant
homology of the \dP9 surfaces is
\begin{equation}
  H_2(B_i,\Z)^{G_1} = 
  \Span_{\Z}\Big\{
  f,\,
  t,\,
  u,\,
  v
  \Big\}
  ,
\end{equation}
where $f$ and $t$ are the $G_1\times G_2$ invariant divisors, see
eq.~\eqref{eq:H2BiInv} and\footnote{At this point it is not obvious
  why we choose $2t+\theta_{11}$ instead of just $\theta_{11}$ for the
  final generator of the $G_1$-invariant cohomology. As we will see
  below, this particular basis choice is better adapted to the
  \Kahler{} cone.}
\begin{equation}
  \begin{split}
    u \eqdef&\;
    \theta_{21}+\theta_{31}+\theta_{41}+3\mu
    = 6f + 6\sigma -2\alpha_1 -\alpha_2
    \\
    v \eqdef&\;
    2t+\theta_{11}
    = 
    -3\alpha_1 + 3 \alpha_3 + 2 \alpha_4 
    + \alpha_5 + 3 \alpha_8        
  \end{split}
\end{equation}
are only $G_1$ but not $G_2$-invariant. As in
\autoref{sec:Cohinvariant}, pulling these back yields a basis for the
$G_1$-invariant divisor classes of the Calabi-Yau threefold. We define
\begin{equation}
  \upsilon_1 \eqdef \pi_1^{-1}(u)
  ,\quad
  \upsilon_2 \eqdef \pi_2^{-1}(u)
  ,\quad
  \psi_1  \eqdef \pi_1^{-1}(v)
  ,\quad
  \psi_2  \eqdef \pi_2^{-1}(v)
\end{equation}
in addition to eq.~\eqref{eq:phitaudef}. As usual, we will not
distinguish between divisors and their duals in cohomology, see
\autoref{fn:divisors}. With this abuse of notation, we obtain the basis
\begin{equation}
  \label{eq:H2XtG1inv}
  H^2\big(\Xt,\Z\big)^{G_1}
  = 
  \Span_{\Z}
  \Big\{ 
  \phi,\,
  \tau_1,\, \upsilon_1,\, \psi_1,\,
  \tau_2,\, \upsilon_2,\, \psi_2
  \Big\}
  .
\end{equation}
All products between these cohomology classes are determined by the
relations
\begin{multline}
  \label{eq:ringXtG1inv}
  H^\even\big(\Xt,\Q\big)^{G_1} = 
  \Q[\phi,\tau_1,\upsilon_1,\psi_1,\tau_2,\upsilon_2,\psi_2]
  \Big/
  \big<
    \phi^2
    ,\, 
    \tau_1 \phi = 3 \tau_1^2
    ,\,
    \tau_2 \phi = 3 \tau_2^2
    ,\, 
    \\
    \phi \upsilon_1 = 3 \tau_1^2
    ,\, 
    \phi \upsilon_2 = 3 \tau_2^2
    ,\, 
    \phi \psi_1 = 6 \tau_1^2
    ,\, 
    \phi \psi_2 = 6 \tau_2^2 
    ,\, 
    \tau_1 \upsilon_1 = 3 \tau_1^2 
    ,\, 
    \tau_2 \upsilon_2 = 3 \tau_2^2 
    ,\, 
    \\
    \tau_1 \psi_1 = 3 \tau_1^2 
    ,\, 
    \tau_2 \psi_2 = 3 \tau_2^2 
    ,\, 
    \upsilon_1 \upsilon_1 = 3 \tau_1^2 
    ,\, 
    \upsilon_2 \upsilon_2 = 3 \tau_2^2 
    ,\, 
    \upsilon_1 \psi_1 = 6 \tau_1^2
    ,\, 
    \upsilon_2 \psi_2 = 6 \tau_2^2 
    ,\, 
    \\
    \psi_1 \psi_1 = 6 \tau_1^2
    ,\, 
    \psi_2 \psi_2 = 6 \tau_2^2
    ,\, 
    (\tau_1-\upsilon_1) (\tau_2-\upsilon_2)
    ,\, 
    (2 \upsilon_1-\psi_1) (2 \upsilon_2-\psi_2)
    ,\, 
    \\
    (2 \upsilon_1-\psi_1) (2 \tau_2-\psi_2)
    ,\, 
    (2 \upsilon_2-\psi_2) (\tau_1-\upsilon_1) 
  \big>
  .
\end{multline}
Using the above relations, we find that any triple intersection can be
rewritten as a multiple of $\tau_1^2\tau_2=3\ptset$. Therefore, the
non-vanishing intersection numbers are
\begin{equation}
  \label{eq:cubicformXtG1inv}
  \begin{aligned}
    \phi \tau_1 \tau_2 =&\; 9 &
    \phi \tau_1 \upsilon_2 =&\; 9 &
    \phi \tau_1 \psi_2 =&\; 18 &
    \phi \upsilon_1 \tau_2 =&\; 9 &
    \phi \upsilon_1 \upsilon_2 =&\; 9 \\
    \phi \upsilon_1 \psi_2 =&\; 18 &
    \phi \psi_1 \tau_2 =&\; 18 &
    \phi \psi_1 \upsilon_2 =&\; 18 &
    \phi \psi_1 \psi_2 =&\; 36 &
    \tau_1^2 \tau_2 =&\; 3 \\
    \tau_1^2 \upsilon_2 =&\; 3 &
    \tau_1^2 \psi_2 =&\; 6 &
    \tau_1 \upsilon_1 \tau_2 =&\; 9 &
    \tau_1 \upsilon_1 \upsilon_2 =&\; 9 &
    \tau_1 \upsilon_1 \psi_2 =&\; 18 \\
    \tau_1 \psi_1 \tau_2 =&\; 9 &
    \tau_1 \psi_1 \upsilon_2 =&\; 9 &
    \tau_1 \psi_1 \psi_2 =&\; 18 &
    \tau_1 \tau_2^2 =&\; 3 &
    \tau_1 \tau_2 \upsilon_2 =&\; 9 \\
    \tau_1 \tau_2 \psi_2 =&\; 9 &
    \tau_1 \upsilon_2^2 =&\; 9 &
    \tau_1 \upsilon_2 \psi_2 =&\; 18 &
    \tau_1 \psi_2^2 =&\; 18 &
    \upsilon_1^2 \tau_2 =&\; 9 \\
    \upsilon_1^2 \upsilon_2 =&\; 9 &
    \upsilon_1^2 \psi_2 =&\; 18 &
    \upsilon_1 \psi_1 \tau_2 =&\; 18 &
    \upsilon_1 \psi_1 \upsilon_2 =&\; 18 &
    \upsilon_1 \psi_1 \psi_2 =&\; 36 \\
    \upsilon_1 \tau_2^2 =&\; 3 &
    \upsilon_1 \tau_2 \upsilon_2 =&\; 9 &
    \upsilon_1 \tau_2 \psi_2 =&\; 9 &
    \upsilon_1 \upsilon_2^2 =&\; 9 &
    \upsilon_1 \upsilon_2 \psi_2 =&\; 18 \\
    \upsilon_1 \psi_2^2 =&\; 18 &
    \psi_1^2 \tau_2 =&\; 18 &
    \psi_1^2 \upsilon_2 =&\; 18 &
    \psi_1^2 \psi_2 =&\; 36 &
    \psi_1 \tau_2^2 =&\; 6 \\
    \psi_1 \tau_2 \upsilon_2 =&\; 18 &
    \psi_1 \tau_2 \psi_2 =&\; 18 &
    \psi_1 \upsilon_2^2 =&\; 18 &
    \psi_1 \upsilon_2 \psi_2 =&\; 36 &
    \psi_1 \psi_2^2 =&\; 36
    .
  \end{aligned}
\end{equation}

The $G_1$-invariant \Kahler{} cone on $B_i$ consists of the potential
\Kahler{} classes in $H^2(B_i,\Z)^{G_1}$. It can be
computed~\cite{Gomez:2005ii} as the dual of the cone of effective
curves on $B_i$. The effective curves are~\cite{MR632841}
\begin{theorem}[Looijenga]
  The cone of effective curves on a \dP9 surface $B$ is generated by
  the following curve classes $e\in H_2(B_i,\Z)$:
  \begin{enumerate}
  \item The exceptional curves ($e^2=-1$). These are the elements of
    the \MWgrp{} $MW(B)$. 
  \item The irreducible components of singular Kodaira fibers
    ($e^2=-2$). 
  \item The ``future cone'' of the positive classes ($e^2\geq 1$). 
  \end{enumerate}
\end{theorem}
For the $\ZZZ$-symmetric \dP9 surfaces $B_1$, $B_2$ that we are
interested in, the \MWgrp{} consists of the $9$ elements given in
eq.~\eqref{eq:B1MWgrp}. Furthermore, the $4I_3$ Kodaira fibers have
$12$ irreducible components $\theta_{10},\dots,\theta_{42}$.  The
positive classes do not yield any extra constraints on the dual cone.
The \Kahler{} cone
\begin{equation}
  \Kcone(B_i)^{G_1} 
  = 
  \Span_{\R_>} 
  \Big\{
  \kappa_1,\,
  \kappa_2,\,
  \kappa_3,\,
  \kappa_4,\,
  \kappa_5,\,
  \kappa_6,\,
  \kappa_7,\,
  \kappa_8
  \Big\}
  \quad\subset H^2\big(B_i,\Z\big)^{G_1}
\end{equation}
turns out to be non-simplicial with edges 
\begin{equation}
  \begin{gathered}
    \kappa_1 \eqdef
    f
    \qquad
    \kappa_2 \eqdef
    t
    \qquad
    \kappa_3 \eqdef
    u
    \qquad
    \kappa_4 \eqdef
    v
    \\
    \begin{aligned}
    \kappa_5 \eqdef&\;
    3t+f-v
    &\qquad
    \kappa_6 \eqdef&\;
    3t+u-v    
    \\
    \kappa_7 \eqdef&\;
    f-u+v
    &
    \kappa_8 \eqdef&\;
    3t+f-u
    .      
    \end{aligned}
  \end{gathered}
\end{equation}
For future reference we note that the intersection matrix of the
\Kahler{} cone generators on $B_i$ is
\begin{equation}
  \begin{array}{c|cccccccc} 
    (-)\cdot(-) & 
    \kappa_1 & \kappa_2 & \kappa_3 & \kappa_4 & 
    \kappa_5 & \kappa_6 & \kappa_7 & \kappa_8 
    \\ \hline
    \kappa_1 & 0&3&3&6&3&6&3&6\\
    \kappa_2 & 3&1&3&3&3&3&3&3\\
    \kappa_3 & 3&3&3&6&6&6&6&9\\
    \kappa_4 & 6&3&6&6&9&9&6&9\\
    \kappa_5 & 3&3&6&9&3&6&6&6\\
    \kappa_6 & 6&3&6&9&6&6&9&9\\
    \kappa_7 & 3&3&6&6&6&9&3&6\\
    \kappa_8 & 6&3&9&9&6&9&6&6
  \end{array}
\end{equation}
We note that $G_1$ and $G_2$ commute. Hence, $G_2$ acts on the
$G_1$-invariant homology and \Kahler{} cone. Using the explicit group
action, see eq.~\eqref{eq:Bg1g2matrix}, one finds
\begin{equation}
  \label{eq:Bg1invg2act}
  g_2
  \begin{pmatrix}
    f \\ t \\ u \\ v 
  \end{pmatrix}
  = 
  \begin{pmatrix}
    1 & 0 & 0 & 0 \\
    0 & 1 & 0 & 0 \\
    1 & 3 & 0 &-1 \\
    0 & 3 & 1 &-1 \\
  \end{pmatrix}
  \begin{pmatrix}
    f \\ t \\ u \\ v 
  \end{pmatrix}
\end{equation}
and 
\begin{equation}
  \Big( 
  \kappa_1, \kappa_2, \kappa_3, \kappa_4, 
  \kappa_5, \kappa_6, \kappa_7, \kappa_8
  \Big)
  \stackrel{g_2}{\mapsto}
  \Big( 
  \kappa_1, \kappa_2, \kappa_5, \kappa_6, 
  \kappa_7, \kappa_8, \kappa_3, \kappa_4
  \Big)
\end{equation}
Using the \Kahler{} cone on the base \dP9 surfaces, the \Kahler{} cone
on $\Xb$ is finally found~\cite{Gomez:2005ii} to be
\begin{equation}
  \begin{split}
    \Kcone\big(\Xb\big)
    =&\; 
    \Kcone\big(\Xt\big)^{G_1}
    \\ =&\;
    \Span_{\R_>} \Big\{
    \phi,\, 
    \tau_1,\, 
    \pi_1^\ast(\kappa_3),\dots, \pi_1^\ast(\kappa_8),~
    \tau_2,\,
    \pi_2^\ast(\kappa_3),\dots, \pi_2^\ast(\kappa_8)
    \Big\}
    .
  \end{split}  
\end{equation}

Let us now return to the instanton counting on $\Xb=\Xt/G_1$. Recall
from eq.~\eqref{eq:HomCohXtZ3} that
\begin{equation}
  H_2\big( \Xb, \Z \big) = \Z^7 \oplus \Z_3
  .
\end{equation}
Using the same trick as in \autoref{sec:quotientA}, we can determine
the prepotential on $\Xb$. We pick restricted \Kahler{} moduli
\begin{equation}
  \omega = 
  t^1_R \phi + 
  t^2_R \tau_1 + t^3_R \upsilon_1 + t^4_R \psi_1 +
  t^5_R \tau_2 + t^6_R \upsilon_2 + t^7_R \psi_2 +
  t^8_R \beta_1
\end{equation}
corresponding to a basis\footnote{Note that the $7$ generators $\phi$,
  $\tau_1$, $\upsilon_1$, $\psi_1$, $\tau_2$, $\upsilon_2$, $\psi_2$
  are the edges of one maximal simplicial subcone of the \Kahler{}
  cone. This ensures again that the Fourier series of the prepotential
  will only contain positive powers.} for the $G_1$-invariant
cohomology, see eq.~\eqref{eq:H2XtG1inv}, and one additional generator
$\beta_1$ which detects the generator of
\begin{equation}
  H_2\big(\Xt,\Z\big)_{G_1,\tors} = 
  \Z_3 = 
  H_2\big(\Xb,\Z\big)_\tors
  ,
\end{equation}
see eq.~\eqref{eq:betadef}. The Fourier transformed variables, which
we will use in the following, are
\begin{equation}
  \label{eq:PQRdef}
  \begin{gathered}
    P \eqdef e^{2\pi\iunit t^1_R}
    ,
    \\
    Q_1 \eqdef e^{2\pi\iunit t^2_R}
    ,\quad
    Q_2 \eqdef e^{2\pi\iunit t^3_R}
    ,\quad
    Q_3 \eqdef e^{2\pi\iunit t^4_R}
    ,
    \\
    R_1 \eqdef e^{2\pi\iunit t^5_R}
    ,\quad
    R_2 \eqdef e^{2\pi\iunit t^6_R}
    ,\quad
    R_3 \eqdef e^{2\pi\iunit t^7_R},
    ,
  \end{gathered}  
\end{equation}
and
\begin{equation}
  b_1 \eqdef e^{2\pi\iunit t^8_R}
  ,
\end{equation}
where
\begin{equation}
  b_1^3=1
  .
\end{equation}
The relations between the restricted variables and the full $19$
variables are
\begin{equation}
  \label{eq:G1quotientVariables}
  \begin{gathered}
    p_0 = P Q_1^5 Q_2^6 R_1^5 R_2^6
    \\
    \begin{gathered}
      q_0 = Q_1^5 Q_2^6
      \\
      \begin{aligned}
        q_1 =&\; Q_1^{-2} Q_2^{-2} Q_3^{-3} b_1  & 
        q_2 =&\; Q_1^{-1} Q_2^{-1}             \\
        q_3 =&\; Q_3^3                       & 
        q_4 =&\; Q_3^2 b_1                   \\
        q_5 =&\; Q_3                         & 
        q_6 =&\; 1                          \\
        q_7 =&\; b_1                         & 
        q_8 =&\; Q_1 Q_3^3              
      \end{aligned}
    \end{gathered}
    \qquad
    \begin{gathered}
      r_0 = R_1^5 R_2^6   
      \\
      \begin{aligned}
        r_1 =&\; R_1^{-2} R_2^{-2} R_3^{-3} b_1^2  & 
        r_2 =&\; R_1^{-1} R_2^{-1}               \\
        r_3 =&\; R_3^3                         & 
        r_4 =&\; R_3^2 b_1^2                   \\
        r_5 =&\; R_3                           & 
        r_6 =&\; 1                            \\
        r_7 =&\; b_1^2                         & 
        r_8 =&\; R_1 R_3^3                
        .
      \end{aligned}
    \end{gathered}
  \end{gathered}
\end{equation}
As done previously for the full quotient, we now substitute these
variables into the formula for the prepotential on the covering space
$\Xt$, see eq.~\eqref{eq:Aprepot}, and divide by $|G_1|=3$. The result
is
\begin{equation}
  \begin{split}
  \label{eq:PrepotXG1fromXt}
  &\FprepotNP{\Xb}
  (P,Q_1,Q_2,Q_3,R_1,R_2,R_3, b_1) 
  =
  \\ &\qquad =
  \frac{1}{|G_1|}~ \FprepotXtNP
  (p,q_0,\dots,q_8,p_0,\dots,p_8 )
  \\ &\qquad =
  \frac{1}{3}
  P\,
  \Ab(Q_1,Q_2,Q_3,b_1)
  \Ab(R_1,R_2,R_3,b_1^{-1})
  +
  O(P^2)
  ,    
  \end{split}
\end{equation}
where 
\begin{multline}
  \Ab(Q_1,Q_2,Q_3,b_1)
  \eqdef
  \ThetaEeight\Big( 
    Q_1^3 Q_2^3 Q_3^6 ;\, 
    Q_1^2 Q_2^2 Q_3^3 b_1^2 ,\, 
    Q_1 Q_2 ,\, 
    Q_3^{-3},\, 
  \\ 
    Q_3^{-2} b_1^2 ,\, 
    Q_3^{-1},\, 
    1,\, 
    b_1^2,\, 
    Q_1^{-1} Q_3^{-3} \Big)
  \,
  P\big( Q_1^3 Q_2^3 Q_3^6 \big)^{12}  
\end{multline}
and the analogous expression for $\Ab(R_1,R_2,R_3,b_1^{-1})$.
Expanding $\Ab(Q_1,Q_2,Q_3,b_1)$ as a power series, we find
\begin{multline}
  \label{eq:AXG1modform}
    \Ab(Q_1,Q_2,Q_3,b_1)
    =
    (1+b_1+b_1^2) 
    \times
    \\
  \begin{aligned}
    \times
    \big(
    1+&\;
    Q_2+
    Q_2 Q_3+
    Q_1 Q_2 Q_3+
    3 Q_1 Q_2 Q_3^2+
    3 Q_1 Q_2^2 Q_3^2+
    Q_1 Q_2 Q_3^3+
    \\ +&\;
    Q_1^2 Q_2 Q_3^3+
    3 Q_1 Q_2^2 Q_3^3+
    3 Q_1^2 Q_2^2 Q_3^3+
    Q_1^2 Q_2 Q_3^4+
    Q_1 Q_2^3 Q_3^3+
    \\ +&\;
    Q_1^2 Q_2^3 Q_3^3+
    9 Q_1^2 Q_2^2 Q_3^4+
    9 Q_1^2 Q_2^3 Q_3^4+
    3 Q_1^2 Q_2^2 Q_3^5+
    \\ +&\;
    3 Q_1^3 Q_2^2 Q_3^5+
    Q_1^2 Q_2^4 Q_3^4+
    9 Q_1^2 Q_2^3 Q_3^5+
    \\ +&\;
    9 Q_1^3 Q_2^3 Q_3^5+
    3 Q_1^3 Q_2^2 Q_3^6+
    3 Q_1^2 Q_2^4 Q_3^5+
    Q_1^2 Q_2^3 Q_3^6+
    \\ +&\;
    3 Q_1^3 Q_2^4 Q_3^5+
    25 Q_1^3 Q_2^3 Q_3^6+
    Q_1^2 Q_2^4 Q_3^6+
    \\ +&\;
    (\text{total degree}\geq 13)
    \big)
    \qquad 
    \in \Z[[Q_1,Q_2,Q_3]] \otimes 
    \Z[b_1]\big/ \left<b_1^3=1\right>
  \end{aligned}
\end{multline}
Finally, we note that we can now compute the prepotential on $X=\Xt/G$
in terms of the prepotential on $\Xb=\Xt/G_1$. One can easily show
that the correct substitution of variables is
\begin{equation}
  \label{eq:G2quotient}
  \begin{gathered}
    P = p
    \\
    \begin{aligned}
      Q_1 =&\; q \\
      Q_2 =&\; b_2 \\
      Q_3 =&\; b_2 
    \end{aligned}
    \qquad
    \begin{aligned}
      R_1 =&\; r \\
      R_2 =&\; b_2^2 \\
      R_3 =&\; b_2^2 
      .
    \end{aligned}    
  \end{gathered}
\end{equation}
Obviously one obtains exactly the same as prepotential as in
eq.~\eqref{eq:PrepotXfromXt}, where we divided out $G=G_1\times G_2$
in one step rather than first $G_1$ and then $G_2$.  However, as we
will show in the companion paper \partB, one can use toric mirror
symmetry to compute any desired term in the prepotential on $\Xb$.
Knowing the above substitution, eq.~\eqref{eq:G2quotient}, will enable
us to find the prepotential on $X=\Xb/G_2$ beyond linear order in $p$,
including its $b_2$ torsion expansion.



\section{Conclusion}
\label{sec:conclusion}

The goal of this paper is to investigate rational curves on the
Calabi-Yau threefold $X$, which is the $G=\ZZZ$ quotient of its
universal cover $\Xt$. Its Hodge numbers and integral homology are
\begin{equation}
  h^{p,q}\big(X\big) = ~
  \vcenter{\xymatrix@!0@=7mm@ur{
    1 &  0  &  0  & 1 \\
    0 &  3  &  3  & 0 \\
    0 &  3  &  3  & 0 \\
    1 &  0  &  0  & 1 
  }}
  , \quad
  H_i\big(X, \Z\big) 
  \simeq
  \begin{cases}
    \Z
    & i=6 \\
    0
    & i=5 \\
    \Z^{3} \oplus \big(\Z_3\big)^2
    & i=4 \\
    \Z^{8} \oplus \big(\Z_3\big)^2
    & i=3 \\
    \Z^{3} \oplus \big(\Z_3\big)^2
    & i=2 \\
    \big(\Z_3\big)^2
    & i=1 \\
    \Z
    & i=0
    .
  \end{cases}
\end{equation}
Interestingly, this is one of the few known examples of Calabi-Yau
manifolds whose degree-$2$ homology has a finite part
(\textdef{torsion}). The prepotential is a function of the $3$ free
generators $p,q,r$ and the $2$ torsion generators $b_1,b_2$. We found
a closed formula for the genus zero prepotential
\begin{equation}
  \FprepotXNP
  (p,q,r,b_1,b_2)
  =
  \bigg( \sum_{i,j=0}^2 p b_1^i b_2^j \bigg)
  P(q)^4
  P(r)^4
  +O(p^2)
  = 
  \sum_{i,j=0}^2 \Li_3(p b_1^i b_2^j) 
  + \cdots
\end{equation}
to linear order in $p$. This allows us to derive part of the instanton
numbers on $X$, distinguishing the torsion part of the curve class in
the integral homology. The corresponding instantons are listed in
\autoref{tab:1qrb1b2Inst} on page~\pageref{tab:1qrb1b2Inst}.

Clearly, we would like to obtain the complete prepotential and not
just up to linear order in $p$. However, this is very difficult to do
directly. In \partB{}~\cite{PartB}, we will use mirror symmetry to
attack this problem. There, we will find a way to obtain the higher
order terms as well. The final result, limited only by computing
power, will be
\begin{equation}
  \begin{split}
    \FprepotXNP&(p,q,r,b_1,b_2) 
    = 
    \FprepotNP{X^\ast}(p,q,r,b_1,b_2)
    \\ =&\;
    \sum_{i,j=0}^2
    \Big(
    \begin{array}[t]{ll}
        \Li_3(p b_1^i b_2^j)
      + 4  \Li_3(p q b_1^i b_2^j)
      + 4  \Li_3(p r b_1^i b_2^j)
      \\~
      + 14  \Li_3(p q^2 b_1^i b_2^j)
      + 16  \Li_3(p q r b_1^i b_2^j)
      + 14  \Li_3(p r^2 b_1^i b_2^j)
      \\~
      + 40  \Li_3(p q^3 b_1^i b_2^j)
      + 56  \Li_3(p q^2 r b_1^i b_2^j)
      + 56  \Li_3(p q r^2 b_1^i b_2^j)
      \\~
      + 40  \Li_3(p r^3 b_1^i b_2^j)
      + 105 \Li_3(p q^4 b_1^i b_2^j)
         + 160 \Li_3(p q^3 r b_1^i b_2^j)
      \\~
      + 196 \Li_3(p q^2 r^2 b_1^i b_2^j)
         + 160 \Li_3(p q r^3 b_1^i b_2^j)
         + 105 \Li_3(p r^4 b_1^i b_2^j)
      \\~
      -2 \Li_3(p^2 q b_1^i b_2^j)
      -2 \Li_3(p^2 r b_1^i b_2^j)
      -28 \Li_3(p^2 q^2 b_1^i b_2^j)
      \\~
      +32 \Li_3(p^2 q r b_1^i b_2^j)
      -28 \Li_3(p^2 r^2 b_1^i b_2^j)
      -192 \Li_3(p^2 q^3 b_1^i b_2^j)
      \\~
      +440 \Li_3(p^2 q^2 r b_1^i b_2^j)
      +440 \Li_3(p^2 q r^2 b_1^i b_2^j)
      -192 \Li_3(p^2 r^3 b_1^i b_2^j)
      \hspace{2ex}\smash{\Big)}
    \end{array}
    \\ &+
    3 \Li_3(p^3 q ) + 3 \Li_3(p^3 r) 
    \\ &+
    9 \Li_3(p^3 q^2 ) + 
    27 \sum_{(i,j)\not=(0,0)} \Li_3(p^3 q^2 b_1^i b_2^j)
    \\ &+
    9 \Li_3(p^3 r^2 ) + 
    27 \sum_{(i,j)\not=(0,0)} \Li_3(p^3 r^2 b_1^i b_2^j)
    \\ &+
    27 \Li_3(p^3 q r ) + 
    81 \sum_{(i,j)\not=(0,0)} \Li_3(p^3 q r b_1^i b_2^j)
    \\ &+ 
    \big(\text{total }p,q,r\text{-degree }\geq 6\big)
    .
  \end{split}  
\end{equation}
This provides some interesting examples of instanton numbers that do
depend on the torsion part of their homology class, see
\autoref{tab:n1n2n3b1b2Inst}.
\begin{table}[htpb]
  \centering
  \renewcommand{\arraystretch}{1.3}
  \newcommand{\s}{\scriptstyle}
  \newcommand{\sss}{\hspace{5mm}}
  \begin{tabular}{@{\sss}c@{\sss}@{\hspace{10mm}}@{\sss}c@{\sss}}
    $n_{(3,n_2,n_3,0,0)}$ &
    $n_{(3,n_2,n_3,m_1,m_2)},~(m_1,m_2)\not=(0,0)$
    \\
    \begin{tabular}{c|ccc}
      \backslashbox{$\mathrlap{n_2}$}{$\mathclap{n_3~}$}
      &
      $0$ & $1$ & $2$ 
      \\ \hline
      $0$ &
      $0$&$\mathemph{3}$&$\mathemph{36}$
      \\
      $1$ &
      $\mathemph{3}$&$\mathemph{108}$
      \\
      $2$ &
      $\mathemph{36}$
    \end{tabular}
    &
    \begin{tabular}{c|ccc}
      \backslashbox{$\mathrlap{n_2}$}{$\mathclap{n_3~}$}
      &
      $0$ & $1$ & $2$ 
      \\ \hline
      $0$ &
      $0$&$\mathemph{0}$&$\mathemph{27}$
      \\
      $1$ &
      $\mathemph{0}$&$\mathemph{81}$
      \\
      $2$ &
      $\mathemph{27}$
    \end{tabular}
  \end{tabular}
  \caption{Some of the instanton numbers $n_{(n_1,n_2,n_3,m_1,m_2)}$ computed 
    by mirror symmetry. The entries marked in
    \textcolor{red}{\textbf{bold}} depend non-trivially 
    on the torsion part of their 
    respective homology class.}
  \label{tab:n1n2n3b1b2Inst}
\end{table}

\section*{Acknowledgments}

The authors would like to thank Albrecht Klemm, Tony Pantev, and
Masa-Hiko Saito for valuable discussions. We also thank Johanna Knapp
for providing a Singular~\cite{GPS05} code to compute the intersection
ring of Calabi-Yau manifolds in toric varieties.
\GrantsAcknowledgements
E.~S. thanks the Math/Physics Research group at the University of
Pennsylvania for kind hospitality.

\appendix

\makeatletter
\def\Hy@chapterstring{section}
\makeatother

\section{Duology}
\label{sec:dualappendix}

\subsection{Poincar\'e Duality and Equalities}
\label{sec:dualpoincare}

For any closed, connected, oriented $d$-dimensional manifold $Y$ there
are non-singular\footnote{A bilinear map is non-singular if, when
  written in terms of integral bases, it is represented by a square
  matrix of determinant $1$.} pairings
\begin{equation}
  \begin{aligned}
    H_k\big(Y,\Z\big)_\free \times H^k\big(Y,\Z\big)_\free 
    \;&\to \Z
    ,&\quad
    (S,\varphi) 
    \mapsto&\; 
    \int_S \varphi
    ,\\
    H^k\big(Y,\Z\big)_\free 
    \times H^{d-k}\big(Y,\Z\big)_\free 
    \;&\to \Z
    ,&
    (\varphi,\psi) 
    \mapsto&\; 
    \int_Y \varphi \wedge \psi
    ,\\
    H_k\big(Y,\Z\big)_\free 
    \times H_{d-k}\big(Y,\Z\big)_\free 
    \;&\to \Z
    ,&
    (M,N) 
    \mapsto&\; 
    M\cdot N
    .
  \end{aligned}
\end{equation}
The consequence is that the corresponding (co)homology groups are of
the same rank. Moreover, if a group $G$ acts orientation-preservingly
on $Y$ then the corresponding (co)homology groups are dual
$G$-representations.

However, the ``best'' version of Poincar\'e duality identifies
homology and cohomology including torsion, and is a map
\begin{equation}
  \label{eq:PD}
  PD:~
  H^k\big(Y,\Z\big) \isolongrightarrow H_{d-k}\big(Y,\Z\big)
  ,\quad
  \varphi\mapsto [Y]\cap \varphi
  .
\end{equation}
This map $PD$ is an isomorphism; by abuse of notation we will denote
the inverse by $PD$ as well. In full generality, the map $PD$ is the
cap-product with the fundamental class. Ignoring torsion, we can also
describe $PD$ on the level of differential forms as follows: Consider
a $(d-k)$-dimensional submanifold $S\subset Y$. Then the $k$-form
$PD(S)$ is the Thom class of the normal bundle $N_{Y|S}$, that is, a
bump $k$-form along the normal directions of $S$. Note that $PD$ does
not involve any duality. If there is an orientation-preserving
$G$-action on $Y$, then $H^k(Y,\Z)\simeq H_{d-k}(Y,\Z)$ are isomorphic
group representations.

\subsection{Tate Duality}
\label{sec:dualtate}

Looking at the result for $\ZZZ$ group (co)homology in
eq.~\eqref{eq:HZ3Z3result}, there seems to be the following relation 
\begin{equation}
  \label{eq:HomCohomRdual}
  H_i\big(G,R^\dual\big)_\tors
  \simeq 
  H^{i+1}\big(G,R\big)_\tors
\end{equation}
between group homology and group cohomology. In fact, this is a
general property known as Tate duality. Recall that the Tate cohomology
groups unify group homology and cohomology into
\begin{equation}
  \Htate^i(G,M) = 
  \begin{cases}
    H^i(G,M) & i>0 \\
    M^G/{(\tr)M} & i=0 \\
    \ker(\tr)/{IM} & i=-1 \\
    H_{-i-1}(G,M) & i<-1
    ,
  \end{cases}
\end{equation}
where $M$ is any $G$-module. If $M$ is $\Z$-torsion free, that is, a
representation of $G$ on a lattice $\Z^n$,
then~\cite{LangGroupCohomology}
\begin{equation}
  \Htate^{ i}\big( G, \Hom(M,\Z) \big) 
  \simeq 
  \Hom\left[ 
    \Htate^{-i}( G, M)
    , \Q/\Z
  \right]
\end{equation}
In particular, setting $M=R$ proves eq.~\eqref{eq:HomCohomRdual}.

\section{Relations Amongst Divisors}
\label{sec:relations}

In \autoref{sec:dP9}, eq.~\eqref{eq:Bbasis} we chose one particular
basis for the homology of the \dP9 surfaces, namely
\begin{equation}
  H_2(B_i,\Z) = 
  \Span_\Z\Big\{
  \sigma,\,
  f,\,
  \theta_{11},\,
  \theta_{21},\,
  \theta_{31},\,
  \theta_{32},\,
  \theta_{41},\,
  \theta_{42},\,
  \mu,\,
  \nu
  \Big\}
  .
\end{equation}
In this appendix we give the expansion of the other curves of interest
in terms of this chosen basis. The expansion of any other curve can be
found using its intersection numbers with the $10$ base curves.

The $9$ sections forming the \MWgrp{}
intersect the vertical divisors according to
eq.~\eqref{eq:thetaaction}, and they do not intersect amongst
themselves. Hence,
\begin{equation}
  \begin{split}
    \sigma =&\;
    \sigma
    , \\
    \mu =&\; 
    \mu
    , \\
    \mu \boxplus \mu =&\;
    -\sigma-f+\theta_{21}+\theta_{31}+\theta_{41}+2\mu
    , \\
    \nu =&\;
    \nu
    , \\
    \nu\boxplus\mu =&\;
    -\sigma-f+\theta_{31}+\theta_{32}+\theta_{41}+\mu+\nu
    , \\
    \nu\boxplus\mu\boxplus\mu =&\;
    -2\sigma-2f+\theta_{21}+\theta_{31}+\theta_{32}
    +2\theta_{41}+\theta_{42}+2\mu+\nu
    , \\
    \nu\boxplus\nu =&\;
    -\sigma-f+\theta_{11}+\theta_{32}+\theta_{41}+2\nu
    , \\
    \nu\boxplus\nu\boxplus\mu =&\;
    -2\sigma-2f+\theta_{11}+\theta_{31}+\theta_{32}
    +2\theta_{41}+\theta_{42}+\mu+2\nu
    , \\
    \nu\boxplus\nu\boxplus\mu\boxplus\mu =&\;
    -3\sigma-3f+\theta_{11}+\theta_{21}+2\theta_{31}
    +2\theta_{32}+2\theta_{41}+\theta_{42}+2\mu+2\nu
    .
  \end{split}
\end{equation}
Finally, the components of $i=1,\dots,4$ distinct $I_3$ Kodaira fibers
intersect as
\begin{equation}
  \begin{array}{c|ccc}
    (-)\cdot (-) & \theta_{i0} & \theta_{i1} & \theta_{i2} \\ 
    \hline
    \theta_{i0} & -2 &  1 &  1   \\
    \theta_{i1} &  1 & -2 &  1   \\
    \theta_{i2} &  1 &  1 & -2   
    .
  \end{array}
\end{equation}
This lets us express the two components $\theta_{12}$, $\theta_{22}$
that are not part of our chosen basis as
\begin{equation}
  \begin{split}
    \theta_{12} =&\;
    3\sigma+3f-2\theta_{11}-\theta_{31}-2\theta_{32}
    -2\theta_{41}-\theta_{42}-3\nu
    , \\
    \theta_{22} =&\; 
    3\sigma+3f-2\theta_{21}-2\theta_{31}-\theta_{32}
    -2\theta_{41}-\theta_{42}-3\mu
    .
  \end{split}
\end{equation}

\section{Image of Group Homology}
\label{sec:image}

The purpose of this appendix is to find the image
\begin{equation}
  \Z_3 \simeq
  H_3\Big( G_{12};\, \Z\Big) 
  \longrightarrow
  H_3\Big( G;\, \Z\Big) 
  \simeq
  \Z_3 \oplus \Z_3 \oplus \Z_3
  \,.
\end{equation}
The obvious way to get an explicit handle on this map is to extend the
inclusion $\Z G_{12}\subset \Z G$ to a chain map of the corresponding
resolutions of $\Z$. Applying $-\otimes\Z$ to the resolution then
makes the image of the homology group clear. 

To write down the resolution, define the following \textbf{t}race and
\textbf{d}ifference maps in the group ring:
\begin{equation}
  t_1\eqdef \sum_{i=0}^2 \big(g_1\big)^i
  \,,\quad
  t_2\eqdef \sum_{i=0}^2 \big(g_2\big)^i
  \,,\quad
  d_1\eqdef 1-g_1
  \,,\quad
  d_2\eqdef 1-g_2
  \,.
\end{equation}
Using these, we write down the following chain map between the
resolutions. From that, one can easily determine the pushforward of
the homology groups as
\begin{equation}
  \label{eq:resolutions}
  \vcenter{
      \xymatrix@M+1mm@C=0mm@R=3mm{
      \Z G_{12} \ar[rr]^{\sum (g_1g_2)^i} \ar[dd]|{\left(
          \cdots
        \right)}
      & &
      \Z G_{12} \ar[rr]^{1-g_1g_2} \ar[dd]|{\left(
          \begin{smallmatrix}
            1 & g_1 & g_1^2 & 1
          \end{smallmatrix}
        \right)}
      & &
      \Z G_{12} \ar[rr]^{\sum (g_1g_2)^i} \ar[dd]|{\left(
          \begin{smallmatrix}
            1 & 1+g_1+g_1g_2 & g_1^2
          \end{smallmatrix}
        \right)}
      & &
      \Z G_{12} \ar[rr]^{1-g_1g_2} \ar[dd]|{\left(
          \begin{smallmatrix}
            g_2 & 1
          \end{smallmatrix}
        \right)}
      & &
      \Z G_{12} 
      \ar@{^{(}->}[dd]
      \\
      \\
      \oplus_5 \Z G
      \ar[rr]_{\left(
          \begin{smallmatrix}
            t_1 & 0 & 0 & 0 \\
            -d_2 & d_1 & 0 & 0 \\
            0 & t_2 & t_1 & 0 \\
            0 & 0 & -d_2 & d_1 \\
            0 & 0 & 0 & t_2
          \end{smallmatrix}
        \right)}
      & &
      \oplus_4 \Z G 
      \ar[rr]_{\left(
           \begin{smallmatrix}
             d_1 & 0 & 0 \\
             d_2 & t_1 & 0 \\
             0 & -d_2 & d_1 \\
             0 & 0 & h_2
           \end{smallmatrix}
        \right)}
      & &
      \oplus_3 \Z G 
      \ar[rr]_{\left(
          \begin{smallmatrix}
            t_1 & 0 \\ 
            -d_2 & d_1 \\
            0 & t_2
          \end{smallmatrix}
        \right)}
      & &
      \oplus_2 \Z G 
      \ar[rr]_{\left(
          \begin{smallmatrix}
            d_1 \\ d_2 
          \end{smallmatrix}
        \right)}
      & &
      \oplus_1 \Z G 
      \\
      \\
      &\hspace{18mm}& &\hspace{13mm}& 
      \hbox{\LARGE$\Downarrow$}
      \ar@{}[rrrr]|(0.6){
        \text{Apply } 
        \left(-\otimes_{\Z G_{12}}\Z\right) 
        \text{ resp. } 
        \left(-\otimes_{\Z G}\Z\right)
      }
      &\hspace{8mm}& &\hspace{3mm}&
      \\
      \Z \ar[rr]^{3} \ar[dd]|{\left(
          \cdots
        \right)}& &
      \Z \ar[rr]^{0} \ar[dd]|{\left(
          \begin{smallmatrix}
            1 & 1 & 1 & 1
          \end{smallmatrix}
        \right)}& &
      \Z \ar[rr]^{3} \ar[dd]|{\left(
          \begin{smallmatrix}
            1 & 3 & 1
          \end{smallmatrix}
        \right)}& &
      \Z \ar[rr]^{0} \ar[dd]|{\left(
          \begin{smallmatrix}
            1 & 1
          \end{smallmatrix}
        \right)}& &
      \Z  
      \ar@{=}[dd]
      \\
      \\
      \Z^5
      \ar[rr]|{\left(
          \begin{smallmatrix}
            3 & 0 & 0 & 0 \\
            0 & 0 & 0 & 0 \\
            0 & 3 & 3 & 0 \\
            0 & 0 & 0 & 0 \\
            0 & 0 & 0 & 3
          \end{smallmatrix}
        \right)}&&
      \Z^4
      \ar[rr]|{\left(
           \begin{smallmatrix}
             0 & 0  & 0 \\
             0 & 3  & 0 \\
             0 & -3 & 0 \\
             0 & 0  & 0
           \end{smallmatrix}
        \right)}&&
      \Z^3
      \ar[rr]|{\left(
          \begin{smallmatrix}
            3 & 0 \\ 
            0 & 0 \\
            0 & 3
          \end{smallmatrix}
        \right)}&&
      \Z^2
      \ar[rr]|{\left(
          \begin{smallmatrix}
            0 \\ 0
          \end{smallmatrix}
        \right)}&&
      \Z
      \\
      \\
      && && 
      \save[]*{
        \hbox{\LARGE$\Downarrow$}
      }\restore
      \ar@{}[rrrr]|(0.27){
        \text{Homology} 
      }
      && && 
      \\
      0 \ar@{=}[d]
      &&
      \Z_3 \ar@{=}[d]
      &&
      0 \ar@{=}[d]
      &&
      \Z_3 \ar@{=}[d]
      &&
      \Z \ar@{=}[d]
      \\
      *+<1mm>{\strut} \save[]*{
        H_4\big(G_{12};\Z\big)
      } \restore
      \ar[dd]
      &&
      *+<1mm>{\strut} \save[]*{
        H_3\big(G_{12};\Z\big) 
      } \restore
      \ar[dd]^{\left(
          \begin{smallmatrix}
            1 & 1 & 1
          \end{smallmatrix}
        \right)}
      &&
      *+<1mm>{\strut} \save[]*{
        H_2\big(G_{12};\Z\big) 
      } \restore
      \ar[dd] 
      &&
      *+<1mm>{\strut} \save[]*{
        H_1\big(G_{12};\Z\big) 
        \qquad
      } \restore
      \ar[dd]^{\left(
          \begin{smallmatrix}
            1 & 1
          \end{smallmatrix}
        \right)}
      &&
      *+<1mm>{\strut} \save[]*{
        H_0\big(G_{12};\Z\big) 
      } \restore
      \ar@{=}[dd]
      \\
      \\
      *+<1mm>{\strut} \save[]*{
        H_4\big(G;\Z\big) 
      } \restore
      &&
      *+<1mm>{\strut} \save[]*{
        H_3\big(G;\Z\big) 
      } \restore
      &&
      *+<1mm>{\strut} \save[]*{
        H_2\big(G;\Z\big) 
      } \restore
      &&
      *+<1mm>{\strut} \save[]*{
        H_1\big(G;\Z\big) 
      \quad
      } \restore
      &&
      *+<1mm>{\strut} \save[]*{
        H_0\big(G;\Z\big) 
      } \restore
      \\
      \big(\Z_3\big)^2 \ar@{=}[u]
      &&
      \big(\Z_3\big)^3 \ar@{=}[u]
      &&
      \Z_3 \ar@{=}[u]
      &&
      \big(\Z_3\big)^2 \ar@{=}[u]
      &&
      \Z \ar@{=}[u]
      \,.
    }
  }
\end{equation}
It is much easier to determine the image under the inclusion
$G_1\subset G$ and $G_2\subset G$. Using the same bases as in
eq.~\eqref{eq:resolutions}, they are
\begin{equation}
  \vcenter{\xymatrix@R=1mm@C+10mm{
      H_3\big(G_1;\Z\big) =
      \Z_3
      \ar[r]^{\left(
        \begin{smallmatrix}
          1 & 0 & 0
        \end{smallmatrix}
      \right)}
      &
      \big(\Z_3\big)^3 =
      H_3\big(G;\Z\big) 
      \\
      H_3\big(G_2;\Z\big) =
      \Z_3
      \ar[r]^{\left(
        \begin{smallmatrix}
          0 & 0 & 1
        \end{smallmatrix}
      \right)}
      &
      \big(\Z_3\big)^3 =
      H_3\big(G;\Z\big) 
      \,.
  }}
\end{equation}


\bibliographystyle{utcaps} \renewcommand{\refname}{Bibliography}
\addcontentsline{toc}{section}{Bibliography} 
\bibliography{Volker,Emanuel}

\end{document}